\documentclass[12pt,a4paper]{book}
\usepackage{mathtools}
\usepackage{graphicx}
\usepackage{color}
\graphicspath{ {images/} }
\usepackage{amsmath}
\usepackage{amssymb}
\usepackage{accents} 
\usepackage{amsbsy}
\usepackage{amsfonts}
\usepackage{amsthm}
\usepackage{color}
\usepackage{fancyhdr}
\usepackage{subfigure}
\usepackage[top=2.8cm, bottom=2.9cm, left=4.0cm, right=2.5cm]{geometry}
\usepackage{graphicx}
\usepackage{slashed}
\usepackage{pdfpages}
\usepackage{notoccite}
\usepackage{cite}
\usepackage[pdftex,
            breaklinks=true,
            pdfauthor={Arun Kumar Pandey},
            pdftitle={Generation and evolution of primodial magnetic field and it's cosmological consequences},
            pdfsubject={Science --> Physics --> Cosmology},
            pdfkeywords={Cosmology, Electoweak Phase transition, Baryon Assymetry of Universe, Hypercharge magnetic field, Chiral plasma, Kinetic theory of plasma, Berry curveture, Magnetohydrodynamics,  CMB},
            pdfproducer={Latex with hyperref},
            pdfcreator={latexmk}]{hyperref}
\makeatletter
\def\cleardoublepage{\clearpage\if@twoside \ifodd\c@page\else
  \hbox{}
  \thispagestyle{empty}
  \newpage
  \if@twocolumn\hbox{}\newpage\fi\fi\fi}
\makeatother
\hypersetup{
      colorlinks=true,
      citecolor=blue,
      linkcolor=blue,
      urlcolor=blue}
  \usepackage[autostyle=false, style=english]{csquotes} 
  \MakeOuterQuote{"}
  \renewcommand{\vec}[1]{\mathbf{#1}}

\begin{document}
\thispagestyle{empty}
\begin{titlepage}
\begin{center}
\Large{\textbf{Origin and dynamics of the Primordial Magnetic field in a parity violating plasma}}\\
\vspace*{1.0 cm}
\large{{A thesis submitted in partial fulfilment of}}\\
\large{{the requirements for the degree of}} \\
\vspace*{0.2 cm}
\Large{\textbf{Doctor of Philosophy}}\\
\vspace*{0.2 cm}
 {\it{by}} \\
\vspace*{0.2 cm}
{\Large{\textbf{Arun Kumar Pandey}}} \\
\vspace*{0.5 cm}
{\large{Under the guidance of}}\\
{\Large \textbf{Prof. Jitesh R. Bhatt}}\\
\vspace*{-0.2 cm}
{\large{Theoretical Physics Division}}\\
\vspace*{-0.2 cm}
{\large{Physical Research Laboratory, Ahmedabad, India.}}\\
\vspace*{0.5 cm}
\begin{figure}[h]
\centering
 \includegraphics[width=3.5cm]{iitgn_logo.pdf}
\end{figure}
{\large{DISCIPLINE OF PHYSICS}}\\
{\small{INDIAN INSTITUTE OF TECHNOLOGY GANDHINAGAR, GUJARAT, INDIA}}\\
{\large{2017}}
\end{center}
\end{titlepage}

\cleardoublepage
%



\pagenumbering{roman}
\setcounter{page}{1}
\thispagestyle{empty}
\phantomsection

\newpage
\phantomsection
\thispagestyle{empty}
\addcontentsline{toc}{chapter}{Abstract}
\markboth{\MakeUppercase{Abstract}}{\MakeUppercase{Abstract}}
\begin{center}
	  {\LARGE \textbf{Abstract}}   
\end{center}
The Universe is magnetized on all scales that we have observed so far: stars, galaxies, cluster of galaxies etc. 
Recent observations indicates that typical magnetic field strength in a galaxy or a galaxy-cluster can have is about a few $\mu$G and its coherent length is around ten kpc. Recent observations suggest that the intergalactic medium (IGM) have magnetic field in the range of ($10^{-16}-10^{-9}$) G with the coherent length scales around Mpc. Though it might be possible to explain the observed magnetic field of the galaxies and stars by some kind of astrophysical process, it is hard to explain the observed magnetic field of an IGM void. There exists an intriguing possibility of relating the origin of these large scale magnetic fields with some high-energy process in the early Universe.  Thus it is of great interests to study the origin, dynamics and constraints on the magnetic field generated by such mechanism. This forms the prime focus of this thesis which considers the above problem in the context of a high-energy parity violating plasma. 

In recent times there has been considerable interest in studying the magnetic field evolution in high-energy parity violating plasma. It is argued that there can be more right-handed particles over left-handed particles due to some process in the early Universe at temperatures $T$ very much
higher than the electroweak phase transition (EWPT) scale ($T \sim$100GeV). Their number is effectively conserved
at the energy scales much above the electroweak phase transitions and this allows one to introduce the chiral
chemical potentials $\mu_R(\mu_L)$. However , at temperature lower than $T\sim 80$~TeV  processes related
with the electron chirality flipping  may dominate over the Hubble expansion rate and the chiral chemical potentials can not be defined. Further the right handed current is not conserved due to the Abelian anomaly in the standard model (SM) and it is the number density of the right handed particles are related with the helicity of the fields as $\partial_\tau (\Delta \mu +\frac{\alpha_2}{\pi}\,\mathcal{H}_B)=0$. (Here $\Delta \mu$, $\alpha_2$ and $\mathcal{H}_B$ represents the chiral chemical imbalance, fine structure constant and magnetic helicity respectively). Therefore even if initially when there is no magnetic field, a magnetic field can be generated at the cost of the asymmetry in the number density of the left and right handed particles in the plasma. Recently, it has been interesting development in incorporating the parity -violating effects into a kinetic theory formalism by including the effect of the Berry curvature. Berry curvature term takes into account chirality of the particles. By using this modified  kinetic equation, we show that chiral-imbalance leads to generation of hypercharge magnetic field in the plasma in both the collision dominated and collisionless regimes. We show that in the collision dominated regime, chiral vortical effects can generate a chiral vorticity and magnetic field. Typical strength of the magnetic field in the collision and collisionless regime are $10^{27}$ G at $10^5/T$ length scale and a magnetic field of strength $10^{31}$ G at $10/T$ length scale at a temperature $T\sim 80$ TeV. We also show that the estimated values are consistent with the present observations.  

We also show that in the presence of chiral imbalance and gravitational anomaly, a magnetic field of the strength $10^{30}$ G can be generated at a scale of $10^{-18}$ cm (much smaller than the Hubble length scale i.e. $10^{-8}$ cm). The idea is that in the presence of the gravitational  anomaly, the current expression for the chiral plasma consists of a vortical current, proportional to square of temperature i.e. $T^2$. The silent feature of this seed magnetic field is that it can be generated even in absence of chiral charge. In this work we have considered the scaling symmetry of the chiral plasma to obtain the velocity  spectrum. Under this scaling symmetry in presence of gravitational anomaly, amount of energy at any given length scale is much larger than the case where only chiral asymmetry is considered. We also show that under such scenario energy is transferred from large to small length scale, which is commonly known as inverse cascade. 

In the context of chiral plasma at high energies, there exist new kind of collective modes [{\it e.g.}, the chiral Alf\'ven waves] in the MHD limit and these modes exist in addition to the usual modes in the standard parity even plasma. Moreover, it has been shown that chiral plasmas exhibit a new type of density waves in presence of either of  an external magnetic field or vorticity and  they are respectively known as Chiral Magnetic Wave (CMW) and the Chiral Vortical Wave (CVW). In this regard we have investigated the collective modes in a magnetized chiral plasma and the damping mechanisms of these modes using first order and second hydrodynamics. Using first order conformal hydro,  we obtained previously derived modes in the chiral plasma. However we show in addition that these modes get split into two modes in presence of the first order viscous term. By using second order conformal magnetohydrodynamics, we show that there are a series of terms in the dispersion relation and these terms are in accordance with the results obtained using AdS/CFT correspondence. We also calculated one of the transport coefficients related with the second order magnetohydrodynamics.  
~\\~\\
{\bf Keywords:} Cosmology, Primordial Magnetic Field, EW phase transition, Left Right symmetric model, Turbulence theory, Kinetic theory, CMB, BBN
%
\newpage
\phantomsection
\addcontentsline{toc}{chapter}{Contents}
\tableofcontents

\newpage
\phantomsection
\addcontentsline{toc}{chapter}{List of Figures}
\listoffigures


\pagenumbering{arabic}
\setcounter{page}{1}
\chapter*{List of terms used in the thesis}\label{list}
\begin{tabular}{ |p{3cm}|p{9cm}|}
	\hline
	Short name/ Symbol  & Definition / meaning \\
	\hline
	IGM & Inter Galactic Medium\\
	ICM & Inter Cluster Medium \\
	EGMF & enter galactic magnetic field \\
	MHD &  Magnetohydrodynamics\\
	CMB  & Cosmic Microwave Background \\
	BBM & Biermann Battery Mechanism \\
	BBN & Big Bang Baryogenesis \\
	CME & Chiral Magnetic Effect \\
	ChMHD & Chiral Magnetohydrodynamics \\
	CVE & Chiral Vortical Effect \\
	CVW & Chiral Vortical Waves\\
	CAW & Chiral \'Alfven Waves\\
	CMW & Chiral Magnetic Waves\\
	CSE & Chiral separation effect\\
	EM & Electromagnetic  \\
	SM & Standard Model \\
	EW & Electroweak \\
	GUT & Grand Unified Theory \\
	QCD & Quantum Chromodynamics\\
	BAU & Baryon Asymmetry of the Universe\\
	\hline
\end{tabular}
\newpage
\begin{tabular}{ |p{3cm}|p{9cm}|}
	\hline
	Short name/ Symbol  & Definition / meaning \\
	\hline
	n D & n dimensional space\\
	kpc/Mpc & kiloparsec/megaparsec\\
	eV & electron volt \\
	keV & kilo electron volt \\
	MeV & mega electron volt \\
	GeV & giga electron bolt \\
	H & Hubble constant \\
	$\rho$ & Matter energy density\\
	$\rho_\gamma$ & Radiation energy density \\
	$\boldsymbol{\omega}$ & vorticity vector \\
	$m_p$ & mass of the ions \\
	$k_B$ & Boltzmann's coefficient \\
	$c$ & speed of light \\
	$\sigma$ & electrical conductivity \\
	$\alpha$ & fine structure constant \\
	$m_e$ & electron mass \\
	$\mu$ & micro \\
	$\hbar$ & reduced Planck constant ($\hbar= h/2\pi$) \\
	$\text{M}_{pl}$ & Planck Mass\\
	$G$ & Gauss\\
	$\Re$ & set of Real number\\
	\hline
\end{tabular}
\cleardoublepage
\newpage
\chapter{Introduction}\label{ch1}
Magnetic fields are present on almost all scales, from stars, galaxies to intergalactic medium. A field strength of few $\mu$~G has been observed in galaxies at kpc (Kilo Parsec) length scale. However in the intra-cluster medium (ICM) a magnetic field of strength of $10^{-16}$ G at a coherence length scale of the order of $(10-100)$ kpc can be present \cite{Kim1991}. One of the unsettled issues of the modern cosmology and astrophysics is the origin of  these magnetic fields. One understanding is that these fields could have been originated during structure formation or in the early stages of Universe. However, one of the intriguing possibility is that they are relic from the early Universe and subsequently amplified and maintained by a dynamo during structure formation. During structure formation, it is necessary to have a seed magnetic field of $B \sim 10^{-20}$~G to explain currently observed magnetic field \cite{Davis1999, ruzmaikin2013magnetic}. However a  magnetic field of the order of $B \sim 10^{-9}$ G can explain the observed galactic magnetic fields \cite{Banerjee2004}. In the former scenario, seed magnetic fields get amplified via dynamo mechanism  \cite{Davis1999, Subramanian1994, Kulsrud1997, Armando1997}, while in the latter scenario it is not necessary to have a dynamo mechanism for amplification. This chapter provides the overview of current understanding of the astrophysical and cosmological magnetic fields. 

\section{Magnetogenesis and magnetic fields} 
Models for generation of the magnetic field is broadly classified into two classes:
%
\begin{itemize}
	\item Astrophysical models
	\item Models based on early universe processes
\end{itemize}	
We can separate creation and amplification mechanism based on the early universe model into three process namely 
\begin{enumerate}
	\item Before recombination
	\item During recombination
	\item After recombination
\end{enumerate}
Astrophysical models mainly comes into the category of processes after recombination. However few authors have discussed generation of primordial magnetic field due to some astrophysical phenomenon during recombination era \cite{Naoz2013}  (for review see the reference \cite{Durrer2013}).
\subsection{Astrophysical models}
Post recombination processes include explanation for the amplification of the seed fields in galaxies by the  dynamo \cite{Grasso2001, Davies2000, ruzmaikin1990} (Review Ref. \cite{Widrow2002}). Models based on astrophysical process can be divided into two categories \cite{Durrer2013}: 
a bottom-up (astrophysical processes) scenario, where the seed field is typically very weak and the observed large-scale magnetic field is transported from local sources within galaxies to larger scales \cite{Kulsrud:2007an}, and a top-down (cosmological processes) scenario where a significant seed field is generated prior to galaxy formation in the early universe on a scale of a cosmological interest  \cite{Kandus2011xy}. The "{\it Biermann Battery process}" (BBM)\cite{Beirmann1950}  is one process, which discuss generation of magnetic field in the prototype galaxies, clusters or on shock fronts \cite{Davies2000, Ando2010, Gnedin:2000ax, Cho2014}). Harrison in \cite{Harrison1970, Harrison1973}, discussed another generation  mechanism. The idea was that, in presence of plasma vortical motion during radiation dominated era, can produce a sufficient large magnetic field.
\subsubsection{Generation from vortical motion of the plasma during radiation dominated era}
Harrison mechanism \cite{Harrison1968, Harrison1967avc, Harrison1970, Harrison1973} for generation of magnetic fields depend on the  transfer of angular momentum between the ion and electron-photon gases in the expanding radiation dominated plasma. This can be achieved by considering 
that rotational velocities  of the ion and the electrons decreases differently in the expanding background before recombination takes place. 
The generating mechanism depends on the transfer of angular momentum between the ion and electron-photon gases in the expanding radiation dominated plasma. To understand this behaviour, let us take a simple example of a spherically symmetric region of radius $r$, uniformly rotating, and containing radiation of uniform density $\rho_\gamma$, and matter of uniform density $\rho$ consisting of ions and non-relativistic electrons. As the eddy expands, both energy in the matter and the radiation in the regime of radius $r$, {\it i.e.} $\rho r^3$ and $\rho_\gamma r^4$ remains constant. Assume that the angular velocities of matter and radiation are $\omega$ and $\omega_\gamma$; then in the absence of interactions the angular momentum proportional to $\rho \omega r^5$ and $\rho_\gamma \omega_\gamma r^5$ are separately conserved. Hence,  $\omega \propto a^{-2}$ and  ${\bf \omega}_{\gamma} \propto a^{-1}$. So as eddy expands, radiation spins down more slowly than matter. Here $a$ is scale factor for the expanding Universe. Above assumption was based on the difference in the effectiveness of Thomson scattering for electrons than ions. The Thomson cross-section is much larger for the electron than for ions and the electrons tend therefore to be dragged along by the photon gas.  In the radiation dominated era when $\rho_\gamma >\rho$, the electron-photon coupling is relatively tight. Therefore there are only two fluids: one positively charged ion gas of density $\rho$ and relatively charged photon gas of density $\rho_\gamma$.The differences in the angular velocities of the ion and electron-photon gases in an expanding eddy generated a magnetic field, which is follows following equation \cite{Harrison1970}
\begin{equation}
	\frac{\partial }{\partial \tau}\left(a^2 \boldsymbol{\omega}+\frac{e}{m_p}{\bf B}\right)=\frac{e}{4\pi \sigma m_p}\nabla^2{\bf B}.
\end{equation}
where $m_p$ and $\tau$ are mass of the ions and conformal time respectively. If we assume that there exist some initial vorticity prior to the decoupling, then above equation can be solved and the magnetic field can be related to the initial vorticity as
\begin{equation}
	{\bf B}\approx -\frac{m_p}{e}\left(\frac{a_i}{a}\right)^2 \,\boldsymbol{\omega}_i
\end{equation}
So, it is clear that if an appropriate value of initial vorticity is assumed, current observed bounds of magnetic field strength can be obtained. In this work, Harrison showed that in an expanding eddy, a magnetic field of strength ${\bf B}=-2\cdot 1\times 10^{-4}~\boldsymbol{\omega}$ G can be generated in the case of a hydrogen plasma. Primordial turbulence, if present at the epoch of recombination, can lead to a magnetic field of strength $10^{-8}$ G on a length scale of $1$ Mpc at current time.
%
In the context of topological defects, it has been argued that the velocity gradients induced as matter flows in the baryonic wakes, can generate vorticity required for the generation of magnetic fields \cite{Vachaspati1991}.
Vilenkin \cite{Vilenkin1978xcr} has also shown, in a different scenario, that the vorticity can be induced in a parity violating processes in the Weinberg-Salam model of the electroweak interactions, as macroscopic parity violating currents may develop a vortical motion in the thermal background.
Even though this process can generate magnetic field at a length scale of cosmological interest, it was realized (see ref. \cite{Vilenkin1978xcr}) that currently observed magnetic field in the proto-galaxies can not be obtained by the processes discussed above. 
As a consequence of the Helmholtz-Kelvin circulation theorem,  small deviation from homogeneity and isotropy can not generate large rotational perturbations near the initial singularity as compared to the scalar and tensor perturbations.
%
Rebhan et. al. \cite{Rebhan:1992} showed that the existence of collisionless matter exist in the very beginning of the Universe, 
can lead to growing modes of the vorticity on super-horizon scale.
The lack of too large anisotropies in the Cosmic Microwave Background (CMB) spectrum constrains the amount of primordial vorticity. 
The upper bound on the strength of magnetic field at a length scale of $L$  at present time, consistent with the CMB observations, is \cite{Rebhan:1992} 
%
\begin{equation}
	B_0(L)<3\times 10^{-18} h^{-2}L^{-3}_{Mpc} ~G,
\end{equation}
and this field acts as a seed for galactic dynamo. 

Another mechanism that can explain the generation  of seed magnetic fields is Biermann battery Mechanism \cite{BIERMANN1950}, which we are going to discuss in the next section.
%
\subsubsection{Biermann Battery process}
The Biermann mechanism, which was first discussed in the context of stellar, exploits the difference between the accelerations of electrons and ions which is triggered by their pressure gradients. This would in return generate electric current and hence the magnetic field.
Force balance on a single electron in the fluid can be given as
\begin{gather}
	n_e m_e \left(\frac{d\, {\bf v}}{dt}\right)=-n_e e {\bf E}- \nabla p_e -\frac{e n_e}{c} ({\bf v}\times {\bf B})
	\label{electronbalanceeq}
\end{gather}
where $e$, $n_e$, $m_e$, and $p_e$ respectively denote charge, number density, mass of the electrons and electron pressure in the fluid. 
%
Electron pressure in the fluid is defined as $n_e k_B T$. 
We can drop any term with $m_e$, as it is very small compared to the others terms. 
In this case, right hand side of the eq.  (\ref{electronbalanceeq}) will be non-zero, while left hand side can be set to zero. %
Taking curl of equation  (\ref{electronbalanceeq}), under above assumption and using Maxwell equation, one can obtain
\begin{equation}
	\frac{d {\bf B}}{dt} =\nabla \times({\bf v}\times {\bf B}) -c\frac{\nabla n_e \times \nabla p_e}{e n_e^2 }\label{bb1.3}
\end{equation}
%
%
Second term on the right hand side of eq. (\ref{bb1.3}) is known as Biermann Battery term and has been well exploited to explain the generation of magnetic field \cite{Beirmann1950}. Note that this term will work only when  $\nabla p_e$ and $\nabla n_e$ are not parallel to each other. Davies and Widrow in \cite{Davies2000}, showed that in presence of vorticity in the plasma, Biermann Battery term gives a non zero value. Consider the evolution of a fluid described by Euler equation
\begin{equation}
	\frac{d{\bf v}}{d t}+({\bf v}\cdot \nabla){\bf v}=-\frac{1}{\rho}\nabla p,
\end{equation}
where $\rho$ is the energy and $p$  is the pressure  density of the fluid. Let's take the curl of this equation, and substitute $\boldsymbol{\omega}=\nabla \times {\bf v}$
\begin{equation}
	\frac{d \boldsymbol{\omega}}{d t}-\nabla \times({\bf v}\times \boldsymbol{\omega})=\frac{\nabla \rho\times \nabla p}{\rho^2}.\label{BBMvorticity}
\end{equation}
In the limit of very small fluid velocity, the above equation takes the following form
\begin{equation}
	\frac{d \boldsymbol{\omega}}{d t}=\frac{\nabla \rho\times \nabla p}{\rho^2}.\label{vorevol1.8}
\end{equation}  
Going back to magnetic field evolution equation (\ref{bb1.3}), in the limit of no background field  and very small $v$
\begin{equation}
	\frac{d{\bf B}}{dt}=-c\frac{\nabla n_e \times \nabla p_e}{e n_e^2 }
\end{equation}
Assuming local charge neutrality and thermal equilibrium, 
we find
\begin{equation}
	\frac{d{\bf B}}{dt}=-\alpha \frac{\nabla \rho\times \nabla p}{\rho^2},
\end{equation}
where $\alpha=m_pc/e(1+\xi)$ with $m_p$ being the proton mass and $\xi$ being the ionization fraction. Therefore, from above two equations, one can write ${\bf B}=\alpha ~ \boldsymbol{\omega}$.
%
%
Under these conditions, the growth of the magnetic fields mimics the growth of vorticity. By this mechanism, magnetic field can be generated with initial conditions $t=0$ and $B=0$. One should note that, Biermann Battery term works only when $\nabla \rho\times\nabla p\neq 0$. The power of the Biermann Battery is realized in the early universe through shocks and subsequent vorticity  \cite{Davies2000}. This idea has been well exploited  to generate the magnetic field on astrophysical scale (Kpc scale).
\subsubsection{Dynamo Mechanism}
There are several alternative scenarios available in the literature, that discuss generation of magnetic seed fields using battery-type mechanism during the post-recombination era. 
Basic understanding about these theories are that the magnetic fields are generated at the cost of the kinetic energy of the turbulent motions of the conductive interstellar medium \cite{Kulsrud:2007an}.
Once seed fields of sufficient strength are generated, they grow due to dynamos for example in the galaxies. For galactic dynamo to work, following ingredients are required
%
%
%
\begin{enumerate}
	\item Small but non vanishing, conductivity of the interstellar medium,
	\item Turbulent motion of the fluid,
	\item The  differential rotation of the galactic system.
\end{enumerate}
Qualitatively, the dynamo action can be understand by a reasonable differential equation. Steenbeck {\it et. al.} in 1966 \cite{Steenbeck:1966} first time derived a differential equation describing dynamo mechanism.Later many authors explored dynamos in different astrophysical objects \cite{Parker1970, vainshtein1992a}.
To bring out the physics, a simple discussion is given below.

The generated magnetic field follows induction equation
\begin{equation}\label{eq:indu123}
	\frac{\partial {\bf B}}{\partial t}=\nabla\times({\bf u}\times{\bf B})+\eta\nabla^2{\bf B}.
\end{equation}
%
The velocity and magnetic fields can be written as ${\bf v}$ and ${\bf B}$ respectively as

\begin{eqnarray}
	{\bf u}=\bar{\bf u}+{\bf v}\nonumber\\
	{\bf B}=\bar{\bf B}+{\bf b} \nonumber
\end{eqnarray}
where $|\bar{\bf B}|\gg |{\bf b}|$ and $|\bar{\bf u}|\gg |{\bf v}|$ represents ensemble average and  
$\eta$ is the magnetic diffusivity. Averaging of the induction equation (\ref{eq:indu123}) over the small scale fluctuations, one can obtain the evolution of mean field as
%
\begin{equation}
	\frac{\partial \bar{\bf B}}{\partial t}=\nabla\times[\bar{\bf u}\times\bar{\bf B}]+\nabla\times[\overline{{\bf v}\times{\bf b}}]+\eta\nabla^2\bar{\bf B}.\label{backgrB}
\end{equation}
We emphasize that for dynamo mechanism to work, the second term on the right hand side is crucial. Evolution of the time-varying component is
\begin{eqnarray}
	\frac{\partial {\bf b}}{\partial t}=\nabla\times[{\bf v}\times\bar{B}]+\nabla\times[\bar{\bf u}\times{\bf b}]+\nabla\times[{\bf v}\times{\bf b}-\overline{{\bf v}\times{\bf b}}]+\eta\nabla^2{\bf b}.\label{overlnineB}
\end{eqnarray}
We can drop term with $\bar{\bf u}\times{\bf b}$, as it can be removed by a Galilean transformation. This equation can be solved for each realization of the turbulent motions, and the ensemble average of ${\bf v}\times{\bf b}$ is to be calculated at each point. Assuming isotropic helical turbulence, one can write $\overline{{\bf v}\times{\bf b}}=-c_{1} \bar{\bf B}+c_2\nabla \times \bar{\bf B}$, where $c_1=(\overline{{\bf v}\cdot\nabla\times{\bf v}})\tau_v/3$ and $c_2=(\overline{{\bf v}\cdot{\bf v}})\tau_v/3$ ~ \cite{Kulsrud1999}. Here $\tau_v$ is the velocity correlation time of the turbulence. The dynamo equation then becomes
\begin{eqnarray}
	\frac{\partial \bar{\bf B}}{\partial t}=\nabla\times[\bar{\bf u}\times\bar{\bf B}] + \nabla\times(\alpha\bar{\bf B})+(\beta+\eta)\nabla^2\bar{\bf B}.
\end{eqnarray}
$c_2$ represents turbulent diffusivity and acts on the mean field. For a fast dynamo, $c_2\gg \eta$. The effect of first term is known as $\omega$-effect and it vanishes if there is no differential rotation. The term with $c_1$, represents the effect of the cyclonic motions on the mean field. $c_1$ effect corresponds to a mean turbulent electromotive force parallel to magnetic field. This effect deforms a straight magnetic field into a helix. This effect survive only when there is finite fluid helicity. It has been shown in Ref. \cite{Kulsrud1999}, that the few modes of the magnetic field grow exponentially. 
Several authors have questioned the operational efficiency of the galactic dynamo, as dynamo mechanism itself could be shut down before the generation of large scale magnetic fields. Such a termination of the dynamo mechanism could happen when small scale magnetic fields are strongly amplified and come to equipartition with the turbulent motions which in turn leads to the no net transfer of kinetic energy into magnetic energy.
This early termination can be due to the strong amplification of small scale magnetic fields which may come to equipartition with the turbulent motions resulting in the end of transfer of kinetic energy into magnetic energy. Furthermore, the origin of the seed fields which are needed by the dynamo amplification, are not explained by the dynamo theory itself, but must be assumed to preexist. 
In a recent work, S. Naoz and R. Narayan \cite{Naoz2013}, showed that the vorticity induced by temperature fluctuations, which are scale dependent, or spatially varying sound  speed in the gas on linear scale can lead to the generation of  primordial fields.  
This process operates at the time of recombination and works even when there is relative velocity between dark matter and baryonic matter. This modifies the predicted spatial power spectrum of the magnetic field. 
They estimated the strength of the magnetic field  to be ($10^{-25}-10^{-24}$) G on a comoving scales $\sim 10$ kpc. 
\subsection{Models based on early Universe processes}
Models based on the astrophysical processes can satisfactorily explain the magnetic field in astrophysical objects but fails to explain large scale magnetic fields due to its large correlation length scale. Therefore we expect that the cosmic magnetic are generated in the early Universe. 
In literature, there are many mechanism available, but one can broadly classify them in following three categories
\begin{enumerate}
	\item Inflationary scenario: when conformal invariance of the electromagnetic interaction are broken \cite{Dolgov1993a, Turner_88, Ratra1992, Gasperini_95, Giovannini2000b, yokoyama2008, Subramanian2010, Kandus2011xy, Durrer2013, Fujita2016}.
	\item Phase transition: Bubble collision during the first order phase transitions in the early universe \cite{Hogan1983}. Topological defects, cosmic strings, domain wall etc are also related with phase transitions \cite{Vachaspati1991}.
	\item Asymmetries in the early Universe \cite{Dolgov1993, Joyce1997} 
\end{enumerate}
%
Based on this categorization, there are many models available, for example during  reheating period after inflation  \cite{Carroll1990, Giovannini2000b}, cosmological defects \cite{Vilenkin:1991zk, Vachaspati1991,Vilenkin1997} phase transitions \cite{Vachaspati_91b, Enqvist1993, Kibble_95, Quashnock_89}, electroweak Abelian anomaly \cite{Joyce1997, Cornwall1997},  temporary electric charge non conservation \cite{Dolgov1993a}, trace anomaly\cite{Dolgov1993a} or breaking gauge invariance \cite{Turner_88}.
Below we discussed some of the important processes, by which cosmic magnetic fields can be generated.
\subsubsection{Generation during Inflation}
Problem with the magnetic fields  generated after inflation and before recombination era was that fields had very small coherence length scale, and so may not have importance on cosmological length scale ($\sim$ Mpc). One of the fundamental reason behind this is that the coherence scale of the fields are within the horizon at the time of magnetogenesis. One possibility is that these fields are generated during inflation. As inflation can inflate the scale of the generated magnetic fields to the horizon size, they provide an ideal set up for the generation of primordial magnetic fields \cite{Turner_88}.
Although model based on inflation explains the existing large coherence length of the magnetic field, yet it fails to generate the enough strength of the magnetic field to seed the galactic dynamos. In particular, B-fields that has survived a period of standard de-Sitter inflation are typically too weak to sustain the galactic dynamo. Turner and Widrow in \cite{Turner_88} showed that  magnetic field of astronomical interest (of large scale $\sim$ Mpc) could be produced during this period of accelerated expansion provided the conformal symmetry of the electromagnetic theory is broken. 
Inflation provides following important ingredients for the production of primeval magnetic fields 
\cite{Turner_88}
\begin{enumerate}
	\item Inflation provides the kinematic means of producing very-long wavelength effects at very early times through micro-physical processes operating on scales less than the Hubble scale. An electromagnetic wave with wavelength larger than Hubble length scale i.e., $\lambda \gtrsim H^{-1}$, appears as static fields. This will leads to large-scale 
	\item If there is mechanism to break the conformal invariance of the electromagnetic action then magnetic field could be excited during the expansion phase of the Universe. Modes with length scale $\lambda\lesssim H^{-1}$ could be excited from the quantum fluctuations. However due to expansion of the Universe length scale also get stretched and the energy density of the modes with length scale $\lambda\simeq H^{-1}$ follows $d\rho/dk\sim H^3$.
	\item During the inflationary period conductivity of the Universe was negligible due to absence of charge particles. However during reheating period huge number of charged particles were produced. Therefore, magnetic fields produced during inflationary period will be frozen.
\end{enumerate}
Turner and Widrow also \cite{Turner_88} proposed three methods of breaking the conformal invariance, which are as follows:
\begin{itemize}
	\item {\it Coupling of the photon to the gravitational field :} 
	A term of following form $R A_\mu A^\mu$, $R_{\mu\nu}A^\mu A^\nu$, $R_{\mu\nu\lambda\kappa}F^{\mu\nu}F^{\lambda\kappa}/m^2$, $R F^{\mu\nu}F_{\mu\nu}$, (where  $A_\mu$ is the electromagnetic field, $F^{\mu\nu}$ is field strength tensor, $R_{\mu\nu\lambda\kappa}$ is Ricci tensor of fourth order,  $R_{\mu\nu}$ is the Ricci tensor, $R$ is the curvature scalar and  $m$  denotes the mass scale required by dimensional considerations) leads to breaking of conformal invariance.
	As a consequence of the conformal symmetry breaking the photons gain an effective time-dependent mass.
	\item {\it Coupling of the photon to a charged scalar field} 
	\cite{Denardo:1980yv, Shore:1979as, Toms:1980sx},
	\item {\it Anomalous coupling of the axions and photons} \cite{Cheng:1995fd}.
\end{itemize}
During inflationary period, quantum vacuum fluctuations of the electromagnetic fields can be excited within the horizon and later when they leave the horizon, they transformed to the classical fluctuations. Due to rapid expansion of the Universe, the existing charged particles, if any, diluted drastically which leads to the negligible conductivity of the Universe. However, during reheating inflaton decays to produce huge amount of charged particles and hence the plasma becomes almost perfect conductor. As a consequence the magnetic flux gets frozen in the plasma. 

One of the most general form of the action may include coupling of the EM fields with inflatons, scalar fields or with charged fields. Moreover if there exist extra-spatial dimensions with scale factor $b(t)$ then one of the  possible action in $(1+3)$ dimension can be written as
\begin{gather}\label{eq:inf123}
	S=\int \sqrt{-g}\, d^4x \, b(t)\left[-\frac{f^2(\phi,R)}{16\pi}F_{\mu\nu}\tilde{F}^{\mu\nu}-g_1 R A^2+g_2~\theta F_{\mu\nu}\tilde{F}^{\mu\nu} -(D_\mu \psi) (D^\mu \psi)^*\right]
\end{gather}
where $g_1$ and $g_2$ are some constant related with the coupling constants. 
This is a qualitative picture of magnetic field generation during the era of inflation and reheating. In the recent past, it has been shown that, the magnetic field generated in this fashion are model dependent and produces very weak seed fields \cite{Giovannini2000b}. 

{\large\bf Example:} To understand the generation of magnetic fields from inflationary process, let us assume that a scalar field $\phi$ is inflaton field and is responsible for the breaking of conformal invariance in equation (\ref{eq:inf123}) (first term in above action is just $\phi(\tau)F_{\mu\nu}\tilde{F}^{\mu\nu}$). Further assume that the electromagnetic field is weak enough that it does not perturb inflaton field and flat FRW expanding background space ($a$ being scale factor). However, in references \cite{Demozzi:2009fu, Parker:1968lp} authors have discussed back reaction of these generated magnetic fields. In reference, \cite{Demozzi:2009fu} authors have given a upper limit on the amplitude of the these field so that it could not spoil the inflation.In the present example, we are interested in a case where these seed fields are very week and do not spoil inflation. For such a case, one can obtain the generalised Maxwell's equation, by using equation (\ref{eq:inf123}) with the assumptions made in the current example. The space component gives
\begin{equation}
	A_i^{\prime\prime}+2\frac{f^\prime}{f}A_i^\prime -a^2 \partial_j\partial^j A_i=0\label{xyz123}
\end{equation}
where prime denotes derivative with respect to conformal time $\tau$ and $\partial^i =g^{jk}\partial_k=a^{-2}\delta^{jk}\partial_k$. In deriving equation (\ref{xyz123}), we have considered
Coulomb gauge for which electromagnetic four vector gauge field $A^\mu$ follows $\partial _i A^i(t,x)=0$ and $A_0(t,x)=0$. Electric and magnetic fields can be written in terms of three space component of the gauge field $A_i$ as: $E_i=\frac{1}{a}A_i^\prime$ and $B_i=\bar{\varepsilon}_{ijk}\delta^{jl}\delta^{km}\partial_l A_m$. 
One can obtain the Maxwell's equation by varying action with respect to $A_\mu$, which is given as
\begin{gather}
	\nabla_\nu F^{\mu\nu}=\frac{4\pi J^\mu}{f_0^2}.
\end{gather}
In this example, value of $f_0$ (here $f_0$ is background value of $f$) normalises the electric charge $e$ as $e/f_0^2$. For a redefined variable $\mathcal{A}=a(\tau)f(\tau)A(\tau, k)$, above equation (\ref{xyz123}) can be reduced to following form
\begin{equation}
	\mathcal{A}^{\prime\prime}(\tau,k)+\left(k^2-\frac{f^{\prime\prime}}{f}\right)\mathcal{A}(\tau,k)=0,
\end{equation}
which is similar to the equation of evolution of harmonic oscillator {\it i.e.}
%
\begin{equation}
	\mathcal{A}^{\prime\prime}(\tau,k)+\omega(\tau)^2\mathcal{A}(\tau,k)=0
\end{equation}
%
For $f=\text{const.}$, above equation reduces to the standard EM action and $\mathcal{A}$ oscillates in time like a standing wave. 

The energy density $T^0_0$ is the sum of magnetic contribution, $T^{0B}_0=-f^2B^2/8\pi$ and electric contribution $T^{0E}_0=-f^2E^2/8\pi$. 
Magnetic and electric energy densities can be defined as: $\rho_B=<0|T^{0B}_0|0>$ and $\rho_E=<0|T^{0E}_0|0>$. 
The spectral distribution of energy in the 
electric and magnetic field is
%
\begin{eqnarray}
	\frac{d\,\rho_E}{d ln~k}& = &\frac{f^2}{2\pi^2}\frac{k^3}{a^4}  \left|\frac{\mathcal{A}(\tau, k)}{f}\right|^2\\
	\frac{d\,\rho_B}{d\, ln\,k} & = & \frac{1}{2\pi^2}\left(\frac{k}{a}\right)^4 k\, |\mathcal{A}(\tau, k)|^2,
\end{eqnarray}
Thus, once we know 
the mode function $\mathcal{A}(k, \tau)$,  the evolution of energy densities $\rho_E$ and $\rho_B$ can be calculated.
During reheating the inflaton fields decays and convert the energy in the inflaton field to radiation.
In the present scenario, the strength of the magnetic fields as observed today can be given as
%
\begin{gather}
	B_0\sim 0.7\times 10^{-10} G\left(\frac{H}{10^{-5} M_{pl}}\right).
\end{gather}
Magnetic field with this strength can, in principle, be generated by appropriately choosing the coupling function.
\subsubsection{Magnetic fields from phase transitions}
There are some other mechanism of magnetogenesis based on the phase transitions in the early Universe, motivated from particle physics.
The first pioneer work based on phase transition was done by Hogan \cite{Hogan1983}. This was based on the idea that phase transition proceeds via bubble nucleation. During this course the bubbles of new phase collides violently leading to turbulence in the plasma. Similar to the "{\it Biermann Battery Mechanism}",  a charge separation of the positive and negative occurs in the plasma during this process \cite{Baym1996gbm} which leads to the generation of electric currents, resulting in the generation of magnetic fields (actually hypermagnetic magnetic fields). In a ref. \cite{Durrer:2003ja}, authors have shown that a primordial magnetic fields, generated due to some causal process in the early Universe, can be suppressed than usually assumed. However in ref. \cite{Kunze:2012ke} it is discussed that the helical fields will suppressed even more. In ref. \cite{Joyce1997}, it was shown that in presence of asymmetry in the left handed and right particles in the early Universe, some modes of the magnetic fields grow exponentially and leads to the generation of helical magnetic fields. This happens in presence of external gauge fields. In the presence of external gauge fields, a current flows in a particular direction. This is known as Chiral  magnetic effect \cite{Kharzeev2014a}.

Another approach to discuss the magnetogenesis during phase transitions is that the phases of the complex  order parameter of the nucleated bubbles are not correlated and so during collision there exists a phase gradient. This gradient leads to the source term for the evolution equation of the gauge field  \cite{Kibble_95}. This idea has been investigated in the context of the Abelian-Higgs model \cite{Enqvist1998aj, Copeland2000st, Saffin1997cj}. In the Abelian Higgs model, the collision of two spherical bubbles leads to a magnetic field which is localized in the interaction region of two bubbles. The strength of the generated magnetic field in this model depends on the velocity of the bubble wall. In the context of Electroweak phase transition, Vachaspati \cite{Vachaspati_91b} showed that the magnetic fields can be generated even if the transition is second order. However, the model parameters were gauge dependent quantities. So it is necessary to give a gauge-invariant definition of the photon in terms of the standard model fields. In an interesting approach \cite{Joyce1997}, authors discussed generation of primordial magnetic fields at the electroweak scale using anomalous magneto-hydrodynamics \cite{Giovannini:2013oga}. The basic assumption in this scenario was that the chemical potential of the pre-existing excess of right-electron converted to the hypercharge fields. The magnetic fields  obtained in this case are sufficiently  strong i.e., $\sim 10^{22}$ G at the electroweak epoch. However the correlation scale is very small and of the order of $10^{-6}H_{ew}^{-1}$. There are ideas related with the generation of magnetic fields at QCD phase transition. As the charge of the strange quarks and down quarks are different, it may be possible that the quarks develop a net charge. Therefore the first order phase transition at QCD scale may develop the different currents for leptons and quarks as the bubble wall moves in the quark-gluon plasma.  In ref. \cite{Quashnock_89} the magnetic field has been estimated to be of the order of $ B \sim \text{G} $ at the time of the QCD phase transition with a typical scale of the order of the meter at the same epoch.

Generation of magnetic field at electroweak phase transition were studied (at temperature $T_c\sim 100~ GeV$) in \cite{Vachaspati_91b, Enqvist1993, Sigl1997, Grasso2001, Durrer2013} and at QCD transitions (at temperature $T_c \sim 150~MeV$) in \cite{Forbes2000, Quashnock1989, Tevzadze2012, Cheng1994a}. Another aspect of the consideration magnetogenesis at phase transitions is that they provides non-equilibrium conditions for processes like baryogenesis \cite{Kuzmin1987, Turok1990} and leptogenesis\cite{Nardi2006, Long2014, Krauss1999} in the early universe. 
\\ \\
{\large \textbf{Helicity of fields and its production:}}\\
It was shown in ref. \cite{Brandenburg2005}, that the generated turbulence during the phase transition can amplify the seed magnetic field further by dynamo action. Model based on phase transitions explain the current observed strength of magnetic fields but they fails to explain the existence of the magnetic fields at large scale. 
Later it was shown that if generated fields have helicity, then the coherence length can increase because of the "{\it inverse cascade}" in presence of MHD turbulence \cite{Brandenburg1996, Banerjee2004, Frisch1975plm}. 
In 3-D MHD, the total (kinetic plus magnetic) energy cascades toward smaller scales, where it is dissipated by viscosity and resistivity. This is known as direct cascading of the total energy. However in the presence of magnetic helicity, energy cascade towards large length scale. This is known as "{\it inverse cascades process}".
There are some other work \cite{Brandenburg:2014mwa},in which authors have shown that energy transfer can occur even in the absence of helical field. This work was later confirmed by other groups \cite{Zrake:2015hda,Linkmann:2015zxa, Reppin:2017uud}. The magnetic helicity density is defined as 
\begin{equation}
	\mathcal{H}_M(t)=\frac{1}{V}\int_V d^3x \,\, {Y}\cdot{\bf B}, \label{hel1}
\end{equation}
where $\textbf{Y}$ is the (hyper)-electromagnetic vector potential and $\textbf{B}=\nabla \times \textbf{Y}$, the corresponding (hyper-)magnetic field. Helicity given above is equivalent to the Chern-Simon number of the particle physics \cite{Cornwall1997}. The time evolution of $\mathcal{H}_M$ is given by
\begin{gather}
	\frac{d\mathcal{H}_M}{dt}= -\frac{1}{\sigma}\int d^3x\,  \,\, {\bf B}\cdot {\bf \nabla}\times {\bf B}. \label{helvar1}
\end{gather}
For perfectly conducting medium, magnetic  helicity remain conserved. Helicity is an ideal invariant and asymptotically conserved within the resistive MHD approximation  \cite{Beck:2013bxa}. Apart from being a conserved quantity, helicity is a interesting for some reasons, given below
\begin{itemize}
	\item Helicity in the MHD and Chern-Simon number, which is related with the topological property of the gauge fields in the field theory, coincides with each other.	
	\item As $\mathcal{H}_M$ is a parity $P$ (parity) and CP-odd variable, the presence of a non-vanishing value of $\mathcal{H}_M$  would be a direct demonstration of breaking of these two symmetries.
	\item The presence of $\mathcal{H}_M$ implies the amplification and reorganization of field configurations towards large scale (which is known as inverse cascading) \cite{Siglunter2002, Vachaspati2001}.	
\end{itemize}
So physically $\mathcal{H}_M$ describes the topology of the magnetic fields lines. Also the inverse cascade allows the magnetic energy to shift from smaller length scale to larger length scales, as system tries to minimize its energy while conserving magnetic helicity \cite{Kahniashvili2013,Kahniashvili2015, Christensson:2000sp, Brandenburg1996}. 
In the last few years, there are several proposed mechanism to produce a helical magnetic fields in the early Universe. One of the important mechanism was discussed by Cornwall \cite{Cornwall1997}. 
In the model, discussed in ref. \cite{Cornwall1997}, it was suggested that a finite baryon number $(B)$ and lepton number $(L)$ stores the information about the initial helicity $\mathcal{H}_M$. These finite baryons $(B)$ and leptons $(L)$ were generated
by some GUT scale baryogenesis mechanism.
Assumptions was that a small amount of the total classically conserved $B+L$ charge, dissipated because of the anomalous processes in the early Universe. In this work, it was shown that $n_{B+L}T^{-3}$ amount of dissipated charged may have converted in to the helicity. The order of the produced helicity from the dissipated charge in natural units is
\begin{gather}
	\mathcal{H}_M \sim \frac{1}{\alpha} (N_B+N_L)\simeq 10^{71}. 
\end{gather}
Above, we have used $B$ and $L$ for baryon number and lepton numbers respectively. However, in the later part of the thesis, we have used $B$ only for magnetic field amplitude. 
\\
{\bf \text{Helicity and Chern-Simon number (CS)}}\\
Cornwall \cite{Cornwall1997} and Vachaspati \cite{Vachaspati2001}, showed in their work that baryogenesis and helicity of the magnetic field are actually connected with each other during EW phase transition. Classically $B$ remain conserved in the EW theory, however it is broken in the quantum theory in the presence of  classical Gauge field configuration.
This is because of anomalous behaviour of EW plasma. In the standard model, the global baryon  current $j^\mu_B=\Sigma_q\frac{1}{3}\bar{q}\gamma^\mu q$ and lepton number currents $j^\mu_L=\Sigma_l(\bar{l}\gamma^\mu l+\bar{\nu}_l \gamma^\mu \nu_l)$ are exactly conserved at the classical level. However, the anomaly equation give
\begin{gather}
	\partial_\mu j^\mu_B= \frac{i N_F}{32\pi^2} \left(-g_2^2 W^{\mu\nu a}\tilde{W}^{a}_{\mu\nu}+g_1^2 Y^{\mu\nu}\tilde{Y}_{\mu\nu}\right),
	\label{baryoncconserv1}
\end{gather}
here $W^{\mu\nu a}$ and $\tilde{W}^{a}_{\mu\nu}$ are the field strength corresponding to the $SU(2)_L$ gauge potential $W^a_\mu$ and it's dual respectively. Dual is defined as
$\tilde{W}^{a}_{\mu\nu}=\epsilon_{\mu\nu\alpha\beta}W^{\alpha\beta a}/2$. Similarly $Y^{\mu\nu}$ and $\tilde{Y}_{\mu\nu}$ are the field strength of $U(1)_Y$ group with coupling constant $g_1$. $N_F=3$  is the number of families and $g_2$ is the coupling constant related to $SU(2)_L$ group. One can also obtain corresponding current expression for each Lepton species as
\begin{gather}
	\partial_\mu j^\mu_{l}= i\frac{1}{32\pi^2} \left(-g_2^2 W^{\mu\nu a}\tilde{W}^{a}_{\mu\nu}+g_1^2 Y^{\mu\nu}\tilde{Y}_{\mu\nu}\right),
	\label{leptonconserv1}
\end{gather}
here $j^\mu_l$ is defined for each lepton species. Even $B$ and $L$ are separately not conserved, $B-L$ remain safe from the anomalous processes. Integrating equation (\ref{baryoncconserv1}) over 4-D space, baryon number changes by $\Delta B=N_F(N_{CS}-n_{CS})$. $N_{CS}$  and $n_{CS}$ are Chern-Simon number defined as
\begin{align}
	N_{CS}=&-\frac{g_2^2}{16\pi^2}\epsilon^{ijk}\int d^3x ~Tr\left[W_k\partial_i W_j+i~\frac{2}{3}~g_2~ W_i W_j W_k\right]\\
	n_{CS}=&-\frac{g_1^2}{16\pi^2}\int d^3x ~\epsilon^{ijk}~Y_k ~\partial_i Y_j \label{chern simon definition}
\end{align}
One can write equation (\ref{baryoncconserv1}) as follows
\begin{equation}
	\partial_\mu j^{\mu}_B= i\frac{N_F}{32\pi^2} \left(-g_2^2 \partial_\mu K^\mu +g_1^2 \partial_\mu k^\mu\right)
\end{equation}
where
\begin{align}
	K^\mu=&2\epsilon^{\mu\nu\alpha\beta}(W^a_\beta \partial_\nu W^a_\alpha-\frac{1}{3}g_2 ~ \epsilon_{abc}~ W^a_\nu W^b_\alpha W^c_\beta)\\
	k^\mu =&2\epsilon^{\mu\nu\alpha\beta} Y_\beta \partial_\nu Y_\alpha
\end{align}
One can also write baryon current as
\begin{equation}
	j^\mu_B=\frac{N_F}{32\pi^2}\left(-g_2^2K^\mu+g_1^2 k^\mu\right)
\end{equation}
Although $N_{CS}$ and $n_{CS}$ are gauge non-invariant objects though  $(N_{CS}-n_{CS})$ at different times remain gauge invariant. Chern-Simon number for the electromagnetic field is usual magnetic helicity. Each Chern-Simon number of the non-Abelian fields defines pure gauge configurations with zero energy and they are separated by barrier of height $E_{sph}\sim M_W/g_2^2$ in field configuration space. Transmission through the barrier happens only via quantum tunnelling at zero temperature and fermion density. This tunnelling is known as instanton \cite{Callancurtis:1978}. The tunnelling at zero temperature and zero fermion density is suppressed exponentially as
$\Gamma\propto e^{-4\pi/\alpha_w}$ \cite{Hooft1976, Hooft1976a}. However at finite temperature, this barrier can be overcome by thermal jumps on the top of the barrier. The field configuration with maximum energy is referred as Sphaleron. If the field configuration with energy $E_{sph}$ is smaller than the temperature of the thermal bath,  {\it i.e.} $E_{sph}< T$, the jump on the barrier are suppressed by exponential factor $\Gamma_{sph}\propto e^{-E_{sph}/T}$. With the change of CS-number or decay of sphaleron configurations, baryons are produced or annihilated. 
The range of temperature in which electroweak number violating processes are in thermal equilibrium in the early universe, can be calculated by comparing rate of sphaleron processes with Hubble expansion rate; {\it i.e.} $\Gamma_{sph}\geq n H$. Here $n \sim T^3$ is the number density of particles of a given type. Therefore sphaleron remain in the equilibrium for
%
\begin{equation}
	\frac{\Gamma_{sph}}{T^3}\geq H(T)=\frac{T^2}{M_{pl}}.
\end{equation}
Using this inequality, we can get temperature bound over which baryon number violating processes are in thermal equilibrium. Bounds are $100 GeV\leq T\leq 10^{12}~ GeV$. So for $T\leq 10^{12} ~GeV$, any preexisting $(B+L)$ is erased exponentially with a typical time scale $\tau_c\sim 2 N_F T^3 /13\Gamma_{sph}$. This can be understood from the following master equation for the evolution of the number density $n_{B+L}$ as \cite{Bochkarev:1987wf}
\begin{equation}
	\frac{d n_{B+L}}{dt}= -\frac{13}{2}N_F \frac{\Gamma_{Sph}}{T^3} n_{B+L}.
\end{equation}
But the baryon number violation within the SM and GUT baryogenesis can occur, if there is 
preexisting $B-L$ asymmetry at the GUT scale, because the sphaleron mechanism decays only $(B+L)$ asymmetry and does not touch the $(B-L)$ asymmetry \cite{Harvey1990jt}. One can see from the following combination for B
\begin{equation}
	B=\frac{B+L}{2}+\frac{B-L}{2}.
\end{equation} 
%
The fact that the combination $B-L$ is left unchanged by the sphaleron transitions opens up the possibility of generating the baryon asymmetry from the lepton asymmetry \cite{FUKUGITA198645}. 
%
It is known that, to generate finite baryon number density at GUT scale, {\it Sakharov conditions} must be satisfied.
However in EW theory, since C and CP are known to be violated by electroweak interactions, it is possible to satisfy all Sakharov's conditions within the SM if the EW phase transition leading to the breaking of $SU(2)_L\otimes U(1)_Y$ is of the first order \cite{Khlebnikov:1988sr}. At temperature T less than the temperature at which EW phase transition occur ({\it i.e.} $T< T_c\sim 100~GeV$) and small densities of the different fermionic charges, the EW symmetry is broken.
And the $U(1)_{em}$ fields are only long range fields \cite{Vachaspati_91b, Sigl1997}. 
At the electroweak epoch the typical size of the Hubble radius is of the order of 3 cm . The typical diffusion scale (defined as $\sqrt{t/\sigma}$, here t is time scale of diffusion and $\sigma$ is the conductivity) of the plasma is of the order of $10^{−9}$ cm. Therefore, over roughly eight orders of magnitude hypermagnetic fields can be present in the plasma without being dissipated  \cite{Giovannini1998}.

In the upcoming section, our main focus will to discuss the evolution of the generated magnetic fields and their correlation length scale from the epoch of magnetogenesis to present time.
\section{Evolution of the Primordial magnetic fields}
To understand the existence of the magnetic field at the galactic scale and in the intergalactic scale, it is crucial to understand the evolution of the magnetic field from the time of generation to the present time. At the time of magnetogenesis, nonlinear interactions of the magnetic field and the plasma fluid can govern the evolution of the magnetic fields. In this case the growth of the peak value of the magnetic field $B_p$ and physical correlation length scale $\lambda_p$ may not evolve adiabatically.  Which means that $B_p \not\propto a^{-2}$ or $\lambda_p \not\propto a$.  In absence of the free charges and small diffusion time of the fields in comparison to the age of the Universe, the stable modes of the fields sustain for a long time. However at high energy for example in the case of QCD plasma, presence of the strong magnetic fields produces a non-trivial instability in the plasma \cite{Akamatsu2013df}. 
There are three effects that play major role in the evolution of magnetic fields in the early Universe  \cite{Banerjee2004}
\begin{itemize}
	\item The effect of viscous diffusion 
	\item The interaction of the fields with the turbulent fluid
	\item The free streaming of photon and neutrino 
\end{itemize}
When a typical scale of the turbulence $\lambda_{T}$ at a certain time $t$ becomes of the order of $\lambda_{T}\simeq v_{T}t$ (here $v_{T}$ is the velocity of the fluid), the coupling of the magnetic fields and turbulence may become important. However in the case when the magnetic field correlation scale is much larger than the turbulence, then magnetic fields grows adiabatically. For the case of $\lambda_{T}\simeq \lambda_B$, inverse cascade starts. Under following scenarios, magnetic fields grows
%
\begin{enumerate}
	\item {\bf Purely inverse cascade case:} the magnetic fields undergo the inverse cascade just after their generation.
	\item {\bf The period when adiabatic and inverse cascade are simultaneously important {\it i.e.} transition case:}  First magnetic fields adiabatically evolve, and subsequently the inverse cascade starts  starts at a temperature $T=T_{TS}$.
	\item {\bf And the purely adiabatic case:} The magnetic fields always evolve adiabatically and never experience the inverse cascade process. 
\end{enumerate} 

To understand these three scenarios, let us first study evolution of a conducting plasma. The time evolution law of a magnetic field in a standard conducting plasma is given as 
\begin{equation}
	\frac{\partial {\bf B}}{\partial t}= \nabla\times ({\bf v}\times {\bf B})+\frac{1}{4\pi\sigma}\nabla^2 {\bf B}
\end{equation}
This equation is known as diffusion equation. In a time $\tau_{diff}(L)=4\pi\sigma L^2$, initial magnetic field configuration will vanished. Here $L$ is the characteristic length scale of the spatial variation of {\bf B}. A magnetic field generated at some time $t$ in the early Universe with coherence length scale $L_0$ will survive till today $t_0$ when  condition $\tau_{diff}(L_0)>t_0$ is satisfied. Length scale $L_0$ corresponds to present time determined by Hubble law.
\begin{equation}
	L_0=L(t_i)\frac{a(t_0)}{a(t_i)},
\end{equation}
where $L(t_i)$ is the length scale at the time of formation of the magnetic configuration. Magnetic fields evolved mostly in the matter dominated Universe and therefore when $\sigma\rightarrow \infty$, the magnetic flux moving with fluid remain frozen in. Magnetic flux can be calculated using 
\begin{equation}
	\frac{d\Phi_S(B)}{d t}=-\frac{1}{\sigma} \int_S \nabla \times(\nabla\times{\bf B})\cdot d{\bf S}
\end{equation}
where $S$ stands for surface integral. On scale where diffusion can be neglected, the field is said to be {\it frozen-in}. For the cae of isotropic expanding Universe, flux conservation implies
\begin{equation}
	B(t)=B(t_i)\left(\frac{a(t_i)}{a(t)}\right)^2.
\end{equation}
Similar to the flux, another quantity which remain conserved is magnetic helicity (defined in eq. (\ref{hel1})). 
In the presence of turbulence, the characteristics of the initially created magnetic field are vastly modified during cosmic evolution between the epoch of magnetogenesis and the present \cite{Brecher1970bg, Baym1997abc}.

In the remaining part of this section, we will focus on general features of the evolution of the magnetized fluids in the turbulent regime. 
Due to the finite conductivity in the early Universe, one can neglect dissipative effects. It is generally believed that the generated magnetic field freely decays without any further input of kinetic or magnetic energy, i.e., as freely decaying MHD (due to viscosity). Lets us first write general equations that describe incompressible MHD with dissipative contributions \cite{Brandenburg1996} 
\begin{gather}
	\frac{\partial {\bf v}}{d t}+({\bf v}\cdot \nabla){\bf v}-({\bf v}_A\cdot \nabla){\bf v}_A={\bf F}\label{euler1}\\ 
	\frac{\partial {\bf v}_A}{d t}+({\bf v}\cdot \nabla){\bf v}_A-({\bf v}_A\cdot \nabla){\bf v}=\frac{1}{\sigma}\nabla^2 {\bf v}_A. \label{diffus1}
\end{gather}
In above ${\bf v}_A(x)$ is known as local Alf\'ven velocity, which is given by ${\bf v}_A(x)={\bf B}(x)/\sqrt{4\pi(\epsilon+p)}$. Other variables are ${\bf v}$, ${\bf B}$, $\epsilon$ and $p$ known as the fluid velocity, magnetic field, mass energy density, and pressure respectively. Equation (\ref{euler1}) and (\ref{diffus1}) are known as Euler and diffusivity equations respectively. The function ${\bf F}$ in (\ref{euler1}) is dissipative term and depends on the scale. 
%
\begin{eqnarray}
	{\bf F} =
	\left\{
	\begin{array}{lr}
		\eta \nabla^2 {\bf v}&\hspace*{1.0 cm} \lambda_{mfp}\ll l\Longrightarrow \text{dissipation due to diffusing particles}\\
		-\alpha {\bf v} & \lambda_{mfp}\gg l\Longrightarrow \text{dissipation due to free streaming}
	\end{array}
	\right.
\end{eqnarray}
here $l$ is some length scale, $\alpha$ and $\eta$ are parameters.  In the early Universe, both of these two regime are important. 
The kinetic Reynolds number define the characteristic of the fluid flow and can be given as 
\begin{eqnarray}
	R_e(l)=\frac{v^2/l}{|{\bf F}|}=
	\left\{
	\begin{array}{lr}
		\frac{v l}{\eta}& \lambda_{mfp}\ll l\\
		\frac{v}{\alpha l} & \lambda_{mfp}\gg l
	\end{array}
	\right.
\end{eqnarray}
Reynolds number represents the relative importance of fluid advective terms and dissipative terms in the Euler equation. Some times it is given by the  ratio of a typical dissipative time scale $\tau_d=(l^2/\eta,1/\alpha)$ to the eddy-turnover time scale $\tau_{eddy}=l/v$. For the turbulent flow {\it i.e.} $R_e(L) \gg 1$ (here $L$ is the coherence scale of the magnetic field), the decay rate of the total energy is independent of dissipative effects and it solely depends on the flow property of the coherence scale and flux lines of the magnetic field will be frozen into the plasma element. 
However in the regime where $R_e(L)\ll 1$, total decay rate depends on the magnitude of the viscosity. Therefore  the flux lines of the magnetic field will diffuse through the plasma. 
In equation (\ref{euler1}), third term on the left hand side 
will establish fluid motions of the order of $v\approx v_A$
within an Alfven crossing time $\tau_A\simeq l/v_A$, at which point back reaction of the fluid flow on the magnetic fields will prevent further conversion of magnetic field into kinetic energy.
The resultant fully turbulent state is characterized by
close-to-perfect equipartition (in the absence of net helicity)  between magnetic and kinetic energy which can be represented by $<{\bf v}^2>\approx <{\bf v}_A^2>$. 
The transport of the fluid energy from the integral scale $L$ to the dissipation scale $l_{diss}$ occurs via a cascading of energy from large scales to small scales, referred {\it direct cascade}. It was shown that at the smallest length scale, magnetic and kinetic energy follows Kolmogorov spectrum ($E_k\propto k^n$). 
\section{Observations and constraints}
One can divide the observed magnetic fields into two components: first one is the uniform component) and another is the non-uniform component.
$$
B_{tot}=\bar{B}+\delta B,
$$
here $B_0$ is the homogeneous part of the total magnetic field ($B_{tot}$) and ($\delta B$) is the non-homogeneous part (where $\delta B\ll B_0$). Depending on the experimental techniques, the total magnetic field or its homogeneous part can be measured. 
Some measurements do measures two important components $B_{\perp}$ and $B_{||}$. Component $B_{\perp}$ and $B_{||}$ are defined as the component that are perpendicular and parallel to the line of sight respectively. Broadly measurement methods can be classified: Faraday rotation measurements, Zeeman effect and Synchrotron emission. However, recently Neronov and Semikoz \cite{Neronov2009} have proposed a new methodology based on Blazar's observations to measure the magnetic fields in the inter galactic medium. When a polarized radio signal passes through a region of space of size $\delta l$ containing a magnetized plasma, the plane of polarization of the wave gets rotated by a amount $\delta \phi\propto \omega_B \frac{\omega_p}{\omega} \delta l$~ \cite{Blasi:1999hu}. Here $\omega_p$ is plasma frequency and $\omega_B$ is Larmour frequency ($\omega_p\propto B_0$). By measuring rotation angle, we can measure  the strength of background magnetic fields. In the case Zeeman effect measurement, the measurement of the splitting of spectral lines (which is characterized by the deviation of the spectral lines $\Delta \nu=\frac{eB_0^{||}}{2\pi m_e}$ (here $m_e$ is the electron mass) \cite{Giovannini:2002sv} gives information about the background magnetic field along the line of sight. The synchrotron emission is sensitive to measure the transverse component $B_{\perp}$, however Faraday Rotation measurement and the Zeeman splitting of spectral lines are sensitive to the longitudinal component of the magnetic field along the line of sight. The key to detection of the magnetic field is the polarization of the emitted light from different sources in the optical, in the infrared, in submillimeter and in the radio wavelengths. The knowledge about the galactic and intergalactic magnetic fields comes from the measurements of the radio observations. The optical polarization occur from the extinction of the magnetic fields while passing through the dust grains in the interstellar medium along the line of sight.
\subsection{Observational bounds}
\begin{figure}[!h]
	\centering
	\includegraphics[scale=0.20]{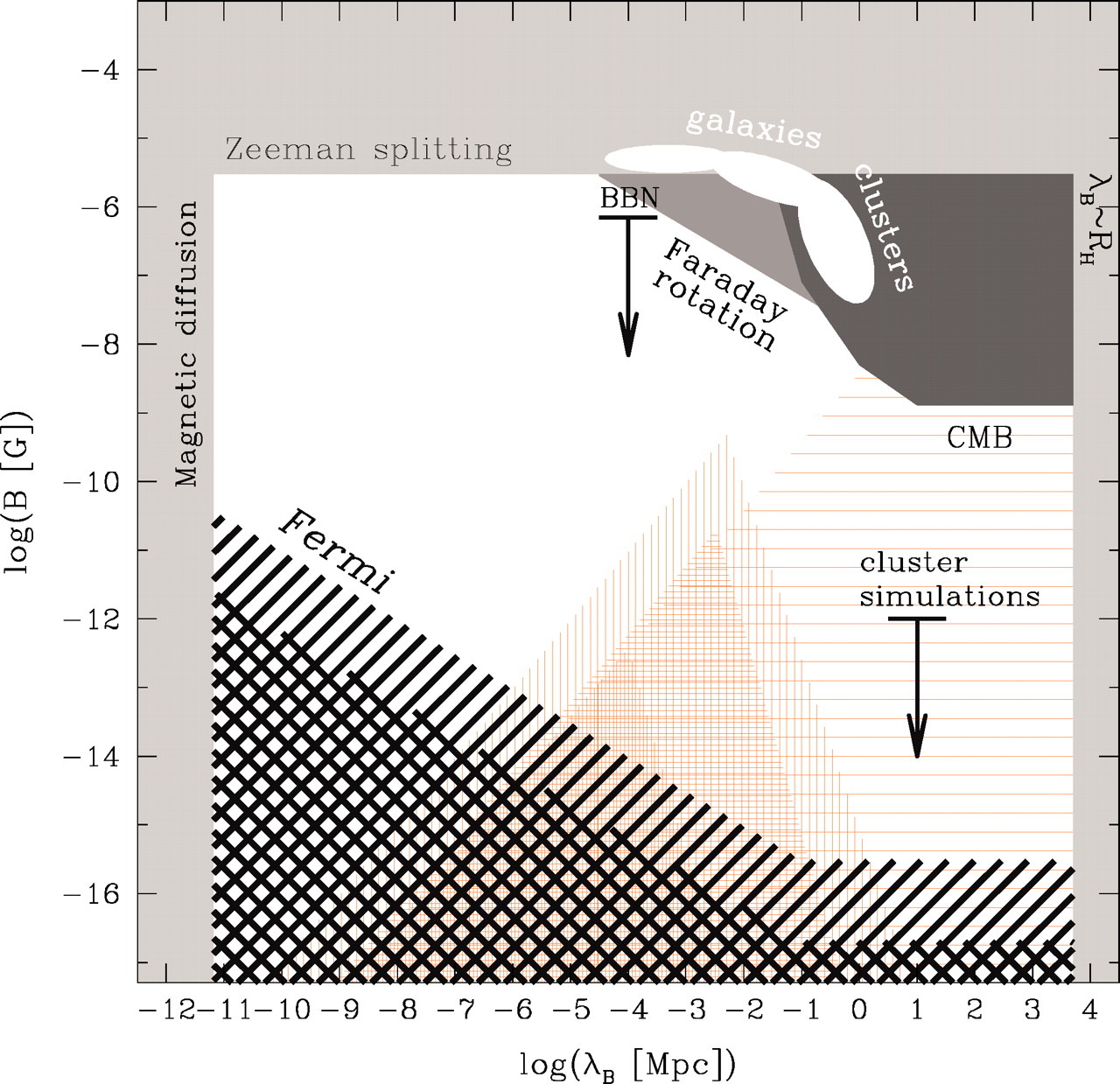}\label{bound1}~~
	\caption{{\bf (Bounds on Extra Galactic Magnetic fields B (EGMFs), 
			from several observations and experiments):} Light, medium and dark gray color shows known bounds on the strength of EGMFs. 
		The bound from the Big Bang Nucleosynthesis is taken from \cite{Grasso2001}. The black hatched region in the lower side of the figure, shows the lower bound on the EGMF derived from observations of 1ES 0347-121(cross-hatching) and 1ES 0229+200 (single diagonal hatching) \cite{Neronov2010}. Orange hatched regions show the allowed ranges of magnetic field and it's correlation scale generated during inflation models (horizontal orange hatching), during EW phase transition (dense vertical hatching), QCD phase transition (medium vertical hatching), and epoch of recombination (light vertical hatching) \cite{Neronov2009}. White ellipses show the range of measured magnetic field strengths and correlation lengths in galaxies and galaxy clusters.
		Reprinted figure from \cite{Neronov2010}, Copyright \textcopyright The American Association for the Advancement of Science}
\end{figure}
Nucleosynthesis provides the earliest probe to measure the magnetic fields as in the presence magnetic field can alters the rate of nucleosynthesis. One of the pioneer work related with the estimate of the magnetic field strength was done by Greenstein in ref. \cite{Greenstein:1951dj} and later it was done in more details by several others in ref.\cite{Matese:1970or, Grasso2001}. The basic idea was that, the rate of $\beta-$decay and rate of expansion of the Universe  can be affect  by the primordial magnetic fields.  In ref. \cite{Kahniashvili:2012dy}, authors have given bounds on the amplitude of the magnetic fields from large scale structure date. Another source to know about the presence of the primordial magnetic fields is the CMB measurements. In a recent observation of PLANCK, it is shown that upper and lower bounds on the peak value of the magnetic field is  $10^{-17}~G \lesssim B_0 \lesssim 10^{-9}$ G \cite{Ade2015}. A summary of the some observations and theoretical predictions has been shown in figure- (\ref{bound1}). The strength of the magnetic field varies with different length scale which is clearly depicted in the figure-(\ref{bound1}). Another work which catches much attentions of ours is a work by the authors of ref. \cite{Fujita2016}. In this work authors have calculated a much tight bound on the strength of the magnetic fields which comes from the early Universe baryon asymmetry.  A magnetic fields with present strength of $10^{-14}~G \lesssim B_0\lesssim 10^{-12}~G$ can be obtained if the magnetic fields were undergone a inverse cascading before the electroweak phase transition. They predicted that beyond this bound any magnetic field can be favoured because of the limited observed baryon asymmetry. 
\begin{figure}[h]
	\centering
	\includegraphics[scale=0.46]{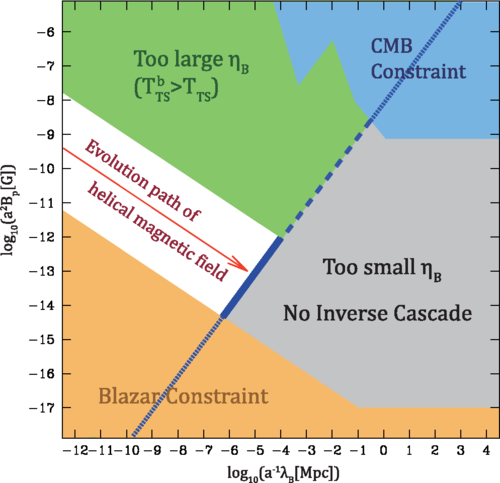}\label{bound2}~~
	\caption{{\bf (a)}.Time evolution of kinetic (dashed lines) and magnetic (solid lines) in the turbulent regime($R_e\gg 1$) without initial helicity. {\bf (b)}.Evolution of magnetic-energy spectra in the turbulent regime ($R_e\gg 1$)  with no initial helicity. Reprinted figure from \cite{Fujita2016}, Copyright \textcopyright APS}
\end{figure}
\section{Objectives of the study}
The universe is magnetized on all scales probed so far. A variety of observations imply that stars, planets, galaxies, clusters of galaxies are all magnetized. The typical strength of the magnetic field ranges from few $\mu$~G (in the galaxies and galaxy clusters) to few $G$ (in planets, like earth) and up to $10^{12}$ G (in the neutron stars). There are several astrophysical and particle models available but no model alone is able to explain the presence of homogeneous magnetic field at all scales. An intriguing possibility is that these observed magnetic fields are a relic from the early Universe, albeit one which has been subsequently amplified and maintained by a dynamo in collapsed objects. So it is quite important to study and construct a model which can explain the generation of magnetic field and its evolution throughout the Universe evolution. 

In our work, we study the generation of magnetic fields above the EW scale due to an anomaly in the primordial plasma consisting of the standard model particles. In our work we also look at the subsequent evolution of the generated magnetic fields in the early Universe. To study the generation of magnetic field, we use kinetic equations modified with Berry curvature. The new kinetic theory framework explains some of the very important features of the plasma with anomaly, like Chiral Magnetic Effects (CME) and Chiral Vortical Effects (CVE). We derive the expression of the Chiral Magnetic and Chiral Vortical currents. We also calculate chiral magnetic and chiral vortical conductivities by using kinetic theory. One of the main feature of our study is that the derived expressions for current and conductivities match with the studies done earlier, in a different context \cite{Joyce1997, Tashiro2012, Son2009}. We also calculate the strength of magnetic field at the EW scale, which comes into the bounds obtained observationally. In another work, we study the generation and evolution of magnetic field in the presence of chiral imbalance and gravitational anomaly which gives an additional contribution to the vortical current. The contribution due to gravitational anomaly is proportional to $T^2$. This contribution to the current can generate seed magnetic field  of the order of $10^{30}$~G at $T\sim 10^9$ GeV,  with a typical length scale of  the order of $10^6/ T$, even in absence of chiral charges (when chiral chemical potential is zero). Moreover, such a system possess scaling symmetry. We show that the $T^2$ term in the vorticity current along with scaling symmetry leads to more power transfer from lower to higher length scale as compared to only chiral anomaly without scaling symmetry. Next we study the evolution of the hydrodynamic excitation in the chiral plasma in the early universe. In this work, we have included the first and second order viscous terms in the hydrodynamic equation to study the effect of these first and second order viscous term on these hydrodynamic excitations. We have calculated few of the second order transport coefficients, and have found that the values of these coefficients fall under current bounds.
\section{Overview of the chapters} 
The thesis is organized as follows: Chapter (\ref{ch2}) contains the theoretical foundation for our work. This chapter contains an introduction to the kinetic theory of plasmas. We also discuss a modification of the kinetic equations in the presence of external magnetic fields.  In Chapter (\ref{ch3}), we discuss the generation of the magnetic field before EW phase transition using kinetic theory. In chapter (\ref{ch4}), we have discussed the generation  of primordial magnetic in presence of the chiral imbalance and gravitational anomaly. In this chapter we have also studied evolution in details. In Chapter (\ref{ch5}), we discuss anomalous magnetohydrodynamics. The final Chapter (\ref{ch6}), contains a summary of our work and the future scope of our work. 
\cleardoublepage
\chapter{Theoretical Foundation}\label{ch2}
Plasma is a many body system of charged and neutral particles, whose behaviour is dominated by collective effects mediated by the electromagnetic force. Here, "collective"  designates phenomena determined by the whole ensemble of particles in the system. The long range behaviour of the electromagnetic force determine the collective aspects of the plasma physics. %
The collective aspects of plasma physics are due to the long-range behaviour of the electromagnetic force. 
Temperature and number density of the charge particles are the one of the basic plasma parameters.
Standard Big Bang cosmology tells us that in the early Universe, temperature was so high that no atoms or molecules could exist. Hence the ionized gas was in the plasma state. 
It is currently believed that almost 99$\%$  of the matter in the Universe is made up of plasma, but such estimates are obviously hard to verify \cite{juliop:1997}. As temperature of the Universe decreases,some fraction of the ionized gas combine to form atoms and we could have partially ionized plasma. As Universe cools down even more, system can be considered as a neutral gas of weakly interacting atoms and molecules. 
A system with density comparable to the $10^{36} ~{\rm meter}^{{-3}}$, is believed to be strongly coupled system. Whenever thermal energies are  larger than interaction energies or of the order of Fermi energy, it is said that system of plasma is strongly coupled. These strongly coupled plasma system have more common property with a liquid than with a standard weakly coupled plasma \cite{Fitzpatrick2011}. Examples of strongly coupled plasmas, are  the interior of giant planets like Jupiter, solid-laser ablation plasmas, plasmas in high pressure arc discharges where the thermal and ionization energies are similar, and cold dusty plasmas. On the other hand, the systems found in space plasma physics, astrophysics, controlled nuclear fusion, and ionospheric physics are all weakly coupled. 

A complete description of a non-relativistic  plasma can be given by tracking individual particles, using the laws of classical mechanics and Maxwell's equations. Exact formulations of a system of very large number of  charged particles (for example plasma systems) are exceedingly complicated, because tracking individual particles is almost impossible. Thus it is customary to formulate approximate description of such systems, that describe  macroscopic properties of plasma. A hierarchy of approximations leads to the three leading plasma theories
\begin{itemize}
	\item Kinetic theory
	\item Multifield theory
	\item Magnetohydrodynamics (MHD) descriptions 
\end{itemize}
One can also divide different regime based on the time scale of interaction of individual particles. This is shown in the following figure (\ref{hydroregime})
\begin{figure}[h]
	\centering
	\includegraphics[scale=0.45]{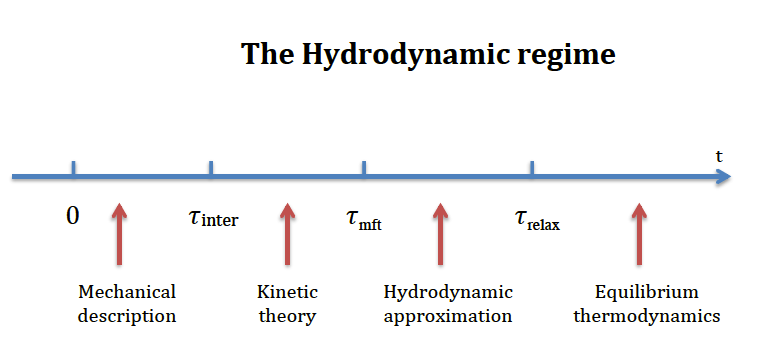}. \label{hydroregime}
\end{figure}
Here $\tau_{inter}$, $\tau_{mft}$ and $\tau_{relax}$ are time scale known as interaction time, mean free time and relaxation time of the plasma. Whenever time scale is less than interaction time of the individual particles, then mechanical description is enough. However when time scale is larger than $\tau_{inter}$ and less than $\tau_{mft}$, then kinetic theory must be used to describe the dynamics of the plasma system. Hydrodynamics approximations work for the times $\tau_{mft}\leq t \leq \tau_{relax}$ and for time $t$ larger than $\tau_{relax}$, system is in equilibrium state.
\subsubsection{Fluid approximations}
Depending on the number of free particles interacting through some coupling, there are number of different ways to describe the dynamics of the system. These approaches can be distinguished in terms of the dimensionless quantity: $\mathcal{R}=\lambda_{DB}/l$, where $\lambda_{DB}$ is the de Broglie wave length associated to each particle and $l$ is the typical inter-particle separation. 
\begin{itemize}
	\item $\mathcal{R}\gtrsim 1$ $\Longrightarrow$ the waveforms of the various particles overlap and a quantum-mechanical description is necessary. In this case, system is described by the N-particle wave function evolving in time following the Schroedinger equation.\\
	\item $\mathcal{R}< 1$ $\Longrightarrow$ then the wave-functions of the different particles are widely separated, the quantum interference is not important, and the individual wave packets evolve according to the Schrödinger equation in an isolated fashion, moving like classical particles. 
\end{itemize}
Above result is known as Ehrenfest theorem. If the system extends over a length scale L so much larger than the typical inter particle separation $l$ (of-course much larger than $\lambda_{DB}$) , that the dynamics of the individual particle can not be followed, not even in statistical terms, the collective dynamics of the system can be approximated by a continuous description in terms of a so called "{\it fluid}". The fluid description is valid only upto the limit of $l/L:=K_n$ (this is known as {\it Knudsen number} and fluid description is valid only for $K_n\ll 1$). When a fluid description is possible, the dynamics can be described in terms quantities averaged over representative "elements", which are large enough to contain a high number of particles, and small enough to guarantee homogeneity over the element. So the velocity of the particle in each element is same and are in thermal equilibrium. This description is given by classical plasma physics. However sometimes quantum mechanics is necessary in limited instances of the usual plasma theory, as for the calculation of nuclear fusion cross sections \cite{Manfredi2005}. A system with very high collision rate, has larger associated coupling constant. Due to this a system of plasma with sufficiently large collisional frequency behaves quantum mechanically. 
So it is relevant to ask, when is quantum and when is classical description of the plasma system are necessary? As mentioned above, extremely dense plasma behave like a quantum ideal gas, due to the exclusion principle.  
It is said that, if the typical length scale $L$ of the system is comparable to the de Broglie wavelength $\lambda_{DB}=\hbar /m v_T$ (where $v_T$ is thermal velocity and is defined as $v_T=(k_B T/m)^{1/2}$, $k_B$ is Boltzmann constant, $\hbar=h/2\pi$ is reduced Planck's constant, m is the mass of the charge carriers), {\itshape i.e.} $\lambda_{DB}\sim L_0$, then the quantum effects should be taken into account.
This can be the case, for instance, in charged particle systems like semiconductor quantum wells, thin metal films, and nanoscale electronic devices in general \cite{Manfredi2005, Jungal2009, Ventra2008}. 
In the later part of this chapter, we will mainly focus on classical plasma. To discuss the system, we need to use kinetic theory, as the length scale of the system under consideration is equivalent to the typical inter particle distance. In the later part of this chapter we have addressed the relativistic kinetic theory for our system (chiral plasma), for which relativistic theory is needed. 

\section{Kinetic theory of a classical plasma system}
Rather than to describe position and velocity of every individual particle in a plasma, kinetic theory divides the plasma into different classes of particles (or species) and describe the evolution of the probability distribution of each species of particles. The classification based on species are done using same mass or same charge. Most of the modern kinetic theory was first introduced by Landau \cite{tagkey1965ii}.
A theory that self-consistently accounts for long range force in plasma is the Lenard-Balesescu-equation \cite{LENARD1960390, Balescu1960}. An important physical effect that kinetic theory captures, but conventional fluid and MHD approximations do not, is  Landau damping \cite{Landau1946}. Landau damping is a process by which waves can either damp, or grow. Waves can damp or grow by different physical mechanisms in  fluid descriptions as well, which are also captured by kinetic theory, but Landau damping is fundamentally a kinetic process. In stable plasma all fluctuations damp, often by Landau damping, and scattering is dominated by conventional Coulomb interactions between individual particles. Landau's and Lenard-Balescu kinetic equations assume the plasma is stable. But plasmas are not always stable. In the presence of a free energy source, fluctuations may grow. The growth of these fluctuations in the system is known as instability. 
Theories that describe the scattering of particles from collective wave motion, typically assume that the instability amplitude is so large that conventional Coulomb interactions are negligible compared to the wave-particle interactions. However the theory of  stable plasma assume that Coulomb interactions dominate. It is interesting to study dynamics of the plasma system in the intermediate regime, where plasma is a weakly  interacting and collective fluctuation amplitude  is sufficiently weak and the collective fluctuations may be, but not necessarily dominant scattering mechanism. In this case the non-linear wave-wave interactions can be regarded to be sub-dominant.  These instabilities can modify the particle distribution functions and hence the amplitude of the fluctuations in the plasma.

For a very large number of particle, it is not realistic to solve the equations of motion for all of the particles. It is more useful to switch to a statistical approach in terms of a distribution function $f(t, {\bf x}, {\bf u})$. Here ${\bf x}$ and ${\bf u}$ are three dimensional position and velocity vector. In the six dimensional phase space $({\bf x}, {\bf u})$, the quantity $f(t, {\bf x}, {\bf u}) dx^3 du^3$ represents particle density in the range of ${\bf x}+d{\bf x}$ and ${\bf u}+d{\bf u}$. The dynamics of the system is then described by the evolution of the distribution function $f(t, {\bf x}, {\bf u})$. This distribution function can change in time either through simple advection in phase space in absence of collisions among the particles or in a more complex way when collisions are present and strongly influence the evolution. 
\subsection{Newtonian Kinetic theory}
Evolution of $f(t, {\bf x}, {\bf u})$ can be described by the {\it Boltzmann Equation}, which represents the foundation of the kinetic theory. There are two important extensions of this equation in which long range forces can play a very important role. In plasma physics, in presence of Coulomb forces, the Boltzmann equation is replaced by the {\it Vlasov-Maxwell equation}, and in presence of long-range gravitational forces, this equation becomes {\it Einstein-Vlasov-Maxwell} equation or {\it Einstein-Vlasov equation}. 
The Newtonian (i.e., non-relativistic) kinetic theory provide the simplest framework to study a system of interacting particles. Total number of particles in the system can be written as
%
\begin{equation}
	N=\int_{\infty}^{-\infty} d^3 x \int_{\infty}^{-\infty} d^3 u ~ f(t, {\bf x}, {\bf u}) =\int f(t, {\bf x}, {\bf u})~ d^3 x~ d^3 u
\end{equation}
It is necessary that the volume element $d^3x$, contain a large number of particles ensuring a small statistical variance and yet small enough with respect to the size of the system so that they can be considered as "points" in phase space and the continuum approach is justified. In a finite time element, $dt$, the particle coordinates change to 
\begin{gather}
	\hat{{\bf x}} ={\bf x} +{\bf u}~ dt  ~~~~~~~~
	and ~~~~~~\hat{{\bf u}} ={\bf u} +\frac{{\bf F}}{m_g} dt 
\end{gather}
Here ${\bf F}$ is an external force and $m_g$ is the mass of the particle. In the absence of collisions, we would have number densities at two different phase space points equal {\it i.e.},
\begin{equation}
	f(t, {\bf x}, {\bf u})~ d^3x~ d^3u = f(t+dt, {\bf x} +{\bf u}~ dt, {\bf u} +\frac{{\bf F}}{m_g} dt)~d^3 x^\prime~d^3 u^\prime.
\end{equation}
Here $d^3 x^\prime~d^3 u^\prime$ represents the volume element at a later time. With collisions, the distribution can change over time $dt$ as\\
\begin{gather}
	f(t+dt, {\bf x} +{\bf u}~ dt, {\bf u} +\frac{{\bf F}}{m_g} dt)  d^3 x^\prime~d^3 u^\prime-f(t, {\bf x}, {\bf u})~d^3 x ~d^3u=\left(\frac{\partial f (t, {\bf x},{\bf u})}{\partial t}\right)_{coll} d^3 x ~d^3u. 
\end{gather}
The Boltzmann equation for the collisional plasma can be obtained from above equation as
\begin{eqnarray}\label{boltzmanncoll1}
	\frac{\partial f}{\partial t}+{\bf u}\cdot \frac{\partial f}{\partial {\bf x}} +\frac{{\bf F}}{m_g}\cdot \frac{\partial f}{\partial {\bf u}}=\left(\frac{\partial f}{\partial t}\right)_{coll}\equiv C_B(f),
\end{eqnarray}
here $C_B(f)$ is the collision operator (or some times referred as collision integral) and depends on the nature of the interaction between particles. In the simplest case in which binary collision occurs with velocity of two colliding particles ${\bf u}_1$ and ${\bf u}_2$, in absence of the external forces, is defined as
\begin{eqnarray}
	C_B(f)= \int d^3 u_2 \int d\Omega ~\frac{d\sigma}{d\Omega}~|{\bf u}_1-{\bf u}_2|~(f_2^\prime f_1^\prime -f_2 f_1), 
\end{eqnarray}
where $f_{1,2}=f(t,{\bf x}, {\bf u}_{1,2})$ and $f_{1,2}^\prime=f(t,{\bf x}, {\bf u}^\prime_{1,2})$ are the distribution functions before and after a collision at time $t$ and position ${\bf x}$, while $d\sigma/ d\Omega $ is the differential cross-section over solid angle $d\Omega$ of the short range interaction responsible for the collisions. Thus Boltzmann equation with the collision integral is a non-linear integro-differential equation. So finding out the collision integral is one of the important problem in the kinetic theory.
From the distribution function obtained from solving equation (\ref{boltzmanncoll1}), it is possible to define the averaged value of the quantity $\Psi$ with respect to the distribution function $f$ as
\begin{equation}
	\langle \Psi \rangle= \frac{1}{n} \int \Psi~ f ~d^3u, \label{averg1}
\end{equation}
where $n$ is the number density, i.e., the number of particles per unit volume, which is defined as
\begin{eqnarray}
	n=\int f ~d^3 u\\ 
	N=\int n ~d^3 x
\end{eqnarray}.
Using above definition of the average quantity (\ref{averg1}), one can write expression for the mean macroscopic velocity ${\bf v}$ as
$$
{\bf v}=\langle {\bf u}\rangle =\frac{1}{n} \int {\bf u} ~ f ~d^3u.
$$
%
Here ${\bf v}$ is referred as fluid velocity.
\subsection{Relativistic Boltzmann theory}
The origin of relativistic kinetic theory dates back to 1911, when Juttner  derived the relativistic Mazwell-Boltzmann equilibrium distribution for a relativistic fluid \cite{Juttner1911}. The relativistic description is valid in the case of high energy plasma system. In the following section, we will discuss basic concepts of the relativistic kinetic theory in the flat space time.	The four space time coordinate point and four-momentum of a particle of rest mass $m$,   can be indicated by $x^\mu$ and $p^\mu= mcu^\mu= (p^0, p^i)$. The four-momentum satisfies $p_\mu p^\mu= m^2 c^2$. As in the Newtonian description, a distribution function $f$ can be defined such that the quantity 
\begin{equation}
	f~d^3 x~ d^3 p=f ~dx^1~ dx^2~dx^3~dp^1~dp^2~dp^3~,
\end{equation}
gives the number of particles in a given volume elements in the six dimensional space. Consider an observer in a frame $\mathcal{O}^\prime$ comoving with the particle, and a second observer $\mathcal{O}$  moving with a speed ${\bf v}$ with respect to observer $\mathcal{O}^\prime$. Lets assume that ${\bf v}$ is aligned to the x-axis. So the proper volume measured by observer in frame $\mathcal{O}^\prime$  is given by
\begin{equation}
	d^3x^\prime =W ~d^3x
\end{equation}
where $W$ is the Lorentz contraction factor between the two frame. Similarly 
\begin{equation}
	\frac{d^3p^\prime}{p_0^\prime}=	\frac{d^3p}{p_0}~.
\end{equation}
Above equations shows that the ratio $d^3p^\prime/p_0$ is a Lorentz invariant (here $p_0^{\prime}=p_0/W$). So using above relations, one can get $d^3x^\prime d^3p^\prime=d^3x d^3p$. So even $d^3x$ is not Lorentz invariant, the product $d^3x d^3p$ is Lorentz invariant. The number of particles measured in two frames must not changed. So
\begin{equation}
	f^\prime(x^\prime , u^\prime)~ d^3x^\prime~ d^3p^\prime = f(x, u)~d^3x~ d^3p~.
\end{equation}
Therefore distribution function $f$ itself must also be Lorentz invariant, i.e., $f^\prime(x^\prime , u^\prime) =f(x, u)$. One can write relativistic Boltzmann equation as
%
\begin{equation}
	p^\mu \frac{\partial f}{\partial x^\mu}+m\frac{\partial (f~F^\mu )}{\partial p^\mu}=\Pi (f)~,
\end{equation}
here $F^\mu$ is the four-force acting on a particle, that may or may not be dependent on the four momentum $p^\mu$. However scalar quantity on the right side of the above equation is the relativistic
generalization of the collision integral and can be written as \cite{cercignani2002:book}
\begin{equation}
	\Pi(f):= \left(\frac{\partial f}{\partial t}\right)_{coll}=\int \frac{d^3p_2}{(p_2)^0} \int d\Omega \sigma (\Omega) \sqrt{(p_1)^\alpha (p_2)_\alpha-m^4 c^4}~(f_2^\prime f_1^\prime -f_2 f_1)~.
\end{equation}
Here $\sigma (\Omega)$ represents differential cross section. The first moment of the distribution function defines the number density current
\begin{equation}
	N^\mu := c\int p^\mu ~f~ \frac{d^3p}{p^0}~.
\end{equation}
Zeroth order of the above 4-vector current density is referred as number density
\begin{equation}
	N^0 =c\int f~d^3p =c~n~,
\end{equation}
and  the spatial (contravariant) components of the 4-vector current density can be given as
\begin{equation}
	N^i= c\int p^i ~f~\frac{d^3p}{p^0}=c^2 \int W~mv^i~f \frac{d^3p}{E}=\int v^i~f~d^3p~.
\end{equation}
The second moment of the distribution function is stress-momentum tensor.
\begin{equation}
	T^{\mu\nu}=c\int p^\mu p^\nu ~f~\frac{d^3 p}{p^0}~.
\end{equation}
This quantity measures the flux of $\mu$-momentum across a surface at $x^\nu$=const., i.e., in the $\nu$-direction. This is relativistic generalization of the classical energy momentum tensor.
\section{Kinetic theory with Berry Curvature}
The relativistic kinetic theory framework discussed above misses the effects of triangle anomalies, responsible for parity and CP violations  \cite{Adler1969, Bell1969}. To discuss such kind of effects, a modified kinetic theory  formalism from the underlying quantum field theory  \cite{Son2013a} has been developed by taking into account the Berry curvature \cite{Berry1984}. In the following section, we shall discuss in brief about the Berry curvature and derivation of the Kinetic equations, modified with Berry curvature.
\subsection{Basics of Berry curvature}
We know that a quantal system can be described by it's Hamiltonian $\hat{H}$. If the Hamiltonian is not a function of time then we can describe the system by its stationary state.
If the system and hence $\hat{H}$, is slowly altered then from the adiabatic theorem, at any instant the system will be in an eigenstate of the instantaneous Hamiltonian. If Hamiltonian is returned to its original form, the system will return to its original state, apart from a small phase factor may be large phase factor in some cases.  This phase  factor is observable by interference experiment.

Let us suppose a cubic box of size $L$, whose ground state is represented by '{\it ket}' state $|n(L(0))\rangle$. Suppose box slowly expands with time, which implies length is a function of time {\it i.e.} $L(t)$. The adiabatic principle states that if the expansion is slow enough, the particle will be in the instantaneous ground state  $|n(L(t_1))\rangle$ of the box of size $L(t_1)$. More generally, if particle Hamiltonian is given by $H(R(t))$, where $R$ is some external coordinate which changes slowly 
in Hamiltonian. So Hamiltonian equation and it's solution can be given as 
\begin{eqnarray}
	|\psi (t)\rangle & = & \text{exp}\left(-\frac{i}{\hslash}\int_{0}^{t} ~E_n(t^\prime) ~dt^\prime\right)|n(t)\rangle\\
	H(t)~|n(t)\rangle & = & E_n(t)~|n(t)\rangle.
\end{eqnarray}
It is to be noted here that if the Hamiltonian $H$ is independent of the time, the above description is true, with the phase factor appropriate to energy $E_n$.  But one can see that the instantaneous energy varies with time and gives the accumulated phase factor shift . To see what is missing in the above ansatz, let us modify it as follows:
\begin{eqnarray}
	|\psi(t)\rangle=c(t)~\text{exp} \left(-\frac{i}{\hslash}\int_{0}^{t} ~E_n(t^\prime) ~dt^\prime\right)|n(t)\rangle,
\end{eqnarray}
here the extra factor $c(t)$ must equal to unity if the old ansatz is right.  Now applying Schrodinger equation to this state
\begin{equation}
	\left(i\hslash \frac{\partial }{\partial t}-H(t)\right)|\psi(t)\rangle=0.
\end{equation}
From this equation one can get
\begin{equation}
	\dot{c}(t)=c(t)\langle n(t)|\frac{d}{dt}|n(t)\rangle.
\end{equation}
Solution of the above equation for $c(t)$ gives
\begin{align}
	c(t)&=c(0) \text{exp}\left(-\int_{0}^{t}\langle n(t^\prime)|\frac{d}{dt^\prime}|n(t^\prime)\rangle dt^\prime\right)= c(0) \text{e}^{i\gamma},
\end{align}
where $\gamma$ is given as
\begin{eqnarray}
	\gamma&=i\int_{0}^{t}~\langle n(t^\prime)|~\frac{d}{dt^\prime}~|n(t^\prime)\rangle ~dt^\prime
\end{eqnarray}
This extra phase factor is known as {\it Berry phase} or the {\it geometric phase}. It is not recognize here that it has a measurable consequences. Since instantaneous kets are defined only up to a phase factor, and we can choose a new set and modify the extra phase. If we choose new phase factor $\xi(t)$, then state can be be written as $|n^{\prime (t)}\rangle =\text{e}^{i\gamma(t)}|n(t)\rangle$. So one can get 
\begin{eqnarray}
	i \langle n^\prime(t)|\frac{d}{dt}|n^\prime(t)\rangle=i\langle n(t)|\frac{d}{dt}|n(t)\rangle -\frac{d\xi(t)}{dt}.
\end{eqnarray}
which suggests that, one can choose this new phase so that they cancel. Let us suppose a case where Hamiltonian return back to its starting value after time $t_1$ so that $H(t_1)=H(0)$. So in this case, it is obvious that one can not get rid of the extra phase factor. 
\begin{eqnarray}
	i \oint\langle n^\prime(t)|\frac{d}{dt}|n^\prime(t)\rangle dt=i\oint\langle n(t)|\frac{d}{dt}|n(t)\rangle -(\xi({t_1})-\xi(0)).
\end{eqnarray}
So the choice of phase factor is quite arbitrary, but it must satisfies the requirement of single value in the region containing the closed loop.  So it is necessary 
that choosing different set of basis do not altered the phase factor. The phase factor should depends only on the path in the parameter space, which explains the name "geodesic phase".  To see meaning of $\gamma \neq 0$, let us consider a system of some nucleus of some heavy object. Let ${\bf R}$ and ${\bf r}$ be the coordinate of some nucleus and electron respectively (electron is orbiting the nucleus). The electron moves under the influence of Coulomb potential created by nucleus. In this case the phase factor can be written as-
\begin{align}
	&\text{exp}\left(-\int_{0}^{t}\langle n(t^\prime)|\frac{d}{dt^\prime}|n(t^\prime\rangle dt^\prime \right)\nonumber \\
	&= \text{exp}\left(\frac{i}{\hslash} i\hslash\int_{0}^{t}\langle n(t^\prime)|\frac{d}{dt^\prime}|n(t^\prime\rangle dt^\prime \right)\nonumber \\
	&= \text{exp}\left(\frac{i}{\hslash} \int_{0}^{t}i\hslash \langle n(R(t^\prime))|\frac{d}{d{\bf R}}|n({\bf R}(t^\prime))\rangle \frac{d{\bf R}}{dt^\prime}dt^\prime \right)\nonumber \\
	& =\text{exp}\left(\frac{i}{\hslash} \int_{0}^{t} \mathcal{A}^n({\bf R})\frac{d{\bf R}}{dt^\prime}dt^\prime\right),
\end{align}
where 
\begin{equation}
	\mathcal{A}^n({\bf R})= i\hslash \langle n({\bf R})|\frac{d}{d{\bf R}}|n({\bf R})\rangle .
\end{equation}
The quantity $\mathcal{A}^n({\bf R})$, is known as Berry potential or Berry connection. Also normalisation condition satisfies i.e., $\langle n({\bf R}(t))| n({\bf R}(t))\rangle =1$. So Berry phase can be written also in the following form
\begin{equation}
	\gamma(t)=i\int_{{\bf R}_{0_{\bf C}}}^{{\bf R}(t)} \mathcal{A}^n({\bf R})\cdot d{\bf R}.\label{berryphase}
\end{equation}
Berry connections depends on the state, in which electronic degree of freedom is in. C in the integral represent integral over a path over which adiabatic process take place and ${\bf R}$ changes from ${\bf R}_0$ to ${\bf R}(t)$. As $\mathcal{A}^n({\bf R})$ is purely imaginary, the Berry phase defined in equation (\ref{berryphase}) is real. For a closed path (for which ${\bf R}_0={\bf R}(t_1)$) of integration, above problem becomes much simpler. For this closed loop, using Stoke's theorem one can write
\begin{align}
	\gamma(C)=i\oint_C \langle n({\bf R})|\nabla_{{\bf R}} n({\bf R})\rangle d{\bf R}=-\int \int_C \Omega^n({\bf R}).d{\bf S}, \label{berrycurvature1}
\end{align}
here we changed notation of derivative from $d/d{\bf R}$ to  $\nabla_{\bf R}$. Also $\Omega^n$ is known as Berry curvature. The right hand side of equation (\ref{berrycurvature1}) is the flux of the Berry curvature on the surface $S$; such flux remains meaningful even on a closed surface (e.g. a sphere or a torus). The key result is that such an integral is quantized \cite{davidj2011}.\\
{\bf \text{Example}:} Let us consider a single chiral fermion described by Hamiltonian $H=\sigma\cdot {\bf \hat{p}}$. So energy eigen equation can be written as (from now on we will use natural units)
\begin{equation}
	H~ u_{{\bf p}}(e) =\pm e ~|p|~u_{{\bf p}},\nonumber
\end{equation}
where 
\begin{equation}
	H = 
	\left[ {\begin{array}{*{20}c}
			\text{cos} \theta & e^{-i\phi}\text{sin}\theta  \\
			e^{i\phi}\text{sin} \theta & -\text{cos} \theta  \\
	\end{array} } \right]\nonumber
\end{equation}
and $\pm$ signs are for two component spinor. $+/-$ signs are for right/left handed fermions respectively. We can construct following form for the spinors
\begin{equation}
	u_p(+)= 
	\left[ {\begin{array}{*{20}c}
			e^{-i\phi}\text{cos} \frac{\theta}{2} \\
			\text{sin} \frac{\theta}{2} \\ %
	\end{array} } \right], ~~~~~~~
	u_p(-)= 
	\left[ {\begin{array}{*{20}c}
			-e^{-i\phi}\text{sin} \frac{\theta}{2}  \\
			\text{cos} \frac{\theta}{2} \nonumber\\
	\end{array} } \right], 
\end{equation}
with this definition one can get Berry correction and Berry curvature for chiral fermions as
\begin{align}
	\mathcal{A}({\bf p})&=-i~ u_p^{\dagger}(+)~\nabla_p~ u_p(+) \nonumber\\
	&=-{\bf \hat{e}}_\phi ~\frac{1}{2|{\bf p}|}~ \frac{\text{cos}(\theta/2)}{\text{sin}(\theta/2)} \nonumber\\
	\Omega_p &=\nabla_p\times\mathcal{A}({\bf p})=\pm \frac{{\bf p}}{2|{\bf p}|^3},\nonumber
\end{align}
where 
\begin{equation}
	\nabla_p={\bf \hat{e}}_p \frac{\partial}{\partial |{\bf p}|}+{\bf \hat{e}}_\theta \frac{1}{|{\bf p}|}\frac{\partial}{\partial \theta}+{\bf \hat{e}}_\phi \frac{1}{|{\bf p}|\text{sin}\theta}\frac{\partial}{\partial \phi} \nonumber.
\end{equation}
One can see for ${\bf p}\neq 0$, the divergence is vanishing i.e., $\nabla \cdot \Omega_p =0$ but is non-vanishing when calculating the total flux on a sphere $\int d^3p \nabla_p\cdot \Omega_p=4\pi\Rightarrow \nabla_p\cdot \Omega_p=2\pi \delta^{(3)}({\bf p})$.\\
{\large \bf \text{Anology to magnetic field}:}
\begin{center}
	\begin{tabular}{l l l}
		\hline
		{\bf \text{Berry part}} & {\bf \text{Magnetic field part}} \\ 
		\hline
		Berry curvature $\Omega({\bf R})$ & Magnetic field ${\bf B}({\bf r})$\\
		Berry connection $\mathcal{A}({\bf R})$& Vector potential ${\bf A}({\bf r})$\\
		Geometric Phase & Ahanonrov-Bohm phase\\
		Chern-Simons number& Dirac monopole\\
		\hline
	\end{tabular}
\end{center}
where Geometric phase, Ahanorov-Bohm phase, Chern-Simons number and Dirac monopole are defined as
\begin{align}
	\text{Geometric phase}&=\oint_C d{\bf R}\cdot \mathcal{A}({\bf R})=\int\int d{\bf S}\cdot \Omega {\bf R},\nonumber\\
	\text{Ahanorov-Bohm phase}&=\oint_C d{\bf r}\cdot A({\bf r})=\int\int d{\bf S}\cdot {\bf B}({\bf R}), \nonumber\\
	\text{Chern-Simons number}&=\oint d{\bf S}\cdot \Omega({\bf R})=\text{integer}\nonumber\\
	\text{Dirac monopole}&=\oint d{\bf S}\cdot {\bf B}({\bf r})=\text{integer}\times\frac{h}{e}\nonumber.
\end{align}
Therefore, a non-zero Berry connection and curvature can be treated as the fictitious vector potential and magnetic field in the momentum space. Therefore, Berry curvature can affect the motion of chiral
fermion in the momentum space and one can write the corresponding action as \cite{Sundaram1999, Duval2006, Xiaodishi2005, Chen2013cv}, 
\begin{equation}
	S(x,p)=\int {dt}[(p^i+e A^i(x))\dot{x^i}-\mathcal{A}^i\bf{(p)}\dot{p^i}-\epsilon_{p}(x)-A^0(x)],
\end{equation}
here $\varepsilon_p(p)$ is energy of the particles and $A_0$ and  $A^i$ are the scalar and magnetic potentials. This equation can also be written as
\begin{equation}
	S(\xi)=\int {dt}[\Sigma_a(\xi)\dot{\xi^a}-H(\xi)],
\end{equation}
Here $\xi^\alpha$ represents thee phase space coordinates ${\bf r}$ and ${\bf k}$ collectively.   
Also $\Sigma_a(\xi)=(p^i+e A_i(x),-{\bf{Q^i(p)}})$, $\xi^a=(x^i, p^i)$ and $H(\xi)=\epsilon_{p}(p)+A^0(x)$. $\mathcal{A}^i ({\bf p})=-i~\bf{u}^\dagger_{p} ~\nabla_{{\bf p}}~ u_{{\bf p}}$ is the Berry connection for chiral fermion. Now the equations of motion of the action read as,  
\begin{equation}
	\Sigma_{ab}~\dot{\xi^b}=-\frac{\partial H{(\xi)}}{\partial \xi^{a}}, 
\end{equation}
Hamilton's equation of motion is,
\begin{equation}
	\dot{\xi^a}=-\{{\xi^a},H(\xi)\}=-\{{\xi^a},{\xi^b}\}\frac{\partial H{(\xi)}}{\partial \xi^{b}}.\label{EOMpoisonbracket}
\end{equation}
One can find easily following relation $\{{\xi^a},{\xi^b}\}=(\Sigma^{-1})^{ab}$.  Using the above equation, we can write down the explicit form of Poisson brackets for variables $x^i, p^i$ with berry curvature as follows,
\begin{align}
	\{{x^i},{x^j}\}&=\frac{\epsilon_{ijk} \Omega_k}{1+e {\bf{B\cdot{\bf \Omega}_{\bf p}}}}, \\ 
	\{{x^i},{p^j}\}&=-\frac{\delta_{ij}+e\Omega_i B_j}{1+e{\bf{B\cdot{\bf \Omega}_{\bf p}}}}, \\
	\{{p^i},{p^j}\}&=-\frac{e\epsilon_{ijk} B_k}{1+e{\bf{B\cdot{\bf \Omega}_{\bf p}}}}, 
\end{align}
where, $B^i=\epsilon^{ijk}\frac{\partial A^k}{\partial x^j}$ and ${\bf \Omega}_{\bf p}={\bf{\nabla_p}}\times \mathcal{A}_{{\bf p}}$. As a result of the modification of the Poisson Bracket, the invariant phase space gets modified \cite{Duval2006, Xiaodishi2005},
\begin{equation}
	d\Gamma=\sqrt{det \Sigma_{ab}}d\xi=({1+e {\bf{B\cdot\Omega}}})\frac{{dp}{dx}}{(2\pi)^3}.
\end{equation}   
Equivalent Liouville's theorem will give,
\begin{equation}
	\dot f_{\bf{p}}-(\Sigma)^{-1}_{ab}\frac{\partial H{(\xi)}}{\partial \xi^{b}}\frac{\partial f_{\bf{p}}}{\partial \xi^{a}}=0,\nonumber
\end{equation}
where, $f_{p}$ is the distribution function of chiral fermion. Taking $H=\epsilon_{p}+A_0$, one can get the following equation,
\begin{align}
	\dot f_{\bf p}+
	\dot{\bf x}\cdot \frac{\partial f_{\bf p}}{\partial {\bf x}}+\dot{\bf p}\cdot \frac{\partial f_{\bf p}}{\partial {\bf p}}=0,
	\label{kineticequation}
\end{align}
where,
\begin{align}
	\dot{\bf x}&=\frac{1}{1 + e {\bf B} \cdot {\bf \Omega}_{\bf p}}\left(\tilde {\bf v} + e\tilde {\bf E} \times {\bf \Omega}_{\bf p}
	+e (\tilde {\bf v} \cdot {\bf \Omega}_{\bf p}) {\bf B}
	\right),\\
	\dot {\bf p}&=\frac{1}{1 + e {\bf B} \cdot {\bf \Omega}_{\bf p}}
	\Big[\left(e \tilde {\bf E} + e\tilde {\bf v} \times {\bf B}
	+ e^2(\tilde {\bf E} \cdot {\bf B}) {\pm \Omega}_{\bf p} \right)\Big].
	\nonumber 
\end{align}
Note that here, $\tilde {\bf{v}} = \frac{\partial \epsilon_{\bf {p}}}{\partial{\bf {p}}}$,
$e{\tilde {\bf {E}}}=e{\bf{E}} - \frac{\partial \epsilon_{\bf {p}}}{\partial \bf{x}}$, 
$\mathbf{\epsilon_{p}}=p(1-e{\mathbf{B\cdot\Omega_{p}}})$ and  ${\mathbf{\Omega_{p}}}=\pm{\mathbf{p}}/{2 p^{3}}$. Above $\pm$ sign respectively  corresponds to the right and left handed particles.  If ${\bf{\Omega_p}}=0$, the above equation reduces to Vlasov equation.
It is easy to check that Eq.(\ref{kineticequation}) gives the anomaly equation as follows,
\begin{equation}
	\partial_t n +\bf{\nabla\cdot j}= e^2 \int \frac{d^3p}{(2\pi)^3} \left(\bf{\Omega_p}\cdot\frac{\partial 	f_{\bf p}}{\partial p}\right)\bf{E}\cdot \bf{B},\label{particlenumber}
\end{equation}
where $n$ and ${\bf j}$ are defined as
\begin{equation}
	n = \int \frac{d^3p}{(2\pi)^3}(1+e{\bf B}\cdot \Omega_{\bf p}) f_{\bf p}, \label{tocdens}
\end{equation}
and 
\begin{equation}
	\label{eq:current}
	{\bf j} = -e\int \frac{d^3p}{(2\pi)^3}
	\left[\epsilon_{\bf p}\frac{\partial f_{\bf p}}{\partial p}
	+ e\left(\bf{\Omega_p}\cdot\frac{\partial f_{\bf p}}{\partial p}\right) \epsilon_{\bf p} {\bf B}
	+ \epsilon_{\bf p} {\bf \Omega}_{\bf p} \times \frac{\partial f_{\bf p}}{\partial x } \right] 
	+ {\bf E} \times {\bf \sigma}.
\end{equation}
here ${\bf \sigma} $ is given by,
\begin{equation}
	{\bf \sigma} = \int \frac{d^3p}{(2\pi)^3} {\bf \Omega}_{\bf p} f_{\bf p}. \label{jindividual}
\end{equation}
The integration on the right hand side of Eq.(\ref{particlenumber}) is not simple because
at $p=0$ there is singularity. At this point motion of particles can be described by quantum
mechanics \cite{Stephanov2012}. To integrate this integral in the regime $|p|<R$, we have exclude the the point surrounding $p=0$ and integrated only in the  classical regime $|p|>R$. In this regime, particles can not be destroyed, they can only enter or exit through the surface at R. 
Therefore, for the region $|p|>R$ we can write Eq.(\ref{particlenumber}),
\begin{equation}
	\partial_t n +\bf{\nabla\cdot j}= e^2 \int_{S^{2}(R)} \frac{dS}{(2\pi)^3}\cdot{\Omega_p}
	~({\bf{E}\cdot\bf{B}})~f_{\bf p},
\end{equation}   
where, $dS$ is the surface element of the sphere. One can see that in the limits of $R\rightarrow 0$,  which implies that $p=0$ and carry out surface integral we get,   
\begin{equation}
	\partial_t n +\bf{\nabla\cdot j}=\frac{e^2}{4\pi^2}
	f_{\bf {p=0}}{\bf{E}\cdot\bf{B}}.
\end{equation} 
Thus total flux remains finite even at $p=0$ due to anomaly (${\bf{E}\cdot\bf{B}}$ term)  \cite{Adler1969, Adler1972}. The presence of $f_{\bf {p=0}}~$ in the above equation shows that there must be some thermal correction. However, it is important to note that at finite temperature one must also consider anti-fermions. Therefore, if we consider both right-handed and left-handed particles/antiparticles and write the same sort of transport equation as above, we can arrive to the following equation,
\begin{equation}
	\partial_{\mu}J^{\mu}=\frac{e^2}{4\pi^2}
	({f^R_{\bf {p=0}}+f^{\bar{R}}_{\bf {p=0}}+f^{L}_{\bf {p=0}}+f^{\bar{L}}_{\bf {p=0}}})({\bf{E}\cdot\bf{B}}),
\end{equation} 
which can be simplified to the following form, 
\begin{equation}
	\partial_{\mu}J^{\mu}=\frac{e^2}{4\pi^2} {\bf{E}\cdot\bf{B}}.
\end{equation}  
Thus, chiral anomaly does not receive any thermal corrections, which is well known result in the literatures of quantum field theory \cite{Itoyama:1982up, Liu:1988ke}. One can also define modified energy density and momentum density of the particle. 
\begin{equation}
	\epsilon = \int \frac{d^3p}{(2\pi)^3} (1+{\bf B} \cdot {\bf \Omega}_{\bf p})
	\epsilon_{\bf p} f_{\bf p}, 
	\qquad
	\pi^i = \int \frac{d^3p}{(2\pi)^3} (1+{\bf B} \cdot {\bf \Omega}_{\bf p})
	p^i f_{\bf p}, 
	\label{eq:density}
\end{equation}
by multiplying Eq.(\ref{kineticequation}) by $\epsilon_{\bf p} \sqrt{\Sigma_{ab}}$ and $p^i \sqrt{\Sigma_{ab}}$ and  performing the integral over momentum ${\bf p}$, we can get energy and momentum conservation laws as follows,
\begin{equation}
	\partial_{\mu} T^{0 \mu} = E^i j^i,  \qquad
	\partial_{\mu} T^{i \mu} = n E^i + \epsilon^{ijk}j^j B^k,
\end{equation}
where,
\begin{eqnarray}
	T^{0i} &= &- \int \frac{d^3p}{(2\pi)^3} \left[(\delta^{ij} + B^i \Omega^j)
	\frac{\epsilon_{\bf p}^2}{2} \frac{\partial f_{\bf p}}{\partial p^j}  + \epsilon^{ijk}\frac{\epsilon_{\bf p}^2}{2} \Omega^j \frac{\partial f_{\bf p}}{\partial x^k} \right],    \nonumber\\
	T^{ij} & =&-\int \frac{d^3p}{(2\pi)^3} p^i 
	\left[\epsilon_{\bf p}(\delta^{jk} + B^j\Omega^k) \frac{\partial f_{\bf p}}{\partial p^k} 
	+ \epsilon^{jkl} \Omega^k \left(E^l f_{\bf p} +
	\epsilon_{\bf p} \frac{\partial f_{\bf p}}{\partial x^l}\right) \right]\nonumber\\ & & - \delta^{ij}\epsilon.  \label{energy-momentum}
\end{eqnarray}
In the above equations, expression of $\epsilon_{\bf p}$ is still not known. It can be determined using the constraint  due to Lorentz invariance, which demands that the energy flux is equal to the momentum density i.e.
\begin{equation}
	T^{0i}=\pi^{i}.  \label{T}
\end{equation}
According to Lorentz invariance above equation is valid at any order of perturbation. Using the expression of for $T^{0i}$ and $\pi^{i}$ from Eqs.(\ref{energy-momentum}) and (\ref{eq:density}), writing down the final equation to the first order in perturbations in the quantities $n_{\bf p}$ and $\epsilon_{\bf p}$, from Eq.(\ref{T}) one can obtain,
\begin{equation}
	\epsilon^0_{\bf p}=p-\frac{{\bf B} \cdot {\hat{\bf p}}}{2p}.
\end{equation}
This is the dispersion relation of chiral fermions near Fermi surface in the presence of magnetic field  \cite{Son:2012zy, Son2013a}.   
We will discuss  application of the modified kinetic theory in detail in next chapter.
\subsection{Relativistic description with anomaly}
The relativistic Boltzmann equation in the presence of external background magnetic field can be written as \cite{Pu2011}
\begin{equation}
	p^\mu \left(\frac{\partial}{\partial x^\mu}-e F_{\mu\nu} \frac{\partial}{\partial p_\nu}\right)f_p = \Pi[f_p]
\end{equation}
where the charge of the particle is $e=\pm 1$. Here $p$ denotes the on shell momentum satisfying $p^2=m^2$ (here m is mass of the particles). Also $\Pi[f_p]$ contains collision terms. In the presence of external electromagnetic field conservation equation becomes
\begin{equation}
	\partial_\mu j^\mu =-C E^\mu B_\mu
\end{equation}
Here $j^\mu$ is the current and $E^\mu=u_\nu F^{\mu\nu}$ and $B^\mu =\frac{1}{2}\epsilon_{\mu\nu\alpha\beta}u^\nu F^{\alpha\beta}$ are electric and magnetic field vectors,
respectively, where $u_\mu$ is the fluid velocity. In this case, one can verify that the equilibrium solution of the
distribution function,
\begin{equation}
	f_p(x)=\frac{1}{\text{exp}[\beta u_\mu(p^\mu-eF^{\mu\nu}x_\nu)-e\beta\mu_0]\pm 1}
\end{equation}
where $\beta=1/T$ (T is the local temperature), $\mu_0$ is the local chemical potential without
electromagnetic field. The plus/minus signs are for Bose or Fermi distributions. One can write above distribution function as follows
\begin{align}
	f_p(x)&= f_{p_{0}}(x)+\delta f_p(x)
\end{align}
where the two parts are given by-
\begin{align}
	f_{p_{0}}(x) &= \frac{1}{e^{({\bf u}\cdot{\bf p}-e\mu_0)/T}\pm 1}\nonumber \\
	\delta f_p(x)&=-f_{p_{0}}(x)[1\pm f_{p_{0}}(x)]\chi_p(x)\nonumber
\end{align}
here $\chi_p(x)=\xi_p(\mu_0, T)({\bf p}\cdot {\bf \omega})+ \xi_B(\mu_0, T)({\bf p}\cdot {\bf B})$. Here $\xi_p$ are coefficients and can be obtained from thermodynamic constraints \cite{Son2009}. 
\cleardoublepage
\chapter{Generation of magnetic fields}\label{ch3}
In this chapter, our aim is to discuss generation of the magnetic field above EW  scale in the presence of Abelian anomaly, using kinetic theory modified with Berry curvature. It has been shown in previous chapter (chapter-(\ref{ch2})) that the Abelian anomaly can be incorporated in the kinetic theory by including Berry phase effect. It is known that in the presence of chiral asymmetry, there can be instability in the chiral plasma, which leads to the creation of turbulence in the plasma and because of this magnetic fields can be generated in the early Universe. 
Typical strength of the generated magnetic field in a regime where collision dominates is $10^{27}$ G at a length scale $10^5/T$. Furthermore, the instability can also generate a magnetic field in the collisionless regime, of the order $10^{31}$ G at a typical length scale $10/T$.  
We show that the estimated values of the magnetic field are consistent with the bounds obtained from current observations.

This chapter is organized as follows. In section-(\ref{Intro:ch3}), we give a brief introduction to this chapter and  in  section-(\ref{3.1genmagew}), we apply modified kinetic theory  to study the primordial magnetic field generation in presence of the chiral asymmetry. We will also calculate the vorticity generation within  the plasma due to the chiral imbalance. Section-(\ref{discuchap3genew}) is dedicated for result and discussion.
\section{Introduction}\label{Intro:ch3}
The dominance of matter over antimatter in the observable
Universe is one of the vexing issue in the standard model of cosmology. This is due to the fact that inflation, a phase of accelerated expansion of the Universe in early time, would wash out any initial asymmetry if present. However, there are certain conditions namely, non-conservation of baryon number (B),  violation of Charge (C) and Parity (P) symmetries, and  occurrence of out of equilibrium processes in the early Universe would lead to the generation of baryon asymmetry in the universe \cite{Sakharov:1967dj}. 
In the simplest GUT models, the generated asymmetry in the baryon number suggest same amount of asymmetry in the lepton number, so that there is no net (B-L) asymmetry  \cite{Sakharov:1967dj, turner, Yoshimura:1978my}. If the Electroweak (EW) phase transition is not of the first order, then any primordial (B+L) asymmetry generated at the GUT scale, will be washed out \cite{FUKUGITA198645}. It may be possible that the Baryon Asymmetry of the Universe (BAU) could actually be reproduced from a lepton flavor asymmetries if one or more generations of leptons were out of equilibrium \cite{Campbell:1992jd, Hooft1976, Kuzmin1987}. Authors of the Ref. \cite{Campbell:1992jd} showed that the BAU could be encoded in a primordial symmetry for the right- handed electron field $e_R$ (and its antiparticles). This can be understood as follows. Out of the 45 chiral states composing the three standard model fermion generations, only right handed electrons is the most weakly coupled to the others. At high temperature electron chiral number densities, in the absence of the reactions that flip the chirality of the interacting particles, remain conserved. At such a high temperature the chirality flipping rates are out of equilibrium, which means that they are smaller than the expansion rate of the Universe ({\it i.e.} $\Gamma_{R}\sim T< H\sim T^2$ )  \cite{Joyce1997}. However at lower temperature, the processes that flip the chirality of the electrons need to be  considered \cite{Cornwall1997}.  At temperature $T\geq 80\,$ TeV, right electron number density  remain in the thermal equilibrium via its coupling with the hypercharge gauge bosons. 
It means that any primordially generated lepton number that occurred as $e_R$ would be decoupled from non-perturbative electroweak effects until  $T = 80 \,$ TeV  reach. 
The only interaction by which it connects to the other standard model states is its Yukawa coupling to the Higgs ($H\rightarrow e_R \bar{e}_L$). This interaction with the Higgs particle can changes its chirality. However a very small electron mass tends to very small Yukawa coupling $h_e\simeq \sqrt{2}m_e /v\simeq 2.1\times 10^{-6}$ \cite{Joyce1997}. 
At temperature below $80$~ TeV,  left and right electron asymmetry evolve parallel through the corresponding Abelian anomalies in the standard model in the presence of a seed hypermagnetic field. 
In this introductory part of the chapter, we have considered only the right handed particles to understand generation of magnetic field qualitatively (assumption is that there is an excess of right handed particles). However for completeness, one need to consider right as well as left handed particles (also their antiparticles). From section- (\ref{3.1genmagew}) and onward of this chapter, we have considered left and right handed particles both (and their antiparticles). 

Current density for the right handed electron satisfies following anomaly equation \cite{Joyce1997}
\begin{equation}
	\partial_{\mu} j^{\mu}_R=-\frac{g_1^{2}y_{R}^2}{64 \pi^2} Y^{\mu\nu} \tilde{Y}^{\mu\nu},
	\label{anomaly}
\end{equation}
and therefore it is coupled to the hypermagnetic field. Here, $Y^{\mu\nu}=\partial^{\mu}Y^{\nu}-\partial^{\nu}Y^{\mu}$ is the field tensor associated with the hyper-charge gauge field $Y^\mu$ and $\tilde {Y}^{\mu\nu}=1/2\epsilon^{\mu\nu\rho\lambda}
Y_{\rho\lambda}$. The coefficient, $g_1$ is a associated gauge coupling, and $y_{R}= -2$ represents the hypercharge of the right-handed electrons. The conservation of the axial current is spoiled due to quantum effects. 
It has been shown that, the term after equality sign on the right hand side of the Eq. (\ref{anomaly}) are related with the Chern-Simon (CS) number $N_{CS}$ (some times referred as helicity). The anomaly Eq. (\ref{anomaly}) relates variation of the number density of the right handed electrons with the Chern-Simon number \cite{Semikoz2005a}. CS number for hypercharge gauge fields, is defined in Eq. (\ref{chern simon definition}).
The magnetic fields generated in the presence of chiral asymmetry can be seen through following equation
\begin{equation}
	\partial_t \left(n_R+\frac{\alpha_1}{\pi}\mathcal{H}_M\right)=0, 
\end{equation} 
which can be obtained easily by  using Eq. (\ref{anomaly}). The finite helicity state can be created even if the initial state has  no helicity ($\mathcal{H}_M=0$) but $n_R\neq 0$. Thus, the magnetogenesis by net non-zero chiral-charges may not require any preexisting seed field. 
It was shown that the chiral imbalance in the early Universe can give rise to a magnetic field of the order of $B\sim 10^{22}$~G at temperature $T\sim$~100 GeV with a typical inhomogeneity scale $\sim 10^{6}/T$\cite{Joyce1997}. 
If these hyper-magnetic fields survives at the time of EWPT, they will give ordinary magnetic fields due to the electroweak mixing $A_{\mu}=\text{cos}\, \theta_w \, \text{Y}_{\mu}$. 

Recently there has been an interesting development in incorporating the parity-violating effects into a kinetic theory formalism \cite{Son:2012zy, Stephanov2012,Son2013xc,Chen2013cv}. In this approach the kinetic (Vlasov) equation is modified with  the Berry curvature term, which takes into account chirality of the particles. 
Incorporation of the parity odd physics in kinetic theory leads to a redefinition of the Poisson brackets which includes contribution from the Berry connection. 
The modified kinetic equation can also be derived from the Dirac Hamiltonian by performing semiclassical Foldy-Wouthuysen diagonalization \cite{Manuel2014a, Silenko:2007wi}. 
The confidence that the new kinetic equation captures the proper physics stems from the fact that the equation is consistent with the anomaly Eq. (\ref{anomaly}) and it also reproduces some of the known results obtained using the quantum field theory with the parity odd interaction \cite{Akamatsu2013df}.
Keeping the above discussion in mind, we believe that it would be highly useful to consider the problem of generation of the primordial magnetic field in presence of the Abelian anomaly by using the Berry curvature modified kinetic theory. In this chapter we incorporate the effect of collisions in the modified kinetic theory and derive expressions for the electric and magnetic resistivities. The new kinetic framework also allows us to calculate the generation of the primordial magnetic field and vorticity. We show that our estimated value of the peak magnetic field actually falls within the constraints obtained from current observations.
\section{Generation of magnetic fields above EW scale}\label{3.1genmagew}
In order to know about the generation of magnetic field, we intend to solve the coupled system of the modified kinetic and the Maxwell's equations in an expanding background. In this chapter,  we have ignored the fluctuations in the metric due to the matter perturbation. We write the underlying equations in a general covariant form, developed in Refs.\cite{Dettmann1993, Holcomb:1989tt, Gailis:1994dfk, Gailis:1995fdt}. This allows one to write the system of kinetic and Maxwell's equations in the expanding background in the form that look similar to their flat space-time form. In this formalism the well developed intuition and techniques of the flat space-time plasma physics can be exploited to study the problem at hand. This can be accomplished by choosing a particular set of fiducial observers (FIDO's) \cite{Dettmann1993} at each point of space-time with respect to which all the physical quantities including hyper-electric and magnetic fields are measured. The line element in an expanding background can be written using the Friedmann-Lema\^{\i}tre-Robertson-Walker metric as
\begin{equation} 
	ds^{2}=- dt^2+a^2(t)(dx^2+dy^2+dz^2)
\end{equation}
where  $x$, $y$ and $z$ represent comoving coordinates and t denotes the proper time seen by observers
at  the fixed $x$, $y$ \& $z$. The quantity a(t) is scale factor. One can introduce the conformal time $\tau$ using the definition $\tau =\int dt/a(t)$ to write this metric as:
\begin{equation}
	\label{eq:FLRWm}
	ds^{2}= a^2(\tau)(-d\tau^2+dx^2+dy^2+dz^2))
\end{equation}
The physical (hyper)-electric field $\boldsymbol{\mathcal{E}}$, (hyper)-magnetic field $\boldsymbol{\mathcal{B}}$ and the current density $\boldsymbol{J}$ are related to the corresponding fiducial quantities by transformations:
$\boldsymbol{E}^*=a^2\boldsymbol{\mathcal{E}}$, $\boldsymbol{B}^*= a^2\boldsymbol{\mathcal{B}}$, $\boldsymbol{J}^*=a^3 \boldsymbol{\mathcal{J}}$.
One can now write the Maxwell's equations in the fiducial frame as \cite{Brandenburg1996, Dettmann1993}: 
\begin{gather}
	\frac{\partial {\boldsymbol{B}}^*}{\partial \tau}+\boldsymbol{\nabla}\times\boldsymbol{E}^*=0 \label{M1}\\
	{\boldsymbol{{\nabla}\cdot E}^*}=4\pi\rho_e ^*\label{M2} \\
	{\boldsymbol{{\nabla}\cdot{B}}^*}=0 \label{M3}\\
	{\boldsymbol{{\nabla}\times{B}}^*}= 4\pi {\boldsymbol{J}^*}+\frac{\partial {{\boldsymbol{E}}^*}}{\partial \tau} \label{M4}
\end{gather}
where $\boldsymbol{B}^*$, $\boldsymbol{E}^*$, $\rho_e$ and $\boldsymbol{J}^*$ are respectively magnetic field, electric field, charge density and current density seen by the fiducial observer. To make our notations simpler,  we will not use star on the variables in the discussions below. The charge $\rho_e$ and the current $\boldsymbol{J}$ in the Maxwell's eqs. (\ref{M1}-\ref{M4}) can be calculated using the Berry curvature modified kinetic equation which is also consistent with the quantum anomaly equation. The modified kinetic equation is given by
\begin{align}
	\frac{\partial {f}}{\partial \tau}+
	\frac{1}{1+e\boldsymbol{\Omega_{\boldsymbol{p}}}\cdot\boldsymbol{B}}
	\bigg[(e\boldsymbol{\Tilde{E}}+e{\boldsymbol{\Tilde{v}}\times\boldsymbol{B}}
	+e^2{(\boldsymbol{\Tilde{E}}\cdot{B}) \boldsymbol{\Omega_p})}\cdot\frac{\partial{f}}{\partial{\boldsymbol{p}}}       \nonumber \\
	+({\boldsymbol{\Tilde{v}}}+e {\boldsymbol{\Tilde{E}}\times\boldsymbol{\Omega_p}}+e({\boldsymbol{\Tilde{v}}\cdot\boldsymbol{\Omega_p})
		\boldsymbol{B}})\cdot 
	\frac{\partial{f}}{\partial{\boldsymbol{r}}}\bigg]=\left(\frac{\partial f}{\partial \tau}\right)_{coll} \label{eqwberry}
\end{align}
Here $\boldsymbol{\Tilde{v}}=\partial \epsilon_{\boldsymbol{p}}/\partial{\boldsymbol{p}}=\boldsymbol{v}$
and $e \boldsymbol{\Tilde{E}}= e \boldsymbol{E}-\partial{\epsilon_{\boldsymbol{p}}}/\partial
\boldsymbol{r}$. $\boldsymbol{\Omega_{p}}= \pm \boldsymbol{p}/(2p^3)$ is Berry curvature. $\epsilon_{\boldsymbol{p}}$ is defined as $\epsilon_{\boldsymbol{p}}=p(1-e\boldsymbol{B}
\cdot\boldsymbol{\Omega_{p}})$ with $p=|\boldsymbol{p}|$. The positive sign corresponds to right-handed fermions where as the negative sign is for left-handed ones. In the absence of Berry correction i.e. $\boldsymbol{\Omega_p}=0$, Eq. (\ref{eqwberry})  reduces to the Vlasov equation when the collision term on the right hand side is absent.

As we have already stated that we are interested in temperature regime $T>T_f (\sim 80 \text{TeV}) \gg T_{EW} (\sim 100 \text{GeV})$, where electrons are ultra-relativistic and the only process that can change electron chirality is its Yukawa interaction with Higgs boson. However, at this temperature this interaction is not strong enough to alter the electron chirality. It is important to note that for temperature smaller than $T_R$, electron mass may play a major role in left-right asymmetry. In Ref.\cite{Grabowska:2015Ksd}, authors have shown that the mass of the electron plays an important role in determining the magnetic properties of the proto-neutron star by suppressing the chiral charge density during the core collapse of supernova at temperature of the order of MeV. However for the present case we ignore the electron mass by considering only $T>T_f$ regime. 

One can use Eq. (\ref{eq:current} given in chapter-(\ref{ch2})), for writing the expression of the current for each species.  Total current can be written as the sum of current contribution from each species i.e., ${\bf j}= \Sigma_{s}{\bf j}_s$. Here index $"s"$ denotes
current contribution from different species of the fermion e.g. right and left handed particles and their antiparticles. ${\bf j}_s$ is defined as:
\begin{align} \label{current1}
	{\bf j}_{s}= - e^s\int 
	\frac{d^3p}{(2\pi)^3}\bigg[\epsilon_{\boldsymbol{p}}^s\frac{\partial {f}_{s}}{\partial \boldsymbol{p}}
	+e^s(\boldsymbol{\Omega_{\boldsymbol{p}}^s}\cdot\frac{\partial{f}_{s}}{\partial{\boldsymbol{p}}})\epsilon_{\boldsymbol{p}}^s{B} 
	+\epsilon_{\boldsymbol{p}}\boldsymbol{\Omega_{\boldsymbol{p}}^s}\times\frac{\partial{f}_{s}}{\partial \boldsymbol{r}}\bigg]
\end{align}
where 
$\epsilon_{\boldsymbol{p}}^s=p(1-e^s\boldsymbol{B} \cdot\boldsymbol{\Omega_{p}}^s)$ with $p=|\boldsymbol{p}|$. Depending on the species, charge $e$, energy of the particles $\epsilon_p$ and Berry curvature $\boldsymbol{\Omega_p}$, form of the distribution function $f$ changes. 

For the right-handed particle, $s=R$, hyper-charge is $e$ while for right-handed anti-particle $s=\bar{R}$ and charge is $-e$ and similarly for other particles. The first term in Eq. (\ref{current1}) is the usual current equivalent to the kinetic theory and remaining second and  third terms are the current contribution due to the Berry correction.
If we follow the power counting scheme as in \cite{Son:2012zy} i.e. $\boldsymbol{Y_{\mu}}=\mathcal{O}(\epsilon)$ , $\partial_r=\mathcal{O}(\delta)$ and considering only terms of the order of $\mathcal{O}(\epsilon\delta)$ in Eq. (\ref{eqwberry}), we will have:
\begin{equation}
	\bigg(\frac{\partial }{\partial\tau}+{\boldsymbol{v}}\cdot\frac{\partial}{\partial \boldsymbol{r}}\bigg)f_s+
	\bigg(e^{s}\boldsymbol{E}+e^{s}(\boldsymbol{v}\times \boldsymbol{B})- 
	\frac{\partial{\epsilon^{s}_p}}{\partial{\boldsymbol{r}}}\bigg)\cdot\frac{\partial f_s}{\partial \boldsymbol{p}}= \bigg(\frac{\partial f_s}{\partial\tau}\bigg)_{\text{coll.}}
	\label{vlasov1}
\end{equation}
Above equation is similar to the Boltzmann equation with one additional term $\frac{\partial \varepsilon_p^s}{\partial{\bf r}}\cdot \frac{\partial f_s}{\partial\tau}$. This term comes only because of inclusion of Berry curvature.
\subsection{Current and polarization tensor for chiral plasma}
Here we assume that the plasma of the standard model particles is in a state of "thermal-equilibrium" at temperature $T>T_R$, where 
masses of the plasma particles can be ignored. We also assume that there exist a left-right asymmetry and there is no large-scale electromagnetic field. Thus the equilibrium plasma is considered to be in a homogeneous and isotropic state which is similar to the assumptions made in Ref.\cite{Joyce1997, Boyarsky2012a}. For a homogeneous and isotropic conducting plasma in thermal equilibrium, distribution function for different species are:
\begin{equation}
	f_{0a}(p)=\frac{1}{exp(\frac{\epsilon_p^0-\mu_a}{T})+1}
\end{equation}
If $\delta f_R$ and $\delta f_{\bar{R}}$ are fluctuations in the distribution functions of the right electron and right-antiparticles around their equilibrium distribution, we can write perturbed distribution functions as
\begin{eqnarray}
	f_{R}({\boldsymbol{r},\boldsymbol{p}}, \tau)= f_{0R}(p)+ \delta f_{R}({\boldsymbol{r, p}}, \tau) \\
	f_{\bar{R}}({\boldsymbol{r,p}},\tau)=f_{0\bar{R}}(p)+ \delta f_{\bar{R}}({\boldsymbol{r,p}},\tau)
\end{eqnarray}
Subtracting equation for $a=\bar{R}$ from $a=R$ and using Eq. (\ref{vlasov1}) one can write:
\begin{align}
	\bigg(\frac{\partial}{\partial\tau}+ {\boldsymbol{v}}\cdot\frac{\partial}{\partial {\boldsymbol{r}}}\bigg)f({\boldsymbol{r,p}},\tau)
	&+ep\frac{\partial(\boldsymbol{{B}\cdot \Omega_p})}{\partial\boldsymbol{r}}\cdot 
	\frac{\partial f_0}{\partial \boldsymbol{p}} 
	+ ({\boldsymbol{{E}.v}})\frac{d {f_0}}{dp} \nonumber \\
	&=\left(\frac{\partial f(\boldsymbol{r,p},\tau)}{\partial\tau}\right)_{coll.} \label{R-R_}
\end{align}
where, $f(\boldsymbol{r,p},\tau)=(f_R-f_{\bar{R}})$ and $f_{0}=f_{0R}+f_{0\bar{R}}$. Here we have used $\frac{\partial f^0}{\partial\boldsymbol{p}}=\boldsymbol{v}\frac{df^0}{dp}$. This equation relates the fluctuations of the distribution functions of the charged particles with the induced gauge field fluctuations.
The gauge field fluctuations can be seen from the Maxwell's electromagnetic eqs. (\ref{M1}-\ref{M4}). Under the relaxation time approximation, the collision term can be written as $({\partial f_a}/{\partial\tau})_{coll.}\approx -\nu_{c}(f_a-f_{0a})$ (one can also look at some studies in chiral kinetic theory with collision in Ref. \cite{Chen:2014cla,Chen:2015gta}). As above differential Eq. (\ref{R-R_}), is linear in the distribution function, one can use Fourier transform to reduce this differential equation into a algebraic form. For this, we have taken Fourier transform of the perturbed quantities namely $\boldsymbol{E}$, $\boldsymbol{B}$ and $f(\boldsymbol{r,p},\tau)$ by considering the spatio-temporal variation of these quantities as
$ exp[-i(\omega\tau-\boldsymbol{k\cdot r})]$. Then using Eq. (\ref{R-R_}) one can get
\begin{equation}
	f_{{\boldsymbol{k}},\omega}=\frac{-e[({{\boldsymbol{v}\cdot \boldsymbol{E}_{\boldsymbol{k}}}})
		+\frac{i}{2p}({\boldsymbol{v}\cdot\boldsymbol{B}_{\boldsymbol{k}}})(\boldsymbol{k\cdot v})]
		\frac{df_0}{dp}}{i({\boldsymbol{k\cdot v}}  -\omega-i\nu_c)}\label{f_{kw}},
\end{equation}
here subscript $k$ with the variables represent their Fourier transformed values. One can calculate, current for the right handed particle and right handed antiparticles in terms of the mode functions using Eq.(\ref{current1}) as:
\begin{align} 
	\boldsymbol{J}_{\boldsymbol{k}R}= & e\int \frac{d^3p}{(2\pi)^3}\bigg[\{{\boldsymbol{v}}-\frac{i}{2p}\boldsymbol{(v\times k)}\}
	f_{\boldsymbol{k}\omega R} \nonumber \\
	&-\frac{e}{2p^2} \{\boldsymbol{B_k} 
	-\boldsymbol{v}(\boldsymbol{v}\cdot \boldsymbol{B}_{\boldsymbol{k}})\}f_{0}
	+\frac{e}{2p}\boldsymbol{B}_{\boldsymbol{k}}\frac{df_{0}}{dp}\bigg]\label{J_{kw}} 
\end{align}
%
In the similar way we can get currents due to left handed particles and left handed antiparticles. 
So the total current can be obtained by adding the contributions from both left and right handed particles and antiparticles by putting perturbations $f_{k\omega}$ for all species in Eq. (\ref{J_{kw}}) and adding as:
\begin{align} \label{j total}
	\boldsymbol{J}_{\boldsymbol{k}} = &-m_D^2 \int \frac{d\Omega}{4\pi} \frac{\boldsymbol{v}(\boldsymbol{v}.
		\boldsymbol{E_{k}})}{i(\boldsymbol{k.v}-\omega-i \nu_c)} \nonumber \\
	&-\frac{h_D^2}{2} \int \frac{d\Omega}{4\pi}\{\boldsymbol{B}_k-\boldsymbol{v}(\boldsymbol{v. B_k})\}\nonumber \\
	&-\frac{i g_D^2}{4}\int \frac{d\Omega}{4\pi} \frac{(\boldsymbol{v \times k})(\boldsymbol{v.B_k})(\boldsymbol{k.v})}{
		(\boldsymbol{k.v}-\omega-i \nu_c)}\nonumber \\
	&-\frac{c_D^2}{2}\int\frac{d \Omega}{4\pi} \bigg\{\frac{\boldsymbol{v} (\boldsymbol{v.B_{k}})(\boldsymbol{k.v})-
		(\boldsymbol{v\times k})(\boldsymbol{v.E_{k}})}{(\boldsymbol{k.v}-\omega-i \nu_c)}+ \boldsymbol{B}_{k}\bigg\},
\end{align}
where $\Omega$ represent angular integrals. In Eq. (\ref{j total}), we have defined $m_D^2 =e^2\int \frac{p^2 dp}{2\pi^2} \frac{d}{dp}( f_{0R}+f_{0\bar{R}}+f_{0L}+f_{0\bar{L}})$,
$c_D^2 =e^2\int \frac{p dp}{2\pi^2}\frac{d}{dp}( f_{0R}-f_{0\bar{R}}-f_{0L}+f_{0\bar{L}})$, $g^2_D=e^2\int
\frac{dp}{2\pi^2}\frac{d}{dp}( f_{0R}+f_{0\bar{R}}+f_{0L}+f_{0\bar{L}})$, $h^2_D=e^2\int \frac{dp}{2\pi^2}
( f_{0R}-f_{0\bar{R}}-f_{0L}+f_{0\bar{L}})$.

Expression for the polarization tensor $\Pi^{ij}$ can be obtained from Eq. (\ref{j total}) by writing the total 
current in the following form $J^{i}_{k}=\Pi^{ij}(k)Y_{j}(k)$ using $\boldsymbol{E_{k}}=-i \omega \boldsymbol{Y_{k}}$ and $\boldsymbol{B_{k}}= i (\boldsymbol{k}\times \boldsymbol{Y_{k}})$. One can express $\Pi^{ij}$ in terms of longitudinal $P^{ij}_{L}= k^{i}k^{j}/k^2$, transverse  $P^{ij}_{T}= (\delta^{ij}-k^{i}k^{j}/k^2)$ and the axial $P^{ij}_{A}=i\epsilon^{ijk}k^k$ projection operators as $\Pi^{ik}=\Pi_{L}P_{L}^{ik}+\Pi_{T}P_{T}^{ik}+\Pi_{A}P_{A}^{ik}$. After performing the angular integrations in Eq. (\ref{j total}) one obtains $\Pi_L$, $\Pi_T$ and $\Pi_A$ as given below
\begin{gather} \label{polarization}
	\Pi_{L} =-m_D^2\frac{\omega\omega^{\prime}}{k^2}[1-\omega^{\prime} L(k)],\dotfill\\
	\Pi_{T} = m_D^2\frac{\omega\omega^{\prime}}{k^2}\bigg[1+\frac{k^2-\omega^{\prime 2}}{\omega^{\prime}}L(k)\bigg],\dotfill\\
	\Pi_{A} =-\frac{h_{D}^2}{2}  \bigg[1-\omega(1-\frac{\omega^{\prime 2}}{k^2})L(k)-\frac{\omega^{\prime}\omega}{k^2}\bigg]
\end{gather}
where, ${\textstyle L(k)=\frac{1}{2k}ln\bigg(\frac{\omega^{\prime}+k}{\omega^{\prime}-k}\bigg)}$, ${\textstyle \omega^{\prime}=\omega+i \nu_{c}}$, ${\textstyle m^2_D= 4\pi \alpha_1 \left(\frac{T^2}{3}+\frac{\mu^2_R+\mu^2_L}{2\pi^2}\right)}$ and ${\textstyle h^2_D= \frac{2\alpha_1\Delta\mu}{\pi}}$ (with ${\textstyle \Delta\mu=(\mu_R-\mu_L)}$).
In the above integrals, we have replaced, $e$ by $\alpha_1$ using relation $e^2=4\pi \alpha_1$. 

One can consider two cases: collision dominated regime (where $\nu_c\neq 0$) and collisionless regime ($\nu_c=0$).
First,  consider case when  $\nu_c\neq 0$. In the limit $\omega\rightarrow 0$, $\Pi_L$ and $\Pi_T$  vanish and
the parity odd part of the polarization tensor $\Pi_A= h_D^2/2\approx \alpha_1 \Delta\mu/\pi $. Note that $\Pi_A$ does not get thermal correction. This could be due to the fact that origin of $\Pi_A$ term is related with the axial anomaly and it is well known that anomaly does not receive any thermal correction \cite{Liu:1988ke, Itoyama:1982up,TGomezNicola:1994vq}. This form of $\Pi_A$ is similar to the result obtained in \cite{Boyarsky2012} using quantum field theoretic arguments at $T\leq 40$~GeV. In this work, authors discussed that the ground state of the plasma of standard model particles can be determined by the temperature and exactly conserved combination of baryon and lepton numbers in the thermal equilibrium. They show that at non-zero values of the global charge, states that are translationally invariant and homogeneous becomes unstable.  In such a situation, in the thermal equilibrium state a large scale magnetic field exists. However, in the kinetic theory approach presented in our work, we have not made any such assumption.  Normal modes for the plasma can be obtained by using expressions for $\Pi_L$, $\Pi_T$ and $\Pi_A$. Using the equation $\partial _{\nu}F^{\mu\nu}= -4\pi J^{\mu}$, we can write the following relation
\begin{equation}
	[M^{-1}]^{ij}Y_{j}(k)=-4\pi J^{i}_{k},
\end{equation}
where, $[M^{-1}]^{ik}=[(k^2-\omega^2)\delta^{ik}-k^ik^k+\Pi^{ik}]$. Dispersion relations can be obtained from the poles of $[M^{-1}]^{ik}$, which are given below
\begin{gather*}
	\omega^2=\Pi_L,\\
	\omega^2= k^2+\Pi_{T}(k)\pm k\Pi_{A}.
\end{gather*}
Once dispersion relations are known, one can study dynamical behavior of the magnetic fields. 
Expression for the total current described by Eq.(\ref{j total})
can be written as $J_{k}^{i}= \sigma_{E}^{ij}{E_k}^{j}+ \sigma_{B}^{ij} B_{k}^{j}$ where $\sigma_{E}^{ij}$ and $\sigma_{B}^{ij}$ are electrical and magnetic conductivities. The integrals involved in Eq. (\ref{j total}) are rather easy to evaluate in the limit $k,\omega \ll \nu_c$ and one can write the expression for $\sigma_E^{ij}$ and $\sigma_B^{ij}$  as:
\begin{gather} 
	\sigma_{E}^{ij} \approx 
	\left(\frac{m_D^2}{3\nu_c}\delta^{ij}+\frac{i}{3\nu_c}\frac{\alpha_1
		\Delta \mu}{\pi}\epsilon^{ijl}k^{l}\right)    \label{sigma E}\\
	\sigma_{B}^{ij} \approx
	-\frac{4}{3}\frac{\alpha_1 \Delta \mu}{\pi} \delta^{ij}  \label{sigma B}   
\end{gather}
Here, we would like to highlight that the Berry curvature correction in the kinetic equation gives us an additional contribution in the expression for $\sigma^{ij}_E$ which was not incorporated in Ref.\cite{Joyce1997}. Note that, first term is the usual dissipative part of the electric current and it contributes to the Joules dissipation. The second term is due to the chiral imbalance and it does not give any contribution to the Joules heating. As we shall demonstrate later, this term is responsible for the vorticity current \cite{Fukushima2008}. One can write the total current as $J^{i}_{\boldsymbol{k}}=\sigma^{ij}_E E^{j}_{\boldsymbol{k}}
+\sigma^{ij}_B B^{j}_{\boldsymbol{k}}$ and the Maxwell's equation:
$i(\boldsymbol{k\times B_k})^{i}= 4\pi J^{i}_{\boldsymbol{k}}$. Here we have
dropped the displacement current term (this is valid when $\frac{\omega}{4\pi\sigma}\ll 1$).
Next by taking vector product of $\boldsymbol{k}$ with the above Maxwell equation one obtains
\begin{align} \label{B_kw vector}
	\frac{\partial \boldsymbol{B}_k}{\partial \tau}+\left(\frac{3\nu_c}{4\pi m_d^2}\right)k^2\boldsymbol{B}_k 
	& -\left(\frac{\alpha_1\Delta\mu}{\pi m_D^2}\right)\left(\boldsymbol{k}\times
	(\boldsymbol{k}\times\boldsymbol{E_k})\right) \nonumber \\
	&+i\frac{4\alpha_1 \nu_c \Delta\mu}{\pi m_D^2} (\boldsymbol{k\times B_k})=0.
\end{align}
This is the magnetic diffusivity equation for the chiral plasma. By replacing $(\boldsymbol{k \times E_k})$ by $-\frac{1}{i}\frac{\partial \boldsymbol{B_k}}{\partial \tau}$ in Eq. (\ref{B_kw vector}), we can solve this equation. Without a loss of generality, consider the propagation vector $\boldsymbol{k}$ in the $z-$direction and the magnetic field having components perpendicular to the $z-$ axis. After defining two new variables: $\tilde{B_k}= (B_k^1+ iB_k^2)$ and $\tilde{B^{\prime}_k}= (B_k^1- i B_k^2)$ one
can rewrite Eq. (\ref{B_kw vector}) as,
\begin{eqnarray}
	\frac{\partial\tilde{B}_{k}}{\partial \tau}+\left[\frac{\left(\frac{3\nu_c}{4\pi m_d^2}\right)k^2-
		\left(\frac{4\alpha_1\nu_c \Delta\mu}{\pi m_D^2}\right)k}
	{(1+\frac{\alpha_1\Delta \mu k}{\pi m_D^2})}\right]\tilde{B}_{k}=0, \label{mode1} \\
	\frac{\partial\tilde{B}_{k}^{\prime}}{\partial \tau}+\left[\frac{\left(\frac{3\nu_c}{4\pi m_d^2}\right)k^2+
		\left(\frac{4\alpha_1 \nu_c \Delta\mu}{\pi m_D^2}\right)k} {(1-\frac{\alpha_1\Delta \mu k}{\pi m_D^2})}\right]
	\tilde{B}_{k}^{\prime}=0.\label{mode2}
\end{eqnarray}
Thus, the magnetic field vector ${\bf B}$ can be decomposed into these new variables $\tilde{\bf B}_k$ \& $\tilde{\bf B}^{\prime}_k$ having definite helicity
(or circular polarization). In the above equation, the effect of Ohmic decay is already taken into account
by including the collisions. It should be noted here that if $\alpha_1\Delta \mu k/\pi m_D^2 \ll 1$ Eq. (\ref{mode1}) is similar to the magnetic field evolution equation considered in Ref.\cite{Joyce1997}. In this limit Eq. (\ref{mode2}) will give a purely damping mode and the dispersion relation reduces to
\begin{equation} \label{dispersr}
	i \omega =\frac{3\nu_c}{4\pi m_d^2}k^2- \frac{4\alpha_1\nu_c \Delta\mu}{\pi m_D^2}k
\end{equation}
The dispersion relation obtained here using kinetic theory matches with the dispersion relation obtained in \cite{Akamatsu2013df}. 

The instability can also occur in the collisionless regime ($\nu_c=0$) \cite{Akamatsu2013df}. In
the quasi-static limit i.e. $\omega\ll k$, one can define the electric conductivity as $\sigma^{ij}_E\approx \pi(m^2_D/2k)\delta^{ij}$ and magnetic conductivity $\sigma^{ij}_B\approx (h^2_D/2)\delta^{ij}$. Note that the above conductivities are independent of collision frequency. Similar to the previous case one can take the propagation vector in the $z-$direction and consider components of the magnetic field in the direction perpendicular to the $z-$ axis. One can write a set of decoupled equations describing the evolution of magnetic field using the variables $\tilde{\bf B}_{k}$ and $\tilde{\bf B_k}^{\prime}$ as:
\begin{eqnarray}
	\frac{\partial\tilde{\bf B}_{k}}{\partial \tau}+\Bigg[\frac{k^2-\frac{4\alpha_1 \Delta \mu k}{3}}{\frac{\pi m_D^2}{2k}}
	\Bigg]\tilde{\bf B}_{k}=0, \label{mode11}\\
	\frac{\partial\tilde{\bf B}_{k}^{\prime}}{\partial \tau}+\Bigg[\frac{k^2+\frac{4\alpha_1
			\Delta \mu k}{3}}{\frac{\pi m_D^2}{2k}}
	\Bigg]\tilde{\bf B_k}^{\prime}=0.\label{mode22} 
\end{eqnarray}
If one replaces $\partial/\partial\tau$ by $-i\omega$ Eqs. (\ref{mode11}) and (\ref{mode22}) gives the same dispersion relation for the instability as discussed in
Ref.\cite{Akamatsu2013df}.
\subsection{Vorticity generated from chiral imbalance in the plasma}
It would be interesting to see if the instabilities arising due to chiral-imbalance can lead to
vorticity generation in the plasma. In order to study vorticity of the plasma, we define the average velocity as:
\begin{equation} \label{vorticity}
	<\boldsymbol{v}> = \frac{1}{\bar n}\int \frac{d^3p}{(2\pi)^3}\boldsymbol{v} (\delta f_R- \delta f_{\bar{R}}+\delta f_L-
	\delta f_{\bar{L}})
\end{equation}
Here we have used the perturbed distribution function in the numerator of the above equation
which is due to the fact that the equilibrium distribution function are assumed to homogeneous and
isotropic and therefore will not contribute to vorticity dynamics. The denominator is the total number density and is defined as in Ref. \cite{turner}:
\begin{align}
	\bar n =&n_{particle}-n_{antiparticle} \nonumber\\
	= & d_f\int_0^{\infty} \frac{d^3p}{(2\pi)^3}\left(\frac{1}{1+exp(\frac{p-\mu}{T})}-\frac{1}{1+exp(\frac{p+\mu}{T})}\right)
\end{align}
which in the case of chiral plasma reduces to $\bar n= \frac{2}{3} T^2 (\mu_R +\mu_L)$. $d_f$ is degree of freedom of the particles and $\mu$ is the chemical potential.  We consider $k, \omega\ll\nu_c$ regime. In this case the perturbed distribution function, say for the right-handed particles, can be written as:
\begin{equation} 
	\delta f_{{\boldsymbol{k}},\omega R}=-\frac{e}{\nu_c}[({{\boldsymbol{v}\cdot \boldsymbol{E}_{\boldsymbol{k}}}})
	+\frac{i}{2p}({\boldsymbol{v}\cdot\boldsymbol{B}_{\boldsymbol{k}}})(\boldsymbol{k\cdot v})]\frac{df_{0R}}{dp} .
	\label{R perturbed}
\end{equation}
If we add the contribution for all the particles species and their anti-particles we can write the numerator in equation (\ref{vorticity}) as
\begin{equation}
	\sqrt{\frac{\alpha_1}{\pi^3}} \frac{1}{\nu_c}\left(\frac{T^2}{3}+\frac{\mu_R^2+\mu_L^2}{2\pi^2}\right)\boldsymbol{E}_k
	\approx \sqrt{\frac{\alpha_1}{\pi^3}} \frac{T^2}{3\nu_c}\boldsymbol{E}_k.
\end{equation}
Above we have neglected $\frac{3}{2\pi^2}\frac{\mu_R^2+\mu_L^2}{T^2}$ in comparison to one, as we have considered $\mu_R/T\ll 1$ and $\mu_L/T\ll 1$. One can write the average velocity as follows
\begin{equation} \label{mu3.38}
	<\boldsymbol{v}_k> =\sqrt{\frac{\alpha^\prime}{\pi^3}} \frac{1}{2\nu_c}\frac{1}{\mu_R+\mu_L}\boldsymbol{E}_k 
\end{equation}
Assuming that the chiral chemical potential and temperature are independent of space and time, we can obtain vorticity by taking curl of the Eq. (\ref{mu3.38}),  which takes the following form
\begin{equation}
	<\boldsymbol{\omega}_k> = i\sqrt{\frac{\alpha^\prime}{\pi^3}} \frac{1}{2\nu_c}\frac{1}{\mu_R+\mu_L}(\boldsymbol{k}\times\boldsymbol{E}_k)
	\label{vorticity1}
\end{equation}
One can find contribution of the vorticity to the total current from eqs. (\ref{sigma E}) and
${J}_{k\omega}^{i}= \sigma_{E}^{ij}{ E_k}^{j}+ \sigma_{B}^{ij} B_{k}^{j}$. 
By using Eq. (\ref{vorticity1}) the vorticity current can be written as
\begin{equation} 
	\boldsymbol{J_{\omega}} \approx -\sqrt{\frac{4\pi\alpha_1}{9}}(\mu_R^2-\mu_L^2)\boldsymbol{\omega}= \xi \boldsymbol{\omega}
	\label{vcurrent}
\end{equation}
\noindent
It is clear from the Eq. (\ref{vcurrent}) that in the absence of any chiral-imbalance there is no vorticity current. Here we note that our definition agrees with the Ref.\cite{Son2009}. Further using Eq. (\ref{vorticity1}), one can eliminate $\left(\boldsymbol{k\times E_k}\right)$ in Eq. (\ref{B_kw vector}) and obtain:
\begin{align}
	\frac{\partial \boldsymbol{B}_k}{\partial \eta}&+\frac{3\nu_c}{4\pi m_d^2}k^2\boldsymbol{B}_k 
	+i \frac{\sqrt{4\pi\alpha_1}\nu_c}{m_D^2} (\mu_R^2-\mu_L^2)(\boldsymbol{k}\times\boldsymbol{\omega_k}) \nonumber\\
	&+i\left(\frac{4\alpha_1\nu_c }{\pi m_D^2}\right)(\mu_R-\mu_L)(\boldsymbol{k\times B_k})=0.
	\label{diffusivity1}
\end{align}
In this equation, second term is the usual diffusivity term, however third and fourth terms are additional terms, which respectively represent vorticity and chiral magnetic effects in the chiral plasma. Therefore Eq. (\ref{B_kw vector}) actually contains terms due to vorticity and magnetic effect. The saturated
state of the instability can be studied by setting $\partial_{\tau} \boldsymbol{B_k}=0$ in Eq. (\ref{diffusivity1}). After taking dot product of Eq. (\ref{diffusivity1}) with fluid velocity $\boldsymbol{v_k}$ in the saturated state, $\partial_{\tau} \boldsymbol{B_k}=0$, leads to,
\begin{equation}
	\left(\boldsymbol{\omega_k}-i\frac{16T\delta }{3}\boldsymbol{v}_k\right)\cdot \boldsymbol{B_k}=0. \label{43}
\end{equation}
where we have defined $\delta= \alpha_1(\mu_R-\mu_L)/T$. We can write expression for magnetic field, which satisfies above Eq. (\ref{43}) as:
\begin{equation}
	\boldsymbol{B_k}= g(\boldsymbol{k}) \boldsymbol{k}\times
	\left[\boldsymbol{\omega_k}-i\frac{16T \delta}{3}\boldsymbol{v}_k\right]
\end{equation}
Where $g(k)$ is any general function, which can be determined by  substituting the above expression for the magnetic field into Eq. (\ref{diffusivity1}) in the case of steady state. On very large length scales, i.e. $\boldsymbol{k}\rightarrow 0$:
\begin{equation}
	g(\boldsymbol{k})= -\frac{3}{32}\sqrt{\frac{\pi^3}{\alpha^{\prime 3}}}\frac{\mu_R^2-\mu_L^2}{(\mu_R-\mu_L)^2}
\end{equation}
Therefore, in this limit, for a very large length scale, $\boldsymbol{k}\rightarrow 0$, magnetic field in the steady state has the form:
\begin{equation}\label{max B with g}
	\boldsymbol{B_k}= -i\sqrt{\frac{\pi^3}{4\alpha_1}}\frac{\mu_R^2-\mu_L^2}{(\mu_R-\mu_L)}\boldsymbol{\omega_k}
\end{equation}
This equation relates the vorticity generated during the instability with the magnetic field in the steady state.

However in the collisionless regime ($\omega\ll k$ and $\nu_c=0$), one can have an instability described by
Eq. (\ref{mode11}) with typical scales $k\sim \alpha_1\Delta\mu$ and $|\omega|\sim \alpha^{\prime 2} T \delta$ \cite{Akamatsu2013df}. Using the expression for electric and magnetic conductivities for modes in this regime one can write the magnetic diffusivity equation as:
\begin{equation}
	\frac{\partial \boldsymbol{B_k}}{\partial \tau}+ \frac{k^2}{4\pi \sigma_1}\boldsymbol{B_k}- i
	\frac{T \delta}{\pi \sigma_1}\left(\boldsymbol{k}\times\boldsymbol{B_k}\right)=0
\end{equation}
where $\sigma_1 =\pi m_D^2/2k$. Here it should be noted that unlike Eq. (\ref{diffusivity1}), the above equation does not have a vorticity term. The last term on the left-hand side arises due to the chiral-magnetic effect.  In the steady state, 
one get $\boldsymbol{\nabla\times B}=(4 T\delta)\boldsymbol{B}$.  This equation resembles the case of magnetic field in a force free configuration of the conventional plasma where the plasma pressure is assumed to be negligible in comparison with the magnetic pressure \cite{Chandrasekhar01041958}.
But for our case no such assumption about the plasma pressure is required.
\section{Discussion}\label{discuchap3genew}
In this chapter, we have applied the modified kinetic theory in the presence of chiral imbalance and obtained equations for the magnetic field generation for both the collision dominated and the collisonless regimes. The instability can lead to generation of the magnetic field at the cost of the chiral imbalance. One can estimate the strength of the generated magnetic field as follows. From Eqs. (\ref{anomaly}) and (\ref{chern simon definition}), one can notice that right-handed electron number density $n_R$ changes with the Chern-Simon number $n_{CS}$ of  the hyper-charge field configuration
as 
\begin{equation}
	\Delta n_R =\frac{1}{2}y^2_R n_{cs}
\end{equation}
Here $n_{CS} \approx \frac{g^{2}_1}{16\pi^2} k Y^2$ and $\Delta n_R=\mu_R T^2 =\frac{88}{783}\delta T^3$\cite{turner}. From this, one can the estimate magnitude of the generated physical magnetic field to be
\begin{equation}
	B^{phy}_Y\approx 10^{-3}\,\left[\frac{\pi^2k}{g^{2}_1\alpha_1T} \right]^{\frac{1}{2}}\left(\frac{\delta}{10^{-6}}\right)^{1/2}T^{2},
	\label{Bfield}
\end{equation}
where we have used $\mathcal{B}_Y \sim k Y$ and $k^{-1}$ is physical length scale, which is related with the
comoving length by $k_{phy}^{-1}=(a/k_c)^{-1}$.

Now consider the regime $\omega,k\ll\nu_c$ where dynamics for the magnetic field is described by Eqs.(\ref{mode1}-\ref{mode2}). Eq.(\ref{mode1}) clearly gives unstable modes if $\left(\frac{T \delta }{\pi m^2_D}\right)k<1$  is satisfied. However, Eq.(\ref{mode2}) gives a purely damping mode if the condition
$\left(\frac{T \delta }{\pi m^2_D}\right)k\ll 1$ is satisfied. One can rewrite this condition as $\left(\frac{T\delta}{3\pi\sigma\nu_c}\right)k\sim \left(\frac{10^{-2}}{3\pi}\frac{\delta}{\nu_c}\right)k\ll1$.Here we have used $m^2_D=3\nu_c\sigma$ with
$\nu_c\sim \alpha^{\prime 2} ln(\frac{1}{\alpha_1})T$ \cite{Baym1997abc} and $\sigma=100T$. Thus for $k\ll\nu_c$ and $ \delta \ll1$, Eq. (\ref{mode2}) can only give purely damped modes. For these values of $k$ and $\delta$,  Eq. (\ref{mode1})  assumes similar form as the equation for the magnetic field dynamics considered in Ref.\cite{Joyce1997}. If one replaces $\frac{\partial}{\partial\tau}$ by $-i\omega$ in Eq.(\ref{mode1}), the dispersion relation for the unstable modes can be obtained. The fastest growth of the perturbation occurs for $k_{max1}\sim \frac{8T \delta}{3}$ and the maximum growth rate can be found to be $\Gamma_1\sim \frac{16}{3\pi}\frac{T^2 \delta^2}{m^2_D}\nu_c$. Here we note that our  $k_{max1}$  differs by a numerical factor from the value of $k$ where there there is a peak in the magnetic 
energy calculated using chiral magnetohydrodynamics \cite{Tashiro2012}. For $\delta\sim 10^{-6}$
and $\alpha_1\sim 10^{-2}$ one can show that $\frac{k_{max1}}{\nu_c}\ll 1$ and $\frac{\Gamma_1}{\nu_c}\ll1$ is satisfied. For these values of $k_{max1}$, $\alpha_1$ and $\delta$ one can estimate the magnitude of the generated magnetic field using Eq. (\ref{Bfield}). We find $\mathcal{B}\sim 10^{26}$~Gauss for $\alpha_1\sim 10^{-2}$ and  the typical length scale $\lambda\sim 10^5/T$. Here we would like to note that the typical Hubble length scale $\sim 10^{13}/T$, which is much larger than the typical length scale of instability. Our estimate of the magnetic field strength $\mathcal{B}$ in the collision dominated regime broadly agree with Ref.\cite{Joyce1997}. Here we note again that Eq. (\ref{mode1}) includes effect of the Ohmic decay due to presence of the collision term. Our analysis shows that Ohmic decay is not important for the instability. Further we have shown that the chiral instability can also lead to generation of vorticity in the collision dominated regime. Typical length scale for vorticity is similar to that of the magnetic field. From Eq.(\ref{max B with g}) magnitude of the vorticity is found to be $\omega_v\sim 10^{-4}B/T$. Next, we  analyse the chiral instability in collisionless regime $\nu_c\ll\omega\ll k$, considering eqs. (\ref{mode11}-\ref{mode22}). Here one finds the wave number $k_{max2}=\frac{8\delta T}{9}$ at which the maximum growth rate $\Gamma_2=\frac{1}{2\pi}\frac{T^3\delta^3}{m^2_D}$ occurs. Now $\frac{k_{max2}}{\nu_c}=\frac{8}{9}\frac{\delta}{\alpha^{2}_1}\ll 1$ and $\frac{\Gamma_2}{\nu_c}\sim \frac{3\delta^3}{8\pi^2\alpha^3_1 ln(1/\alpha_1)}\ll 1$ this puts constraints on the allowed values of
$\delta$. For $\delta\sim 10^{-1}$, $\alpha_1\sim 10^{-2}$ and $T\sim T_f$, one can estimate magnitude of the magnetic field to be $10^{31}$~Gauss. Typical length scale for the magnetic field $\lambda_2\sim 10/T$ and which is much smaller than the length scale in the collision dominated case. This is expected as the typical length scale associated with kinetic theory are smaller than the hydrodynamical case.

The upper and lower bounds on the present observed magnetic field strength from PLANCK 2015 results \cite{Ade2015} and blazers \cite{Neronov2010} are between  $10^{-17} G -10^{-9} G$. However, recently in Ref.\cite{Fujita2016} it has been shown that  if the magnetic field is helical and created before the electroweak phase transition then it can produce some baryon asymmetry. This can put more stringent bounds on the magnetic field ($10^{-14} G -10^{-12} G$). Since the magnetic fields and the plasma evolution are coupled, the produced magnetic field may not evolve adiabatically i.e. like $a(\tau)^{-2}$ due to the plasma processes like turbulence. Similarly the magnetic  correlation length $\lambda_B\propto k^{-1}_{max}$ may  not be proportional to $a(\tau)$. Typical values of $\lambda_B$ for the collision dominated and collisionless regimes in our case are $10^{5}/T$ and $10/T$ respectively. At EW phase transition, this length is of the order of few centimetres One of the most important length scale is the length scale of turbulence. Which can be written in the following form
\begin{align}
	\lambda_T \approx \frac{\mathcal{B}_p}{\sqrt{\varepsilon^{ch}+p^{ch}}}\tau
	\sim \frac{\mathcal{B}_p}{\sqrt{\varepsilon^{ch}+p^{ch}}}H^{-1},
\end{align}
where $\mathcal{B}_p$ is the physical value of magnetic field and $\varepsilon^{ch}$ and $p^{ch}$ are respectively energy and pressure densities of the charged particles. 
For $\lambda_B\gg \lambda_T$ the effect of turbulence can be neglected. However,
the maximum value of the magnetic field (for $\nu_c=0$) is about $10^{31}~G$ in our case and 
this gives $\lambda_T\approx 10^{6}/T$. Thus we have  $\lambda_B\ll \lambda_T$ and following Ref.\cite{Fujita2016} we assume that the generated magnetic fields will undergo inverse cascade soon after their generation. One can relate $\mathcal{B}_p$ and $\lambda_B$ that are undergoing the process of inverse cascade with the present day values of magnetic field $\mathcal{B}_0$ and the correlation length $\lambda_0$ using the following eqs. \cite{Fujita2016}:
\begin{align}
	\mathcal{B}_p^{IC}(T)\simeq  9.3 \times 10^9 ~G \left(\frac{T}{10^2 GeV}\right)^{7/3}\left(\frac{\mathcal{B}_0}{10^{-14} ~G}\right)^{2/3} 
	\times\left(\frac{\lambda_0}{10^2 ~pc}\right)^{1/3}\mathcal{G}_B(T)\\ 
	\lambda_B^{IC}(T)\simeq 2.4\times 10^{-29} Mpc\left(\frac{T}{10^2 ~GeV}\right)^{-5/3}\left(\frac{\mathcal{B}_0}{10^{-14}G}\right)^{2/3}
	\times \left(\frac{\lambda_0}{1 pc}\right)\mathcal{G}_{\lambda}(T)
\end{align}

Where 
\begin{align}
	\mathcal{G}_B(T)&=(g_*^{total}(T)/106.75)^{1/6}(g_*^{ch}(T)/82.75)^{1/6}(g_{*s}(T)	/106.75)^{1/3}\\ \mathcal{G}_{\lambda}(T)&=(g_*^{total}(T)/106.75)^{-1/3} (g_*^{ch}(T)/82.75)^{-1/3}(g_{*s}(T)/106.75)^{1/3}.
\end{align}
Here $g_*^{ch}(T)$ and $g_*^{total}(T)$ are the number of degree of freedom of the  $U(1)$ charged particles 
in the thermal bath. From these equations one can see that for collision dominated case $\mathcal{B}_p\simeq 10^{26}~G$, 
can be achieved when $\mathcal{B}_0\simeq 10^{-12}~G$ and $\lambda_0\simeq 100~ Kpc$. 
However in collisionless regime a value of  $\mathcal{B}_0\simeq 10^{-11}~G$ and 
$\lambda_0\simeq 1 Mpc$ at temperature $T=80 ~TeV$ gives the values that we have found in our estimates 
for the peak value of the magnetic field. Thus the values of the magnetic field and the correlation length scale 
estimated by us can be consistent with the current bounds obtained from CMB  observation and necessary for current observed baryon asymmetry 
Since the values of $\mathcal{B}_p$ and $\lambda_B$ for the collision dominated case are similar to those given in the 
Ref.\cite{Giovannini1998}, we believe that they are also consistent with BBN constraints.

In conclusion we have studied the generation of magnetic fields due to the anomaly in primordial plasma consisting of the standard model particles. We have applied the Berry curvature modified kinetic theory to study this problem. The effect of collisions in the kinetic equation was incorporated using the relaxation time approximation. We find that the chiral instability can occur in presence of the dissipation in both collision dominated and collisionless regimes. We find that in the collision dominated case the chiral instability can produce a magnetic field of the order of $10^{27}$  Gauss with the typical length scale $10^5/T$.  These results are in broad agreement with Ref.\cite{Joyce1997}. However in this work authors have used heuristic kinetic equation and the collision term was not explicitly written in the kinetic equation. While the expression for the total current included the Ohm's law. We have obtained expressions for electric and magnetic conductivities using the modified kinetic theory. We find that expression for electric conductivity in chiral plasma has a non-dissipative term in addition to the standard Ohmic term. It is shown that this new term is related to the vorticity current term found in the chiral magnetohydrodynamics \cite{Giovannini:2013oga}. Further we have also studied the chiral instability in the collisionless regime. It is shown that in this regime magnetic field of strength $10^{31}$~Gauss can be generated at length scale $10/T$. These length scales are much smaller than the length scale of the magnetic field in the collision dominated regime.
Further the obtained values of magnetic-field strength and the length scale are shown to be consistent with the recent constraints from CMB data. We have also shown that in the collision dominated regime results of kinetic theory agree with the hydrodynamic treatment.
\cleardoublepage
\chapter{Chiral Battery, scaling laws and magnetic fields}\label{ch4}
In this chapter, we study the generation and evolution of magnetic field in the presence of chiral imbalance and gravitational anomaly which gives an additional contribution to the vortical current. The contribution due to gravitational anomaly is proportional to $T^2$. This contribution to the current can generate seed magnetic field  of the order of $10^{30}$~G at $T\sim 10^9$ GeV,  with a typical length scale of  the order of $10^6/ T$, even in the absence of chiral charges (when chiral chemical potential is zero). Moreover, such a system possess scaling symmetry. We show that the $T^2$ term in the vorticity current along with scaling symmetry leads to more power transfer from lower to higher length scale as compared to only chiral anomaly without scaling symmetry.
\section{Introduction}
\label{sec:ch-review}
%
Chiral fluid carries both electric charges $j^{\mu}_{elec}$ (which can be written in terms of vector current $j_V^{\mu}=\bar \psi \gamma^{\mu}\psi$ as $j_{elec}^\mu=e j_V^{\mu}$, where $e$ is electric charge) and chiral charges $j^{\mu}_A=\bar \psi \gamma^{\mu}\gamma^{5}\psi$. Above $\psi$ represents the fermions, $\gamma^\mu$ are the Dirac matrices and $\gamma_5$ is defined as $\gamma_5=i\gamma^0 \gamma^1 \gamma^2 \gamma^3$. Sometimes it is convenient to work in the basis of the left handed and right handed currents, which are defined as $j^{\mu}_R=\frac{1}{2}(j^{\mu}_V+j^{\mu}_A)$ and $j^{\mu}_L=\frac{1}{2}(j^{\mu}_V-j^{\mu}_A$). 

The dynamics of a  plasma consisting of chiral fermions in presence of
background fields is governed by the following hydrodynamic equations \cite{Yamamoto2015}
\begin{eqnarray}
	\nabla_\mu\, T^{\mu\nu} = F^{\nu\lambda}\, j_\lambda \, ,
	\label{eq:StressCons}
\end{eqnarray}
\begin{equation}
	\nabla_\mu\, j^\mu  =  0 \, ,
	\label{eq:VCons}
\end{equation}
\begin{eqnarray}
	\nabla_\mu\, j^\mu  =  C\,E_\mu\,B^\mu
	\label{eq:ACons}
\end{eqnarray}
where the vector current $j^\mu = j^\mu_R +j^\mu_L$ and chiral current $j_5^\mu = j^\mu_R -j^\mu_L$ with $j_{R(L)}$ being the current density for right (left) handed particles. 
Here $\nabla_\mu$ represent covariant derivative with respect to the curved space. Greek index runs from $0$ to $3$. In the present work, we have used Latin index for the three-space coordinate points ($1-3$). Here $F^{\mu\nu}$ is the field strength and C is the anomaly coefficient.  $E^{\mu}$ and  $B^{\mu}$ are electric and magnetic four vectors, respectively, and they are measured with respect to the comoving observer. These are purely spatial vectors and are defined as: $E^{\mu}=F^{\mu\nu}u_\nu$, $B^{\mu}=\frac{1}{2}\epsilon^{\mu\nu\alpha\beta}u_\nu F_{\alpha\beta}$
and $F_{\mu\nu}=\nabla_{\mu} A_{\nu}-\nabla_{\nu} A_{\mu}$. 
The Maxwell equations read 
\begin{equation}
	\nabla_\nu F^{\mu\nu}=j^{\mu}~~~~ and ~~~~~ F_{[\mu\nu,\lambda]}=0 \label{maxwell}
\end{equation}
In the state of local equilibrium, the
vector current $j^\mu_v$ and the chiral current $j^\mu_5$ can be expressed
in terms of the four velocity of the fluid $u^\mu$, energy density $\rho$,
vector charge density $n_v$ and axial charge density $n_5$. These quantities respectively takes the following form
%
%
%
\begin{eqnarray}
	j^\mu_{\rm v} = n_{\rm v}\,u^\mu  + \sigma E^\mu + \xi_{\rm v}\omega^\mu +
	\xi_{\rm v}^{B}B^\mu\, ,
	\label{eq:VectorCurrent}
\end{eqnarray}
\begin{eqnarray}
	j_5^\mu = n_5\,u^\mu 
	+ \xi_{\rm 5}\omega^\mu + \xi_{5}^{B}B^\mu \, ,
	\label{eq:AxialCurrent}
\end{eqnarray}
where $n_{\rm v,5} = n_{R} \pm n_L$,  $\xi_{\rm v,5} = \xi_R \pm \xi_L$, 
$\xi^{B}_{\rm v,5} = \xi^{B}_R \pm \xi^{B}_L$, 
$\omega^\mu = \epsilon^{\mu\nu\sigma\delta}\,u_\nu \partial_\sigma\,u_\delta$ is the 
vorticity four vector, $E^\mu = F^{\mu\nu}\,u_\nu$, and 
$B^\mu = 1/2\,\epsilon^{\mu\nu\sigma\delta}\,u_\nu\,F_{\sigma\delta}$. In presence of chiral imbalance and gravitational anomaly, consistency with second law of thermodynamics ($\partial_\mu\,s^\mu \geq 0$, with $s^\mu$ being the entropy density) demand that the coefficients, for each right and left
particle, have following form \cite{Son2009,Landsteiner2011, Neiman2011}
\begin{eqnarray}
	\xi_i& = & C\,\mu_i^2\,\left(1-\frac{2\,n_i\,\mu_i}{3\,(\rho+p)}\right)\, +
	\frac{D\,T^2}{2}\left(1-\frac{2\,n_i\,\mu_i}{(\rho+p)}\right)
	\label{eq:xi} \\
	\xi^{B}_i & = & C\,\mu_i\,\left(1-\frac{n_i\,\mu_i}{2\,(\rho+p)}\right)\, -
	\frac{D}{2}\left(\frac{n_i\,T^2}{(\rho+p))}\right)\, ,
	\label{eq:xiB}
\end{eqnarray}
The constants $C$ and $D$ are related to those of the chiral anomaly and 
mixed gauge-gravitational anomaly
as $C=\pm 1/4\pi^2$ and $D=\pm1/12$ for right and left handed chiral particles
respectively.
It is now easy to see from eq.(\ref{eq:xi}) and eq.(\ref{eq:xiB}) that for 
$\mu_i/T \ll 1$, which is the scenario in the early universe, 
$\xi_i \sim \frac{D}{2} \, T^2$ and $\xi_i^{B} \sim C\, \mu_i$. Note that,
$\xi_{\rm v} = 0 = \xi_{\rm v}^{B}$. 
However, $\xi_5 \sim DT^2$ and $\xi_5^{B} = C\,(\mu_R - \mu_L) = C\,\mu_5$. 
We would like to emphasize here that the $\xi_5$ is independent of chiral 
imbalance and depends only on the temperature. Thus, we expect such a term to  
dominate at high temperature. We exploit this dominance to generate the magnetic 
field in the early Universe. 

In an expanding background, the line element can be described by
Friedmann-Robertson-Walker metric, given in the chapter (\ref{ch3}) in equation (\ref{eq:FLRWm}),
\begin{eqnarray}
	ds^2 =a^2(\tau)\left(-d\tau^2 + \delta_{ij}dx^i\,dx^i\right)\, ,
\end{eqnarray}
where $a$ is the scale factor and $\tau$ is the conformal time. We choose $a(\tau)$ to have dimensions
of length, and $\tau$ , $x^i$ to be dimensionless. Using the fact that the scale factor $a = 1/T$, we
can define the conformal time $\tau = M_*/T$, where $M_* = \sqrt{90/8\,\pi^3\,g_{\rm eff}} M_P$ with
$g_{\rm eff}$ being the effective relativistic degrees of freedom that contributes to the energy density
and $M_P = 1/\sqrt{G}$ being the Planck mass. We also define the following comoving variables 
\begin{eqnarray}
	\vec B_c  =  a^2(\tau)\, \vec B(\tau)\, ,~~~
	\mu_c =  a(\tau)\,\mu\, ,~~~\sigma_c = a(\tau)\,\sigma\,\nonumber \\,~~~
	T_c  = a(\tau)\,T ,~~~
	x_c = x/a(\tau)\, .
\end{eqnarray}
In terms of comoving variables, the evolution equations of fluid and
electromagnetic fields are form invariant \cite{Holcomb:1989tt,Dettmann1993, Gailis:1995fdt}. Thus, in the discussion below, we will work with the above defined comoving quantities and omit the subscript $c$.

Using the effective Lagrangian for the standard model, one can derive the 
generalized Maxwell's equation \cite{Semikoz:2011tm},  
\begin{equation}
	\nabla \times {\bf B} = {\bf j}\, ,
	\label{eq:GenMax}
\end{equation}
with $\vec j = {\bf j}_{\rm v} + {\bf j}_5$ being the total current. In the above equation, we 
have ignored the displacement current. Taking $u^\mu = (1,\vec v)$ 
and using eq.(\ref{eq:VectorCurrent}) and eq.(\ref{eq:AxialCurrent}), one can show
that
\begin{eqnarray}
	& j^0 & = n = n_{\rm v} + n_5 \, \nonumber \\
	&{\bf j} & = n{\bf v}+ \sigma({\bf E} +{\bf v}\times {\bf B})  + \xi \, \boldsymbol{\omega} + \xi^{B}\,{\bf B}  \label{eq:totCurrent}\, ,
\end{eqnarray}
with $\xi = \xi_{\rm v} + \xi_5$ and $\xi^{B} = \xi^{B}_{\rm v} + \xi^{B}_5$. For our analysis, we assume that the velocity field is divergence free,
{\it i.e.} $\nabla\cdot{\bf v}=0$.  
Taking curl of eq.(\ref{eq:GenMax}) and using the expression for current from
eq.(\ref{eq:totCurrent}), we get
\begin{eqnarray}
	\frac{\partial {\bf B}}{\partial \tau} =  \frac{n}{\sigma}\,\nabla\times \vec v
	+\frac{1}{\sigma}\,\nabla^2{\bf B} + \nabla\times(\vec{v}\times {\bf B}) 
	+ \frac{\xi}{\sigma}\, \nabla\times\vec\omega
	+  \frac{\xi^{B}}{\sigma}\,\nabla\times{\bf B}  \, .
	\label{eq:magevol}
\end{eqnarray}
In obtaining the above equation, we have also assumed that chemical
potential  and the temperature are homogeneous.
Note that, in the limit $\sigma\rightarrow \infty$, we will obtain magnetic
fields which are flux frozen. There are other conditions in the cosmological
and astrophysical settings in which the advection term
$\nabla\times({\bf v}\times {\bf B})$ can be ignored compared to other terms and
reduces the above equation to linear equation in $v$ and $B$. We ignore
advection term in our analysis.
\section{Chiral battery
}
\label{sec:seed-mag}
In absence of any background magnetic fields ${\bf B} = 0$, eq.(\ref{eq:magevol}) reduces to 
\begin{eqnarray}
	\frac{\partial {\bf B}}{\partial \tau}  = \frac{1}{\sigma}n\nabla \times  {\bf v}+ \frac{1}{\sigma}\,\xi\, \nabla\times\boldsymbol{\omega}\, .
	\label{eq:NoBMaxEq}
\end{eqnarray}
Note that the two terms on the right hand side of eq.(\ref{eq:NoBMaxEq}) no longer
depend on $B$. This situation is similar to that of the Biermann battery mechanism where 
$\nabla \rho\, \times\, \nabla p$ term is independent of $B$ \cite{BIERMANN1950}. For Biermann 
mechanism to work, $\nabla \rho$ and $\nabla p$ have to be in different directions 
which can be achieved if the system has vorticity.
In our scenario, it is interesting to note that even if $n=0$, $T^2$ term in 
$\xi$ will act as a source to generate the seed magnetic field at high temperature.
On the other hand, in presence of finite chiral imbalance (such that $\mu/T \ll 1$) in the early universe, $T^2$ term in $\xi$ will still be the source to generate the seed magnetic field, but non-zero $\xi^{(B)}$ will lead to an instability in the system. An order of magnitude estimate gives the first term of eq.(\ref{eq:NoBMaxEq}) to be  $ \frac{\alpha}{L}\,\left(\frac{\mu}{T}\right)\,T^2\,v$ while the second term is of the order $\frac{\alpha}{L^2}\,D\,T\,v$. On comparing the two terms we get a critical length $\lambda_{\rm c} \sim (D/T)(\mu/T)^{-1}$ below which first term will be sub-dominant.
\subsection{Mode decomposition}
In order to solve eq.(\ref{eq:NoBMaxEq}), we decompose the divergence free vector fields in the polarization modes, $\varepsilon^{\pm}_i$,
\begin{equation}
	\boldsymbol{ \varepsilon}^\pm({\boldsymbol k}) = \frac{\boldsymbol{e}_1({\boldsymbol k})\,\pm \,i\boldsymbol{\,e}_2({\boldsymbol k})}{\sqrt{2}}\, \exp(i\,{\boldsymbol k}\cdot {\boldsymbol x})\, .
\end{equation}
These modes are the divergence-free eigenfunctions of the Laplacian operator
forming an orthonormal triad of unit vectors  $(\boldsymbol{e}_1, \boldsymbol{e}_2, \boldsymbol{e}_3 = \boldsymbol{k}/k)$.
It is evident that $\nabla\cdot\boldsymbol{\varepsilon}^\pm = 0$ and $\nabla\times\boldsymbol{\varepsilon}^\pm = \pm\,k\,\boldsymbol{\varepsilon}^\pm$. For our work $\boldsymbol{\varepsilon}^{\pm*} (-\boldsymbol k) = \boldsymbol{\varepsilon}^\pm (\boldsymbol k)$.
We decompose the velocity of the incompressible fluid as:
\begin{equation}
	\boldsymbol v (\tau, \boldsymbol x) = \int \frac{d^3\,k}{(2\pi)^3}\,
	\left[
	\tilde{v}^+(\tau, \boldsymbol k)\boldsymbol{\varepsilon}^{+}(\boldsymbol k) +
	\tilde{v}^-(\tau, \boldsymbol k)\boldsymbol{\varepsilon}^-(\boldsymbol k)
	\right]\, 
	\label{eq:Vdecomp}
\end{equation}
where, tilde denote the Fourier transform of the respective quantities. Assuming
no fluid helicity and statistically isotropic correlators,
\begin{eqnarray}
	\langle  \tilde{{\bf v}}^{\pm*}(\eta, {\bf k})\, \tilde{{\bf v}}^{\pm}(\tau, {\bf q})\rangle & = &(2\,\pi)^3 \, \delta^{(3)}({k} - {\bf q})|{\bf v}(\tau,{\bf k})|^2 \\
	\langle  \tilde{{\bf v}}^{+*}(\tau, {\bf k})\,\tilde{{\bf v}}^{-}(\tau, {\bf q})\rangle & = & \langle \tilde{{\bf v}}^{-*}(\tau, {\bf k})\, \tilde{{\bf v}}^{+}(\tau, {\bf q})\rangle = 0\, ,
\end{eqnarray}
the kinetic energy density can be given as
\begin{eqnarray}
	\frac{1}{2}\langle|\boldsymbol v (\tau, \boldsymbol x)|^2\rangle & = &
	\frac{1}{2} \int d\log\,k\,E_{\boldsymbol v}(\tau,{\bf k}) \\
	&=& \int\frac{d^3k}{(2\pi)^3}
	\left[|\boldsymbol v^+(\tau,\boldsymbol k)|^2\, + \,|\boldsymbol v^-(\tau,{\bf k})|^2\right] \, .
	\label{eq:ke-spec}
\end{eqnarray}
Similar to the velocity field, we can decompose the magnetic field as,
\begin{equation}
	\boldsymbol B (\tau, \boldsymbol x) = \int \frac{d^3\,k}{(2\pi)^3}\,
	\left[
	\tilde{B}^+(\tau, \boldsymbol k)\boldsymbol{\varepsilon}^{+}(\boldsymbol k) +
	\tilde{B}^-(\tau, \boldsymbol k)\boldsymbol{\varepsilon}^-(\boldsymbol k)
	\right]\,
\end{equation}
The magnetic energy density is then given by
\begin{eqnarray}
	\frac{1}{2}\langle|\boldsymbol B (\tau, \boldsymbol x)|^2\rangle & = &
	\frac{1}{2} \int d\log\,k\,E_{\boldsymbol B}(\tau,k)\\
	& = & \int\frac{d^3k}{(2\pi)^3}
	\left[|\boldsymbol B^+(\tau,\boldsymbol k)|^2\, + \,|\boldsymbol B^-(\tau,\boldsymbol k)|^2\right] \, .
	\label{eq:mag-en-den}
\end{eqnarray}
Mode decomposition of  eq.(\ref{eq:NoBMaxEq}) gives
\begin{eqnarray}
	\frac{\partial \tilde{B}^+}{\partial \tau} & = & \frac{1}{\sigma}\left(n\,k\, + \xi\, k^2\,\right)\tilde{v}^+ \, ,
	\label{eq:Bpl}\\
	\frac{\partial \tilde{B}^-}{\partial \tau} & = & \frac{1}{\sigma}\left(-n\,k\, + \xi\, k^2\,\right)\tilde{v}^-
	\label{eq:Bmi} \, .
\end{eqnarray}
In absence of external fields, the energy momentum conservation equation
$\nabla_\mu\, T^{\mu\nu} = 0$ implies that  $n$, $\sigma$ and $\xi$ are
constant over time. Thus, the solution to eq.(\ref{eq:Bpl}) and eq.(\ref{eq:Bmi}) is
\begin{equation}
	\tilde{B}^\pm(\tau, \boldsymbol k) =  \frac{1}{\sigma}\left(\pm
	n\,k\, + \xi\, k^2\,\right)
	\int_{\tau_0}^{\tau}\,d\tau'\tilde{v}^\pm(\tau', \boldsymbol k)\,.
	\label{eq:sol}
\end{equation}
where $\tau_0 = M_*/T_0$ and we have set $T_0 = 10^{10}$ GeV for our work. Multiplying eq.(\ref{eq:Bpl}) by $\tilde{B}^{+*}$ and eq.(\ref{eq:Bmi}) by $\tilde{B}^{-*}$, the ensemble average of the combined equation leads to
\begin{eqnarray}
	\frac{\partial |\tilde{B}^+|^2}{\partial \tau} & = &  \frac{2}{\sigma}\left(n\,k\, + \xi\, k^2\,\right)\langle\tilde{B}^{+*}\tilde{v}^+\rangle
	\label{eq:BplAvg}\\
	\frac{\partial |\tilde{B}^-|^2}{\partial \tau} & = &  \frac{2}{\sigma}\left(-n\,k\, + \xi\, k^2\,\right)\langle\tilde{B}^{-*}\tilde{v}^-\rangle\, .
	\label{eq:BmiAvg}
\end{eqnarray}
On using the solution obtained in eq.(\ref{eq:sol}) and the velocity decomposition given in eq.(\ref{eq:Vdecomp}) we obtain
\begin{equation}\langle\tilde{B}^{\pm*}(\tau,\boldsymbol k)~\tilde{v}^\pm (\tau,\boldsymbol k')\rangle  = 
	\,\frac{1}{\sigma}\left(\pm n\,k\, + \xi\, k^2\,\right)
	\int_{\tau_0}^{\tau}  d\tau '\langle
	\tilde{v}^{\pm*}(\tau',\boldsymbol k)~~\tilde{v}^\pm(\tau,\boldsymbol k')
	\rangle \, .
	\label{eq:BVcrossCorrel}
\end{equation}
It is reasonable to expect that any $k$ mode of the fluid velocity  to be
correlated on the time scale $\tau_c \equiv |\tau -\tau'| \sim 2\pi/k\,v$ and 
uncorrelated over longer time scales, where $v$ represents the average velocity of the fluid within the correlation scale. Thus, we can write
\begin{equation}
	\langle
	\tilde{v}^{\pm*} (\tau',\boldsymbol k)~~\tilde{v}^\pm(\tau,\boldsymbol k')\rangle  =
	\left\{ \begin{array}{ll}
		(2\pi)^3\langle v^\pm(\tau, \boldsymbol k)^2\rangle \delta^{(3)}(\boldsymbol k -\boldsymbol {k'}) & \mbox{ for $ \tau_c < \frac{2\pi}{k \,v}$}  \\
		\\
		0 & \mbox{ for $\tau_c > \frac{2\pi}{k\,v}$ } \end{array} \right.
	\label{eq:correl-time}
\end{equation}
On using the above unequal time correlator in eq.(\ref{eq:BVcrossCorrel}), we obtain
\begin{equation}
	\langle
	\tilde{B}^{\pm*}(\tau',\boldsymbol k)\tilde{v}^\pm(\tau,\boldsymbol k')\rangle = 
	\frac{(2\pi)^3}{\sigma}|v^\pm|^2\left(\pm n k+ \xi k^2\right)~ f(\tau,k) \delta^{(3)}(\boldsymbol k -\boldsymbol {k'})
	\label{correl-time1}
\end{equation}
where $f(\tau,k) = \tau -\tau_0 $ for $\tau - \tau_0 \leq 2\pi/(k\,v)$ and zero otherwise.
\subsection{Scaling properties in chiral plasma}
\label{sec:scaling}
It has been shown that ChMHD equations for electrically neutral plasma with
chirality imbalance  have
scaling symmetry under following transformation \cite{Yamamoto:2016xtu},
\begin{align}
	\boldsymbol x\rightarrow \ell x\, , \hspace{0.5cm}
	\tau\rightarrow \ell^{1-h}\,\tau\, , \hspace{0.5cm}
	\boldsymbol v\rightarrow \ell^h\, \boldsymbol  v, \nonumber \\
	\boldsymbol B\rightarrow \ell^{h}\boldsymbol B\,, \hspace{0.5cm} 
	\sigma\rightarrow \ell^{1+h}\,\sigma \label{eq:scalingsymm}
\end{align}
with $\ell>0$ is the scaling factor and $h\in \Re$ is the  parameter, under certain conditions \cite{Yamamoto:2016xtu}. With the above scaling, 
the fluid velocity spectrum is given by
\begin{equation}
	E_v(\boldsymbol k,\tau)\,k^{1+2\,h} = \psi_v(\boldsymbol k,\tau)
	\label{FluidVelSpec}
\end{equation}
where $\psi_v$ satisfies
\begin{equation}
	\psi(k/\ell, \ell^{1 - h}\, \eta) = \psi(k,\tau)\, .
	\label{eq:psi_v}
\end{equation}
On differentiating eq.(\ref{eq:psi_v}) with respect to $\ell$ and taking $\ell = 1$ leads to the following differential equation \cite{Olesen:1996ts}
\begin{equation}
	-k\frac{\partial\psi_v}{\partial k} + (1-h)\, \eta\,\frac{\partial\psi_v}{\partial \tau} =0\, ,
\end{equation}
which can be solved to obtain the form of $\psi$. The solution to the above differential equation can be given as
\begin{equation}
	\psi_{v}(\tau, k) \propto \,k^m\,\tau^{m/(1-h)}\, ,
	\label{ScalePsi}
\end{equation}

where $m$ is a constant and taken as parameter in this work whose value is fixed by assuming Kolmogorov spectrum for $k>k_i$ and white noise spectrum for $k<k_i$, where $k_i$ is the wave number in the inertial
range. From
eq.(\ref{FluidVelSpec}) and eq.(\ref{ScalePsi}) , we get
\begin{equation}
	E_v(\tau, k) = v_i^2\,\left(\frac{k}{k_i(\tau)}\right)^{-1-2h+m}\,\left(\frac{\tau}{\tau_0}\right)^{\frac{m}{1-h}}\,.
	\label{eq:ke-spec1}
\end{equation}
where $v_i$ is some arbitrary function which encodes the information about
the boundary conditions.
Therefore, the $k$ and $\tau$ dependence of fluid velocity can be given by
\begin{equation}
	v(k, \tau)  = v_i\,\left(\frac{k}{k_i(\tau)}\right)^{(-1-2h+m)/2}\,
	\left(\frac{\tau}{\tau_0}\right)^{\frac{m}{2(1-h)}}\,.
	\label{eq:vel}
\end{equation}
We are interested in an electrically neutral plasma with chiral charge. For
such a system $h = -1$ \cite{Yamamoto:2016xtu}.
However, in the radiation dominated epoch the fluid velocity does not change for
$k<k_i(\tau)$. Assuming white noise on such scale, which implies $E_v \propto k^3$, we obtain
\begin{equation}
	k_i(\tau) = k_i(\tau_0)\, \left(\frac{\tau}{\tau_0}\right)^{1/3}
\end{equation}
Substituting the value of $k_i(\tau)$ in eq.(\ref{eq:vel}) we get
\begin{equation}
	v(k,\tau) = v_i\,\left(\frac{k}{k_i(\tau_0)}\right)^{(1+m)/2}\,
	\left(\frac{\tau}{\tau_0}\right)^{(m-2)/6}\,.
\end{equation}
In absence of kinetic helicity in the fluid, eq.(\ref{eq:ke-spec}) implies
$|v^+|= |v^-| = \pi k^{-3/2}\, v$.
\subsection{Evolution of Magnetic field}
Once seed field is generated it may start influencing the subsequent evolution
of the system at a certain mode. In that case, there will be a current flowing parallel to these generated magnetic fields. Therefore, we need to consider the effect of CME in the evolution eq.(\ref{eq:magevol}). We can decompose this eq.(\ref{eq:magevol}) in terms of polarization to obtain
\begin{equation}
	\frac{\partial |\tilde{B}^\pm|^2}{\partial \tau}  =  \frac{2}{\sigma}\left(-k^2\pm \xi^{(B)} k\right) |\tilde{B}^\pm|^2
	+\frac{2}{\sigma^2}\left(\pm n\,k\, + \xi\, k^2\,\right)^2\, |v^\pm|^2\, f(\tau,k)
	\label{eq:b-modes}
\end{equation}
It is evident from the above equation that when the first term dominates over
the second, then
\begin{equation}
	|B^\pm|^2 =|B_0^\pm|^2 \exp\left(\frac{2\,\tau}{\sigma}\,k_{\rm ins}^2\right)\,
	\exp\left(\frac{2\,\tau}{\sigma}(k\mp k_{\rm ins})^2\right) \label{eq:modespm}
\end{equation}
where $k_{\rm ins} = \xi^{(B)}/2$.
From eq.(\ref{eq:b-modes}), evolution of  magnetic energy can be found as
\begin{eqnarray}
	\frac{\partial E_B}{\partial \tau} =\frac{-2\, k^2}{\sigma}E_B+\frac{2\,\xi^{(B)} \, k}{\sigma}\mathcal{H}_B +\frac{2}{\sigma^2}(n^2k^2+\xi^2 k^4)E_v f(\tau,k)
	\label{eq:spectTotalB}
\end{eqnarray}
where $E_B$, $E_v$ and $\mathcal{H}_B$ are respectively magnetic energy, kinetic energy and magnetic helicity which affect the evolution of $\mu$ through the following equation
\begin{equation}
	\frac{d\mu_5}{d\tau} = -2 \alpha \int \frac{d\ln k}{k} \, \frac{\partial \mathcal{H}_B}{\partial \tau} \, - \Gamma_f\, \mu_5\, ,\label{eq:chemhel}
\end{equation}
where $\Gamma_f$ is the chirality flipping rate. The flipping depends on interaction of the right handed particles with the Higgs and their back reaction \cite{Campbell:1992jd}.  At temperature $T\geq 80$ TeV the chirality flipping rate is much slower than the Hubble expansion rate. Therefore, eq.(\ref{eq:chemhel}) can be rewritten as
\begin{eqnarray}
	\frac{\partial}{\partial\tau}\left[ \mu_5 + 2\alpha\int \frac{d\ln k}{k}{\mathcal H}_B\right] = 0\, .
\end{eqnarray}
The term in the bracket is constant in time. Therefore, the helicity will be produced at the cost of chiral 
imbalance. However, when temperature drops to an extent that the flipping rate becomes comparable to the 
expansion rate of the Universe, maximally helical field will be generated. In order to get the complete picture
of generation and evolution of the magnetic field,
we solve eq.(\ref{eq:b-modes}) along with eq.(\ref{eq:chemhel}).
\section{Results and discussion}
\label{sec:result}
\begin{figure}[!h]
	\begin{center}
		\begin{tabular}{c c}
			{\includegraphics[width=2.6in,height=2in,angle=0]{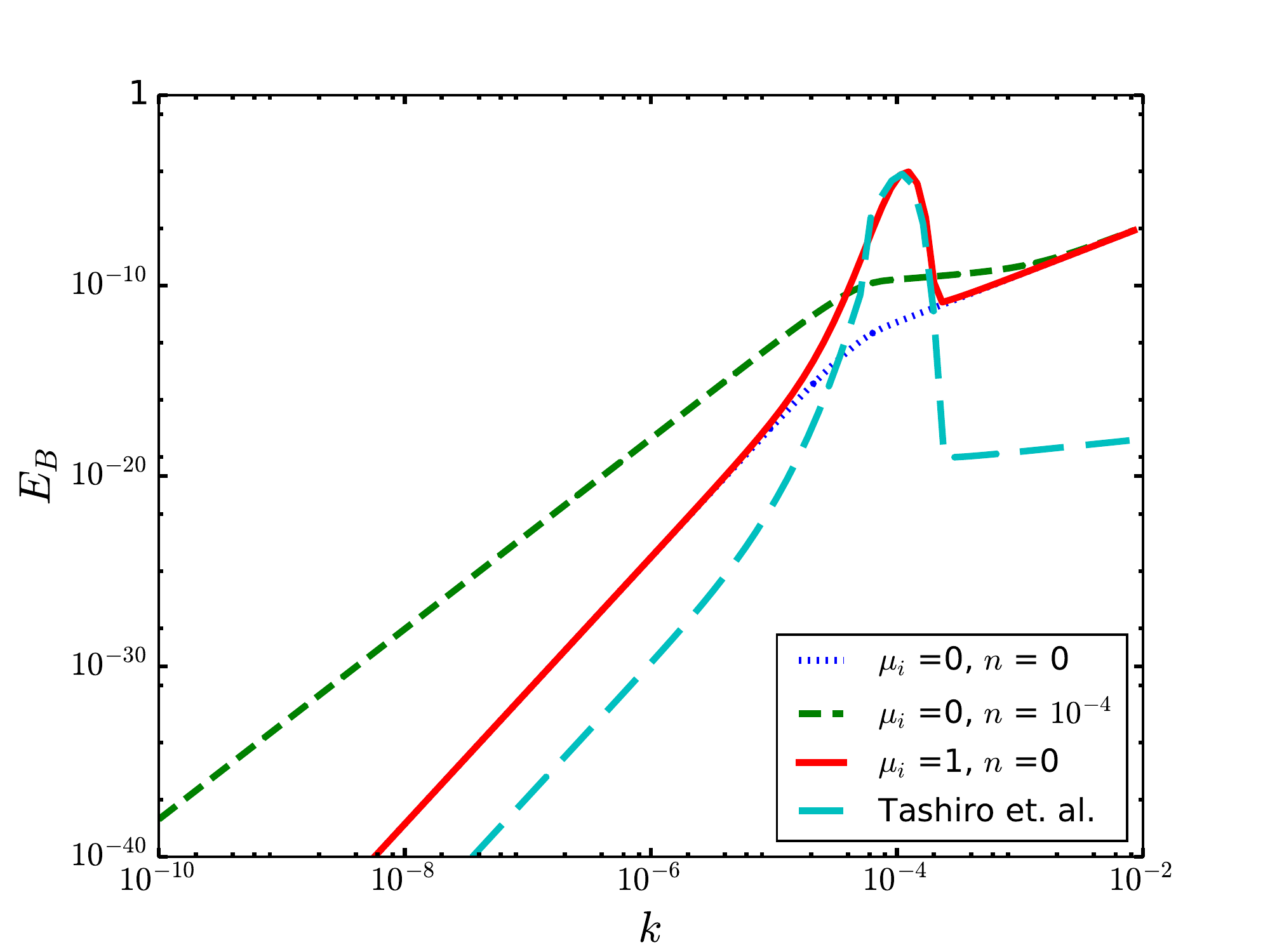}}\label{fig:comp-tashiro}&
			{\includegraphics[width=2.6in,height=2in,angle=0]{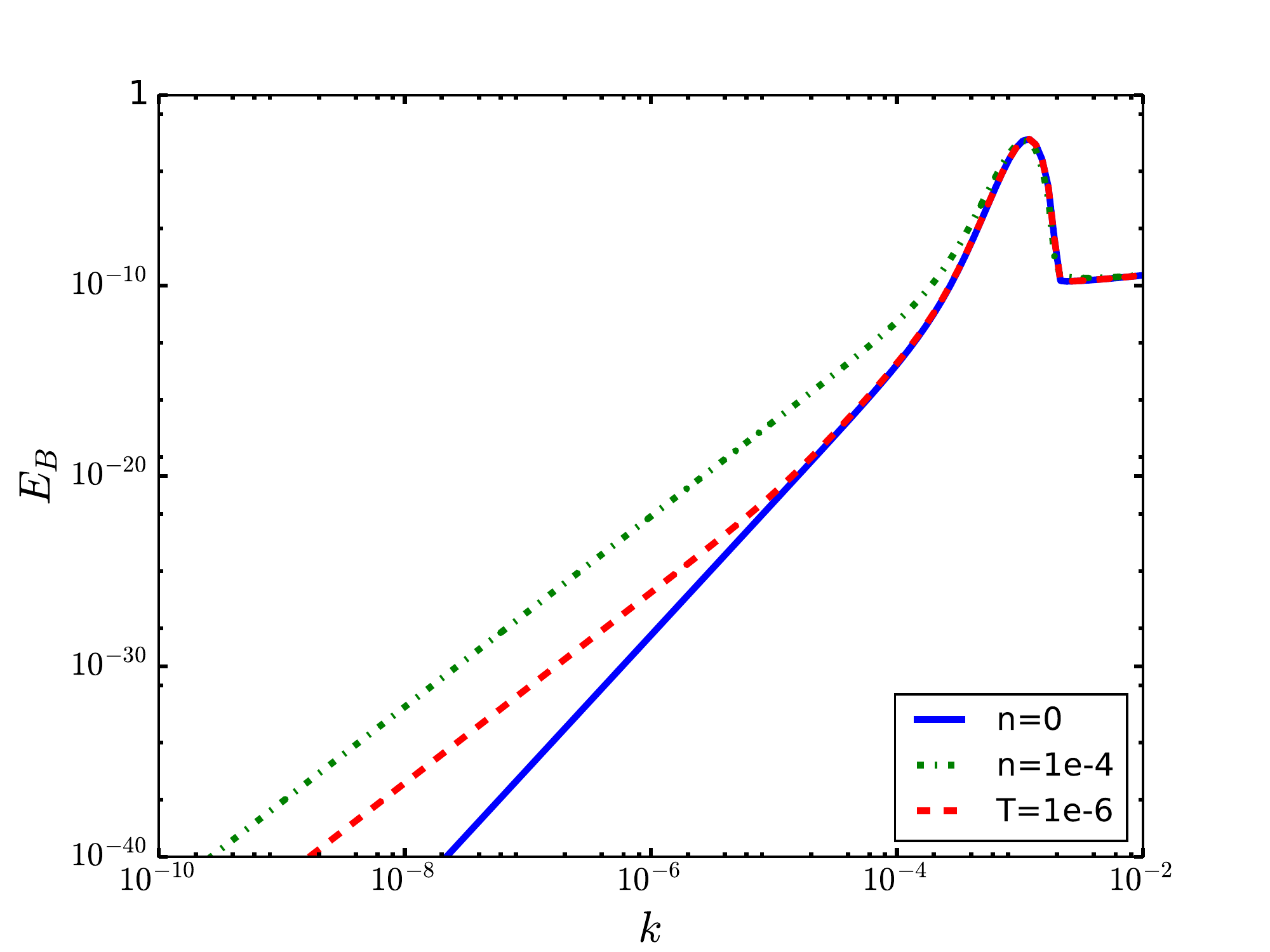}\label{fig:mag_spec}}\\
		\end{tabular}
		\caption{Magnetic energy spectrum as a function of $k$ is shown: (a) when	$n=0$, magnetic energy spectrum $E_B$ goes as $~k^7$ at large length scale. However, the
			power is more than what authors of ref. \cite{{Tashiro2012}} has found for a plasma with chiral asymmetry for a given $k$ value. (b) when $n \neq 0$,
			$E_B$ goes as $~k^5$ at large length scale. In both the figures we have set $T=10^9$ GeV and $v=10^{-4}$.}
	\end{center}
\end{figure}

In the early Universe, much before electroweak symmetry breaking, collision of
bubble walls during first order phase transition at  GUT scale, can generate 
initial vorticity in the plasma. With these assumptions, we
have shown that magnetic field can be generated, in presence of either of the two terms terms on the right hand side of eq.(\ref{eq:NoBMaxEq}).
It is also important to note that, second
term has purely temperature dependence ($\propto D\, T^2\, \vec{\omega}$)
which comes due to gravitational anomaly.
In absence of any background field, the two term $n\,k$ and $\xi\,k^2$ in the right hand side of eq.(\ref{eq:b-modes}) can be source of the seed field.
Further, when $n=0$, only $\xi\,k^2 $ term acts as a source to
generate the seed magnetic field. However,  there may not be any instability in the system 
if $\xi^{B} = 0$ (Fig(\ref{fig:comp-tashiro})). It is also evident form 
eq.(\ref{eq:b-modes}) and eq.(\ref{eq:mag-en-den}) that the magnetic energy 
spectrum will go as
$\sim k^7$ at large length scale (Fig(4.1 a)). This
feature is similar to \cite{Tashiro2012}. However, for a given $k$ the power
transferred to larger length scale is more with the scaling symmetry
(see Fig(4.1 a). On the other hand, when $n \neq 0$, then
$n\,k$ term dominates over $\xi \, k^2$ term at larger length scale (see figure 4.2 b). In the intermediate range, there is a competition between the two terms on the right hand side of eq.(\ref{eq:NoBMaxEq}).
Consequently, the energy spectrum goes as $\sim k^5$ at large length scales (see Fig.(4.1 b). We estimate the strength of the magnetic field produced to be: $\frac{B}{T^2}\,\approx\,\frac{\alpha}{D} \left(\frac{\mu}{T}\right)^2 \,T\, L(\tau)$, where $L(\tau)\,=\,v\,\tau\,\approx \, \frac{D}{T}\,\left(\frac{\mu}{T}\right)^{-1}$. For temperature $T=10^9$~ GeV and $\mu/T\approx 10^{-6}$, the strength of the magnetic field is of the order of $10^{30}$ G at a length scale of the order of $10^{-18}$ cm, which is much smaller than the Hubble length ($10^{-9}$ cm) at that temperature.
\begin{figure}[!h]
	\begin{center}
		\begin{tabular}{c c}
			{\includegraphics[width=2.6in,height=2in,angle=0]{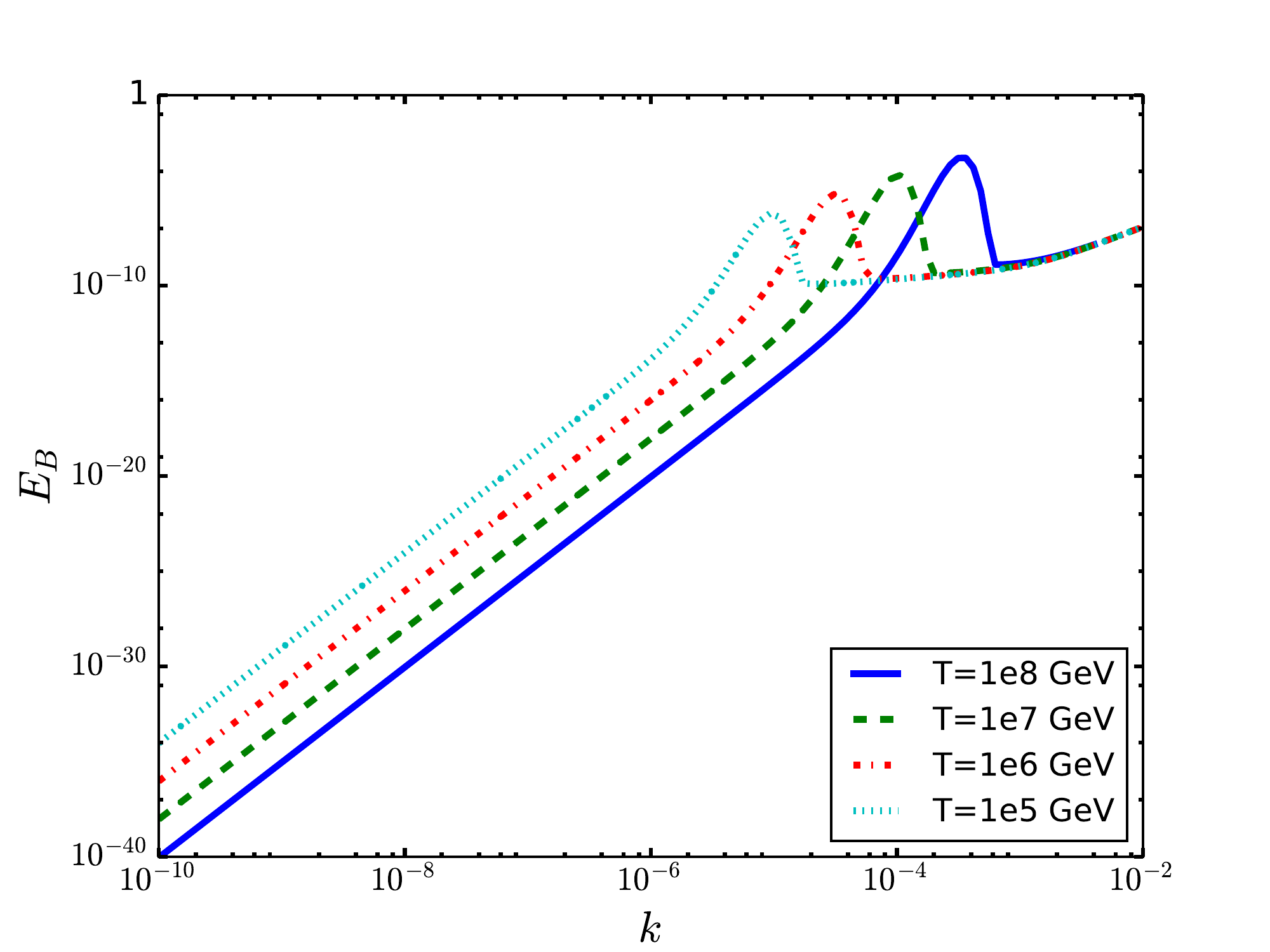}\label{fig:Eb_T}}&
			{\includegraphics[width=2.6in,height=2in,angle=0]{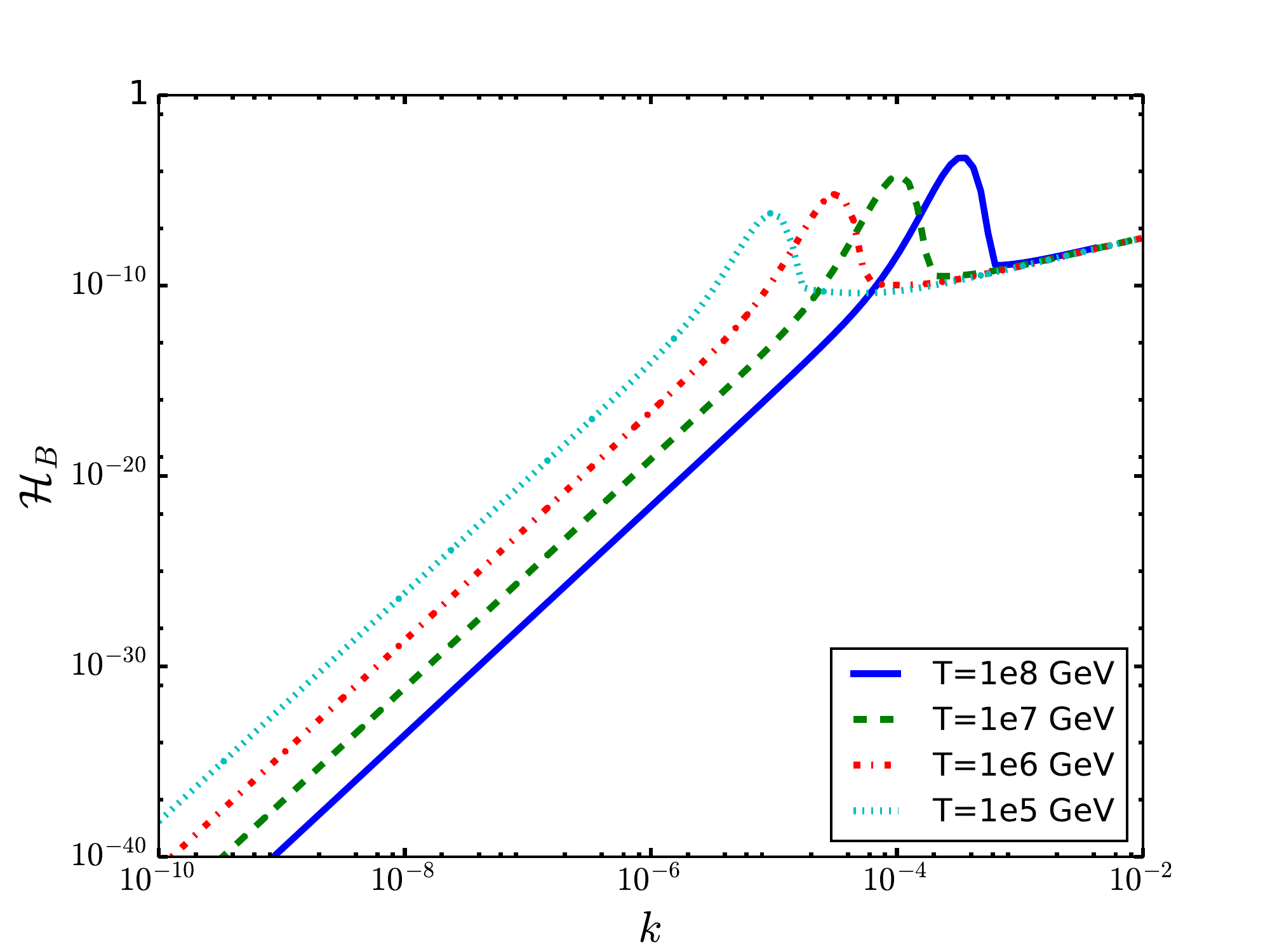}\label{fig:helicity}}\\
		\end{tabular}
		\caption{(a) Evolution of magnetic energy density $E_B$ as a function of $k$ and (b) Evolution of magnetic 
			helicity density $\mathcal{H}_B$  as a function of $k$ for different temperatures. 
			As temperature ($T\sim 1/\tau$) decreases, peak shifts from higher to lower $k$ values which 
			shows inverse cascading {\it i.e.} transfer of magnetic energy from small to large length scale. In both the figures we have set $v=10^{-4}$
			and $n = 10^{-4}.$ }
	\end{center}
\end{figure}

It is evident from eq.(\ref{eq:b-modes}) that once the magnetic field of
sufficiently large strength is produced, it starts influencing the subsequent
evolution of the system and the magnetic energy grows rapidly.
This phenomena is known as the chiral plasma instability
\cite{Akamatsu2013df, Akamatsu:2014yy}. In this regime, we solve 
eq.(\ref{eq:b-modes}) and obtain mode expression in (\ref{eq:modespm}), which
clearly shows that the modes of the magnetic fields grow exponentially  for
the wave number $k\simeq k_{ins}/2\approx C\mu_5/2$. As mentioned earlier,
for $T\geqslant 80$ TeV chiral flipping rate is slower than the expansion rate of the Universe, 
one can safely ignore $\Gamma_f\,\mu_5$ term in eq.(\ref{eq:chemhel}). However, 
the magnetic helicity generated due to $DT^2$ will drive the evolution of 
chemical potential. Below $80$ TeV, flipping rate becomes comparable to the expansion 
rate of the Universe and the chirality of the right particles changes to the left handed.
So to know the complete dynamics of the magnetic field energy, we have solved 
the coupled equations (\ref{eq:b-modes}) and (\ref{eq:chemhel}) simultaneously. 
In figure (4.2 a) we have shown the variation of magnetic energy spectrum
$E_B$ with $k$ for different temperature.
Since, the chemical potential is a decreasing function of time, so it is
expected that with decrease in temperature the instability peak will shift
towards the smaller $k$ which implies the transfer of energy from small to
large length scale. Similarly, we have shown the magnetic helicity spectrum in
figure (4.2 b).
\section{Conclusion}
In the present chapter, we have discussed the generation of magnetic field due
to gravitational anomaly which induces a term $\propto T^2$ in the vorticity
current. The salient feature of this seed magnetic field is that it will
be produced at high temperature irrespective of the whether fluid is charged
or neutral. 
In ref. \cite{Yamamoto:2016xtu}, authors have shown that the equations of 
chiral magnetohydrodynamics (ChMHD), in absence of other effects like charge 
separation effect and cross helicity, follow
unique scaling property and transfer energy from small to large length scales
known as inverse cascade.
\chapter{Damping of the generated magnetic field}\label{ch5}
In chapters (\ref{ch3}) and (\ref{ch4}), we have shown that the spontaneous magnetization can occur in a high energy plasma at temperature $T\geq 80$~TeV in the presence of an excess amount of right handed electrons. Therefore it is natural to study the collective modes in a magnetized chiral plasma and the damping mechanisms of these modes. Similar problem has been studied in the literature of cosmology albeit for  a low energy plasma \cite{Jedamzik1998,  Subramanian1998, Brandenburg1996}. In ref. \cite{Jedamzik1998} Jedamzik et. al. have examined evolution of the collective modes in a magnetized plasma in the FRW background. In this work the authors study the effects of finite viscosity, resistivity and heat conductivity on the collective modes. It has been shown that the Alf\'ven and slow magnetosonic waves survive below the Silk mass scale.

In the context of chiral plasma at high energies, there exist new kind of collective modes [{\it e.g.}, the chiral Alf\'ven waves in ref. \cite{Yamamoto2015} ] in the MHD limit and these modes exist in addition to the usual modes in the standard parity even plasma \cite{landau_59bk}. Moreover, it has been shown that chiral plasmas exhibit a new type of density wave in presence of either an external magnetic field or vorticity and  they are respectively known as Chiral Magnetic Wave (CMW) \cite{Kharzeev2011, Newman:2005hd} and the Chiral Vortical Wave (CVW) \cite{Jiang2015}. Here we would like to note that there are some attempts in studying the damping of MHD waves of these new modes \cite{Abbasi2015a, Abbasi2015} in the context of quark-gluon plasma. In ref.\cite{Kalaydzhyan2017}, the authors have studied damping of thermal chiral vortical and magnetic waves. In these work the first order dissipative MHD dynamics was used. However it should be emphasized here that the first order dissipative hydrodynamics is known for its difficulties in dealing with the acausal behaviour and unphysical instability.  These issues with the first order hydrodynamics can be resolved by using the causal hydrodynamics \cite{Stewart1977, Israel:1979wp, Denicol2007}.  The causal hydrodynamics for the parity violating MHD has been developed in refs. \cite{Romatschke2010, Kharzeev2011a, Erdmenger2009, Banerjee2011}.
In this chapter we study normal modes in the chiral plasma using the causal MHD equations \cite{Kharzeev2011a}. We believe this kind of study was not systematically done earlier.

This chapter is divided in the following sections and subsections: in section-(\ref{ch4:Intro}) we have given a brief introduction and motivation for this chapter; in section-(\ref{waves-Ch}) we have discussed waves in the non-dissipative chiral plasma. 
Section-(\ref{sec:CVChydro}) and its subsections contains a very brief introduction to the conformal viscous hydrodynamics. Last section-(\ref{sec-2result}) contains summary  and discussion of this chapter.
\section{Introduction}\label{ch4:Intro}
The motion of a relativistic fluid can be described by relativistic hydrodynamics for the fluid near equilibrium. Relativistic hydrodynamics play important role in exploring various areas of cosmology, astrophysics and nuclear physics. The fundamental variables are  local fluid velocities $u^{\mu}(x)$, temperature $T(x)$, all conserved charges or their chemical potentials $\mu_s(x)$ (here 's' stands for different species of the fluid and $x$ represents space-time dependence of the variables). The conservation equation for stress tensor $T^{\mu\nu}$ and the currents $J^{\mu}_s$ determine evolution of the fluid. The energy momentum tensor and the currents can be expanded in terms of  the above mentioned variables, as a series of the {\it Kundsen number}  \cite{Israel:1979wp, Betz:2008me, Betz:2009zz}. {\it Kundsen number} is defined as a ratio of mean free path $\lambda_{\rm mfp}$ to the characteristic length scale $l$ of the system under consideration, i.e., $K=(\lambda_{\rm mfp}/l)$. Each term in the series expansion of K is multiplied by a coefficient, also known as a transport coefficient \cite{Kovtun2012}. These transport coefficients are functions of temperature and chemical potentials and are generally obtained from experiments.

The zeroth order of this expansion  represents contribution for the ideal fluid. From the first order, the Navier-Stokes (NS) equations are obtained. There are several representations for the  second order terms. For example, the theory of the conformal fluid \cite{Baier2008}, Israel-Stewart (IS) theory \cite{Stewart1977, Israel:1979wp} and others \cite{Denicol2007}.
It has been shown that the first order theory does not satisfy causality and it do not explain the presence of the unphysical behaviour in the plasma \cite{Olson:1990th, Koide:2010pw,Olson:1989dp, Koide:2006ef}. To demonstrate, we note that the dispersion relation obtained from the velocity diffusion equation is $\omega=-i\frac{\eta_0}{\varepsilon_0+p_0}k^2$ and the speed of diffusion is $v_T=\left|\frac{d\omega}{dk}\right|=2 \frac{\eta_0}{\varepsilon_0+p_0}k$ (here $\eta_0$, $\varepsilon_0$ and $p_0$ are background shear viscosity, energy density and pressure of the fluid). For $k\rightarrow \infty$, the diffusion speed goes to infinity and therefore violates causality \cite{Koide:2006ef}. Therefore the relativistic Navier-Stokes equation is only valid for the short wavenumber.  
Muller, Stewart and Israel in their series of papers \cite{Stewart1977, Israel:1979wp} addressed the causality problem. It was shown that one can avoid problems related to the causality, if we include second and higher order terms of thermodynamic variables.  With the second order hydrodynamic terms, for $k\rightarrow \infty$, longitudinal and transverse modes have group velocity, given by
\begin{align}
	v_L^{max}&
	=\sqrt{v_s^2+\frac{4}{3}\frac{\eta_0}{(\varepsilon_0+p_0)\tau_\pi}+\frac{\zeta_0}{(\varepsilon_0+p_0)\tau_\Pi}}, \\
	v_T^{max}&
	=\sqrt{\frac{\eta_0}{(\varepsilon_0+p_0)\tau_\pi}}.
\end{align}
From the above equations, it is clear that there is no causality problem for sufficiently large values of relaxation times  $\tau_\pi$  and $\tau_\Pi$ 
\cite{Romatschke2010}. However problem with this model is that,  we still don't know the transport coefficients  $\tau_\pi$  and $\tau_\Pi$. Later, it was realized that the theory of Muller-Israel-Stewart does not give a complete picture of evolution of the relativistic fluid under viscous hydrodynamics and it does not address second order terms properly. The generalization to the second and higher order terms has been addressed by many authors (see following Ref. \cite{Romatschke2010, Kharzeev2011a, Erdmenger2009, Banerjee2011} and references there in). These works are focused on constructing all possible second order derivative terms of the hydrodynamical degrees of freedom and then evaluating transport coefficients. 
In literature, few coefficients has been determined from underlying microscopic description.  Transport coefficients from microscopic theory are calculated using kinetic theory \cite{york:2009ag, Arnold:2000dr}, Kubo formalism \cite{Landsteiner2011}, fluid/gravity correspondence \cite{Bhattacharyya:2008jc} and diagrammatic methods \cite{Mañes:2013jl}. In Ref. \cite{Kharzeev2011a}, the authors have shown that out of a total 24 terms of second order derivative expansions, only 18 terms are possible for the parity odd fluid in the presence of external electric/magnetic fields and vorticity in the fluid. It was shown that the number of possible terms can be limited by using some symmetry property namely, conformal symmetry and discrete symmetries (charge conjugation and parity). Earlier in refs. \cite{Erdmenger2009, Banerjee2011},  the number of possible terms were constrained for the second order conformal viscous fluid in the absence of external electric/magnetic fields in the 4-D space. 
Few terms of the second order viscous chiral hydro are odd under the  parity transformation and are related with the quantum anomalies of the microscopic theory \cite{Son2009} and can be explained by quantum field theory \cite{Manuel2014a,Son2013a}. 
The phenomenon of quantum anomaly was first studied in holographic waves of strongly coupled Supersymmetric Yang-Mills (SYM) plasma  in the presence of non-zero chiral chemical potential in the chiral fluid \cite{Sahoo:2009yq, Policastro:2001sd}.
It was shown that the vorticity as well magnetic effect term is allowed not only from symmetry, but also from the hydrodynamic constraint ($\partial_{\mu} s^{\mu} \geq  0$, where $s^{\mu}$ is an entropy density tensor). It has been shown that the coefficients of the CME \cite{Vilenkin1980} and CVE \cite{Vilenkin1979} terms in the current are related with the anomaly coefficient at non zero chiral chemical potential.
%

In the above backdrop, in this chapter we discuss the first and second order conformal hydrodynamics for chiral plasma to study the hydrodynamic excitations. By linearization of the first and second order ChMHD equations, we have discussed some of the interesting aspect of the chiral plasma. 
\clearpage
\section{Waves in the non dissipative chiral plasma}\label{waves-Ch}
\subsection{Chiral Magnetic Waves (CMW)}\label{waves-CMW}
In this subsection, we will discuss the presence of a new type of collective excitation in the plasma, due to the coupling of the electric and chiral charge. These waves are generated in the presence of the external magnetic field, when CME \cite{Vilenkin1980} and {\it Chiral Charge Separation} effects (CSE) interplay \cite{Kharzeev2011}. 
Due to this, they carry several important new transport properties of the hot QCD plasma.
For simplicity let us consider single flavor massless QCD with chiral symmetry. The axial symmetry is violated via triangle anomaly of the global chiral symmetry. Let us assume that because of the presence of the axial and vector chemical potentials ($\mu_{A/V}$), there are two currents in the chiral plasma, namely vector and axial current  
%
\begin{eqnarray}
	{\bf j}_V=\frac{N_c e}{2\pi^2} \mu_5 {\bf B}, \label{eq:jmuV}\\
	{\bf j}_A=\frac{N_c e}{2\pi^2} \mu_V {\bf B} \label{eq:jmuA}.
\end{eqnarray}
Here $N_c$ is the number of colour charges and ${\bf B}$ is the external magnetic field. One can establish a close connection between CME and CSE by the method of dimensional reduction appropriate in the case of strong magnetic field \cite{Basar2010}.
The simple relations $j_V^0=j_A^1$ and $j_A^0=j_V^1$ between the vector and axial currents in the dimensionally reduced (1+1) theory imply that the density of the baryon charge (which is proportional to $\mu_R+\mu_L$) must induce the axial current and the density of axial charges must induce the current of electric charge (CME). This can be seen as follows: Consider a local perturbation in electric charge density, which implies from equation (\ref{eq:jmuA}), perturbation in the axial current. 
In turn the perturbation in the axial current, would induce a local perturbation in the $\mu_5$. 
Thus according to equation (\ref{eq:jmuV}), perturbations in the electric current. The resulting fluctuations  in the electric charge density complete the cycle and lead to the excitations that combine the density waves of electric and chiral charges. This density wave is known as "{\it Chiral Magnetic Waves}". In the matrix form the above two currents can be written as follows:
\begin{equation}
	\left(\begin{array}{c} {\bf j}_V \\ {\bf j}_A \end{array}\right) 
	=\frac{N_c e{\bf B}}{2\pi^2} \left(\begin{array}{cc} 0 & 1\\ 1 & 0 \end{array}\right)
	\left(\begin{array}{c} \mu_V \\ \mu_5 \end{array}\right).
\end{equation}
Let us assume that there is no charge density on average, and plasma is neutral. Now linearization of the above equation gives
\begin{equation}
	\left(\begin{array}{c} \mu_V \\ \mu_5 \end{array}\right) 
	=\left(\begin{array}{cc} \frac{\partial \mu_V}{\partial j}_V^0 & \frac{\partial \mu_5}{\partial j}_V^0\\ \frac{\partial \mu_V}{\partial j_A^0} & \frac{\partial \mu_5}{\partial j_A^0} \end{array}\right)
	\left(\begin{array}{c} {\bf j}_V^0 \\ {\bf j}_A^0 \end{array}\right)
	+\mathcal{O}((j^0)^2).
\end{equation}
In the chiraly symmetric phase, above equation will reduced to 
\begin{equation}
	\left(\begin{array}{c} {\bf j}_V \\ {\bf j}_A \end{array}\right) 
	=\frac{N_c e{\bf B}\alpha}{2\pi^2} \left(\begin{array}{cc} 0 & 1\\ 1 & 0 \end{array}\right)
	\left(\begin{array}{c} j_V^0 \\ j_A^0 \end{array}\right),
\end{equation}
where $\alpha$ is defined as $\frac{\partial \mu_V}{\partial j_V^0}=\frac{\partial \mu_5}{\partial j_A^0}=\alpha$. This equation represents the decoupled current expressions for vector and axial currents. Therefore, in the chiral basis, the current expression for right handed and left handed currents can be written as 
\begin{equation}
	{\bf j}_{L,R}= \mp \frac{N_c e {\bf B} \alpha}{2\pi^2} j_{L,R}^0. \label{eq: jzeroth}
\end{equation}
In above equation, $\pm$ sign depends on the chirality of the plasma. 
Right hand side of equation (\ref{eq: jzeroth}) is the leading order contribution to the chiral magnetic conductivity. However the next leading order correction will be $\partial^2$ or $\omega^2\sim k^2$ in the frequency/momentum space. One of the possible terms are diffusion contribution (which represents the diffusion term $-D\,\nabla j^0$, with $D$ as diffusion constant). With the addition of diffusion term, above current will be modified to following form
\begin{eqnarray}
	{\bf j}_{L,R}= \mp \frac{N_c e {\bf B} \alpha}{2\pi^2} j_{L,R}^0-D_L\frac{{\bf B}({\bf B}\cdot \nabla)}{B^2}j_{L,R}^0 +.., 
\end{eqnarray}
with $D_L$ is longitudinal diffusion constant. One can consider transverse part too. But in the discussion below, for simplicity, we have considered only the longitudinal part. Next using $\partial_\mu j^\mu_{L,R}=0$, and considering ${\bf B}=b \hat{x}$ and only longitudinal gradient $\partial_1$, we have
\begin{eqnarray}
	\left(\partial_0\mp \frac{N_c e B\alpha}{2\pi^2}\partial_1- D_L \partial_1^2\right)j_{L,R}^0=0.
\end{eqnarray}
This describes a directional wave, or chiral wave, of charge densities whose direction of motion is correlated with the magnetic field. The velocity of motion of the wave is
\begin{eqnarray}
	v_x =\frac{N_c e B\alpha}{2\pi^2}.
\end{eqnarray}
The dispersion relation of the wave can be written as
\begin{eqnarray}
	\omega =\mp v_x k-iD_L k^2 +....
\end{eqnarray}
The first term in the dispersion relation, which makes the mode propagating instead of simply diffusing is of interest and these modes are known as chiral magnetic waves \cite{Kharzeev2011}. This wave exists only when triangle anomaly and external magnetic field are present.
\subsection{Chiral Vortical Waves (CVW)}\label{waves-CVW}
A similar situation, as in the chiral magnetic wave, can also happen when vector and axial current/densities (contributed from the vorticity effect in the plasma) couple together. The vector and axial currents are given as
\begin{eqnarray}
	{\bf j}_V & = &\frac{1}{\pi^2}\mu_V \mu_5 \boldsymbol{\omega}\\
	{\bf j}_A & = & \left[\frac{1}{6}T^2+\frac{1}{2\pi^2}(\mu_V^2+\mu_5^2)\right]\boldsymbol{\omega}
\end{eqnarray}
In the similar way as done in the previous subsection, we can write the expression for the right/left handed particles as:
\begin{eqnarray}
	{\bf j}_{L,R} = \mp \left(\frac{1}{12}T^2+\frac{1}{4\pi^2}\mu_{l,R}^2\right)\boldsymbol{\omega}
\end{eqnarray}
Now by combining the continuity equations
\begin{equation}
	\partial_t n_{L,R}+\nabla\cdot {\bf j}_{L,R}=0,
\end{equation}
we obtain
\begin{eqnarray}
	\partial_t n_{L,R}=\pm \frac{1}{4\pi^2}\omega \partial_x (\mu_{L,R}^2)=\pm \frac{\omega \mu_{L,R}}{2\pi^2} \partial_x \mu_{L,R},
\end{eqnarray}
where we have set the vorticity along spatial x-direction {\it i.e.} $\boldsymbol{\omega}\,=\,\omega \,\hat{x}$ with $\hat{x}=-{\bf x}/x$. Now let us define susceptibilities for the corresponding densities as $\chi_\mu= \partial n_{L,R}/\partial \mu_{L,R}$. Let us suppose a case, where background fluid is not neutral i.e. $\mu_0 \neq 0$.In this case, one can linearise above equation using $\delta n=\chi_0 \delta\mu$. So
\begin{eqnarray}
	\partial_t (\delta n) + \frac{\mu_0 \omega}{2\pi^2 \chi_{\mu_0}} \partial_x (\delta n) = 0.
\end{eqnarray}
This is just a wave equation describing a propagating mode with a gapless dispersion relation
\begin{eqnarray}
	V_{\omega}=\frac{\mu_0 \omega}{2\pi^2 \chi_{\mu_0}}
\end{eqnarray}
This is the chiral vortical wave (CVW) corresponding to the right handed wave with wave speed $V_\omega$ \cite{Jiang2015}. One can get left handed wave modes in the similar way but propagation velocity in the negative direction of $\boldsymbol{\omega}$. The CVW obtained above is essentially a hydrodynamic density wave arising from slowly varying vector and axial density fluctuations that are coupled together through vortical effects. 
\subsection{Chiral Alf\'ven Wave (CAW)}
At high temperature authors of the ref. \cite{Yamamoto2015} have shown that a new type of gapless collective excitation can be induced in the presence of the external magnetic fields. They showed that these waves are transverse in nature and they can sustain even in the fluids, where $\nabla \cdot {\bf v}$ {\it i.e.} incompressible fluids. Physical argument for this is: let us assume that an external magnetic field is present in the fluid along the $\hat{z}$-direction, and the perturbation in the velocity fields is in the $\hat{y}$-direction, such that ${\bf v}=v(z)\hat{y}$ with $\partial_z v(z)< 0$. In this case vorticity in the plasma will be in the positive {\bf x}-direction. Therefore the local current due to vorticity will be ${\bf j}\propto T^2 \boldsymbol{\omega}$ \cite{Landsteiner2011}. As a result, there will be a Lorentz force $({\bf j}\times{\bf B})$ in the  {\bf y}-direction. This is in the opposite direction of fluid velocity and so it will act as a restoring force and makes the perturbed velocity {\bf v} to oscillate. This is the basic understanding of the waves in the chiral plasma. We have seen that stress-energy tensor and the axial currents follow following conservation equations (\ref{eq:StressCons})-(\ref{eq:ACons}) 
\begin{gather}
	\nabla_{\mu} T^{\mu\nu}= F^{\nu\lambda}j_{\lambda_V}, \label{conservT}\\
	\nabla_{\mu}j^{\mu}_A=CE^{\mu}B_{\mu},\label{conservJ}\\
	\nabla_{\mu}j^{\mu}_V=0 \label{conserv_vector},
\end{gather}
here energy momentum tensor $T^{\mu\nu}$ for ideal fluid can be given as
\begin{equation}
	T^{\mu\nu}_{ideal}=(\varepsilon+p)u^\mu u^\nu+p g^{\mu\nu}. \label{energymomentumideal}
\end{equation}
The total current in the anomalous plasma at very high temperature, where chemical potential is very small (in spatially flat space) is 
\begin{eqnarray}
	j^\mu = nu^\mu +\xi \omega^\mu +\xi^{B}B^\mu,
\end{eqnarray}
where $\xi$ and $\xi^{B}$ are defined in Eqs. (\ref{eq:xi}) and (\ref{eq:xiB}).
Now one can obtain evolution equations is given as:
\begin{eqnarray}
	(\partial_\tau +{\bf v}\cdot {\bf v})\varepsilon+(\varepsilon+p)\nabla\cdot {\bf v}=0\\
	({\bf v}\cdot \nabla)p=0\\
	(\varepsilon+p)(\partial_\tau+{\bf v}\cdot {\bf v}){\bf v}=-\nabla p+{\bf j}\times{\bf B}\\
	\partial_\tau n+\nabla\cdot j=0
\end{eqnarray}
These equations can be simplified for the case where $\varepsilon$, $p$ and $n$ are independent of space and time. 
\begin{eqnarray}
	(\varepsilon+p)(\partial_\tau+{\bf v}\cdot {\bf v}){\bf v}=(n{\bf v}+\xi \boldsymbol{\omega})\times{\bf B}\\
	\nabla\cdot {\bf v}=0.
\end{eqnarray}
By linearization (considering terms upto linear order in perturbations) one can get following equation
\begin{eqnarray}
	(\varepsilon+p)\, \partial_\tau \,{\bf v}=\xi\, \boldsymbol{\omega}\times{\bf B}
\end{eqnarray}
Where $\xi=DT^2/2$ and using $\boldsymbol{\omega}\times{\bf B}=({\bf B}\cdot \nabla)\,{\bf v}-\nabla ({\bf B}\cdot {\bf v})$, one can obtain
evolution equation 
\begin{equation}
	\partial_\tau {\bf v}=V_T\partial_z {\bf v}.
\end{equation}
Here $V_T= \frac{D T^2}{2}\frac{B}{\varepsilon+p}$. This will give the dispersion relation $\omega=-V_T k_z$. Clearly the wave propagates in the opposite direction of the magnetic field and is known as Chiral Alf\'ven Wave \cite{Yamamoto2015}. Since the wave velocity is proportional to the gauge gravitational anomaly coefficient $D$ for chiral particles, they are absent in the normal fluids.
\section{Conformal viscous  chiral hydrodynamics} \label{sec:CVChydro}
If the action of a theory is invariant under conformal or Weyl transformations of the metric $g_{\mu\nu}(x)\rightarrow g^*_{\mu\nu}= e^{-2w(x)}g_{\mu\nu}$, where $w(x)$ is an arbitrary scalar function of space time, 
the energy momentum tensor for the fluid follows: $\nabla_\nu T^{\mu\nu}_{total}=0 \Leftrightarrow \nabla^*_\nu T^{*\mu\nu}_{total}=0$ and trace is zero, i.e., $T^\mu_\mu=0$. This means that energy momentum tensor of the fluid remains invariant under conformal transformations.  Here $T^{*\mu\nu}=e^{-6w(x)} T^{\mu\nu}$ and $\nabla^*_{\nu}$ is defined with respect to conformal metric $g^*_{\mu\nu}$.
Most general forms of the energy momentum tensor upto second order derivative expansion can be written as
\begin{equation}
	T^{\mu\nu}=T_{ideal}^{\mu\nu}+\tau^{\mu\nu}_{(1)}+\tau^{\mu\nu}_{(2)}.
\end{equation}
Last two terms $\tau^{\mu\nu}_{(1)}$ and $\tau^{\mu\nu}_{(2)}$ are the first and second order derivative terms. $T^{\mu\nu}_{ideal}$ is the ideal part of the total energy momentum tensor and its most general form is given by (\ref{energymomentumideal}).
%
The first order energy momentum tensor follows conformal symmetry and  can be written as
\begin{gather}
	\tau^{\mu\nu}_{(1)}=-2\eta \sigma^{\mu\nu},
	\label{firstorder1}
\end{gather}
where $\eta$ is shear viscosity and $\sigma^{\mu\nu}$ is defined in terms of the covariant derivative of the velocity field as
\begin{equation}
	\sigma_{\mu\nu} =\frac{1}{2}(\nabla_\mu u_\nu+\nabla_\nu u_\mu).
\end{equation}
The covariant derivative of the velocity field can be written as  $\nabla_\mu u_\nu= \sigma_{\mu\nu}+\omega_{\mu\nu}$. Here $\omega_{\mu\nu}=\frac{1}{2}(\nabla_\mu u_\nu-\nabla_\nu u_\mu)$. In Ref. \cite{Kharzeev2011a}, the authors have shown that out of 24 possible terms at second order derivative terms, only 18 terms are possible. 
The most general form of the second order terms of energy momentum tensor for the chiral fluid in presence of external electric/magnetic field can be written as \cite{Kharzeev2011}
\begin{gather}
	\tau_{(2)}^{\mu\nu}= \lambda_1 ~ \Pi^{\mu\nu}_{\alpha\beta}\nabla^\alpha \omega^\beta +\lambda_2~
	\Pi^{\mu\nu}_{\alpha\beta} \omega^\alpha \nabla^\beta \bar{\mu}
	+\lambda_3~\Pi^{\mu\nu}_{\alpha\beta}
	\epsilon^{\gamma\delta\eta\alpha}\sigma^\beta_\eta u_\gamma \nabla_\delta \bar{\mu} \nonumber \\ +\lambda_4~
	\Pi^{\mu\nu}_{\alpha\beta}\nabla^\alpha B^\beta+\lambda_5 ~\Pi^{\mu\nu}_{\alpha\beta} B^\alpha \nabla^\beta \bar{\mu}+\lambda_6~ \Pi^{\mu\nu}_{\alpha\beta}  E^\alpha B^\alpha\nonumber\\
	+\lambda_7~\Pi^{\mu\nu}_{\alpha\beta}\epsilon^{\gamma\delta\eta\alpha}\sigma^{\beta}_\eta u_\gamma E_\delta +\lambda_8~
	\Pi^{\mu\nu}_{\alpha\beta} \omega^{\alpha} E^{\beta}, \label{secondorder}
\end{gather}
where the coefficients $ \lambda_1,  \lambda_2 ....,  \lambda_8$ are transport coefficients and $\Pi^{\mu\nu}_{\alpha\beta}=\frac{1}{2}\large[P_{\alpha}^{\mu} P_{\beta}^{\nu}+P_{\beta}^{\mu} P_{\alpha}^{\nu}-\frac{2}{3}P^{\mu\nu} P_{\alpha\beta}\large]$. Here $P^{\mu\nu}=(g^{\mu\nu}+u^\mu u^\nu)$ is the projection operator and $\bar \mu$ is defined as $\mu_5/T$.  
Here $\omega^{\mu}=1/2 \epsilon^{\mu\nu\alpha\beta}u_\nu \nabla_\alpha u_\beta$ vorticity four vector.

In our discussion, we will use conformal variables in a conformally flat space time, defined in Chapter (\ref{ch3}) just above Maxwell's equations (\ref{M1})-(\ref{M4}). Also in this coordinate system, fluid velocity is measured in the local comoving frame of the fluid with $u^{\mu}=(1, \textbf{v})$ (non-relativistic). Here we have considered electrically neutral plasma.
\subsection{Chiral magnetohydrodynamics with first order viscous terms}
The energy momentum tensor of first order viscous hydrodynamics can be written as
\begin{equation}
	T^{\mu\nu}= T^{\mu\nu}_{ideal}+\tau^{\mu\nu}_{(1)}, \label{energymomentumtensor1}
\end{equation}
where first and second term on the right hand side of the above equation are respectively the ideal fluid part and first order viscous part of the energy momentum tensor. 
Also the currents for the right handed and left handed particles with the anomalous effects  can be written  in terms of thermodynamic variables as 
\begin{gather}
	j^{\mu}_{R,L}= n_{R,L}u^{\mu}+\nu^{\mu}_{R,L}.
\end{gather}
Here $n_{R,L}$ is charge density of the left and right handed particles and $\nu^{\mu}_{R,L}$ is contribution to the current from the chiral imbalance. In the Landau frame $\tau^{\mu\nu}_{(1)}$ and $\nu^{\mu}$ for each species require $u_\mu \tau^{\mu\nu}=u_\mu\nu^\mu=0$.  For each species, $\nu_{R,L}^\mu$ is given by \cite{Son2009}
\begin{equation}
	\nu^\mu_{R,L}=n_R u^\mu -\frac{1}{2}\sigma E^{\mu} +\xi_{R,L}\omega^{\mu}+\xi_{R,L}^{B}B^\mu \label{nuRL1}
\end{equation}
and $\tau_{(1)}^{\mu\nu}$ is given in $\text{eq.}$(\ref{firstorder1}). $\xi_{R,L}$ and $\xi_{R,L}^B$ are transport coefficients corresponding to CME and CVE. They are functions of chemical potentials of the right- and left-handed particles of the chiral plasma. Therefore one can define vector current and axial current for a neutral chiral plasma
\begin{gather}
	j_V^{\mu}=-\sigma E^{\mu}+\xi_V\omega^{\mu}+\xi_V^{B}B^\mu, \label{vectorcurrent} \\
	j_5^{\mu}=n_5u^\mu+\xi_5\omega^{\mu}+\xi_5^B B^\mu, \label{axialcurrent}
\end{gather}
where $\xi_{v, 5}=\xi_R\pm\xi_L$ and $\xi_{v,5}^B=\xi_R^B\pm\xi_L^B$. The first term in $j_V^\mu$ is conventional term and describes how electric current flows in the direction of the electric field. However, other two terms are the Chiral Vortical and Chiral Magnetic currents respectively. 

Let us first assume that there is no turbulence in the plasma and it is in local thermal equilibrium. In that case one can easily get following dispersion relation using Maxwell's equations (\ref{maxwell}) and current expression (\ref{vectorcurrent})
\begin{equation}
	-i\omega =-\frac{k^2}{\sigma}	+\frac{\xi^B}{\sigma}k.
\end{equation}
We have taken spatio-temporal variation as $\text{exp}[-i(\omega\tau-\textbf{k}\cdot\textbf{r})]$. Clearly, for $k<\xi^B$ there will be instability in the plasma and modes will grow exponentially \cite{Akamatsu2013df, Bhatt2016, Bhatt2015} and for $k \approx \xi^B/2$, mode will grow maximally. One can get $|\omega|$  for maximally growing mode which is $|\omega_{max}|\approx \frac{\xi_0^{B^2}}{4\sigma_0}$. This instability persist only for non-zero value of the parameter $\xi^B$ (which is related with the chiral asymmetry in the plasma). This kind of situation can occur in the early universe above the electroweak scale \cite{Giovannini:2013oga, Giovannini1998}. 

Next we include the effects of turbulence in the problem and we do linear analysis of the  \text{Eqs.}~\eqref{conservT}, (\ref{firstorder1}) and (\ref{energymomentumtensor1}) 
using $\varepsilon =\varepsilon_0 +\delta \varepsilon$, $p=p_0 +\delta p $, $ n=n_0+\delta n$, $u^\mu =\delta u^\mu = (0, \delta \textbf{v})$  (here we have taken $\gamma \approx 1$), 
$T^{\mu\nu} =T^{\mu\nu}_0 +\delta T^{\mu\nu}$. Subscript "0" with the quantities shows background values and are homogeneous and isotropic. In our calculations, we assumed that the chemical potential and temperature are independent of space-time. With these considerations, equations (\ref{conservT}) will give
%
\begin{gather}
	\partial_\tau \delta \varepsilon +\nabla\cdot \left[(\varepsilon_0+p_0)\delta\textbf{v}\right]=0 \label{24.00}
	\\
	\partial_\tau\left[\left(\varepsilon_0+p_0\right) \delta \textbf{v}\right] 
	+v_s^2 \nabla \delta \varepsilon -\eta_0\left[\nabla^2 \delta \textbf{v}+\nabla (\nabla\cdot \delta \textbf{v})\right] \nonumber \\ 
	+\frac{1}{3}\eta_0\nabla \left(\nabla\cdot \delta \textbf{v}\right)
	= \xi^B_0 \left(\textbf{B}_0\times \delta \textbf{B}\right)+
	\left(\nabla\times\delta \textbf{B}\right)\times \textbf{B}_0\label{25.00}.
\end{gather}
To obtain the above expression, we have used $\delta p = v_s^2~ \delta \varepsilon$. Here $v_s$ is sound speed and $\textbf{B}_0$ is a background magnetic field. We take the Fourier transform of the perturbations using spatio-temporal variation in eq. \eqref{25.00}. 
From the Eqs. \eqref{25.00}, we have obtained the following equation
\begin{gather}
	-i \omega\, \delta \textbf{v}+i\frac{v_s^2}{\omega}\, \textbf{k}\,(\textbf{k}\cdot \delta \textbf{v})+\frac{\eta_0}{\varepsilon_0+p_0}
	\left[k^2\, \delta \textbf{v} +\textbf{k}\,(\textbf{k}\cdot \delta \textbf{v})\right] \nonumber\\
	-\frac{1}{3}\frac{\eta_0}{\varepsilon_0+p_0}\textbf{k}\,(\textbf{k}\cdot \delta \textbf{v})
	=\frac{\xi_0^B}{\varepsilon_0+p_0}\left(\textbf{B}_0\times \delta\textbf{B}\right) \nonumber\\
	+\frac{i}{\varepsilon_0+p_0}(\textbf{k}\times\delta \textbf{B})\times \textbf{B}_0 \label{eq13}
\end{gather}
In the dimensionless form, eq. \eqref{eq13} takes the following form
%
\begin{gather}
	\left(-i\tilde\omega +\frac{\eta_0 \sigma_0}{\varepsilon_0+p_0}\tilde{k} ^2\right)~\delta \tilde{\textbf{v}}+iv_s^2 
	\frac{4\sigma_0^2}{\tilde{\omega}\xi_0^{B^2}}\tilde{\textbf{k}}(\tilde{\textbf{k}}\cdot \delta \tilde{\textbf{v}})+\frac{2}{3}\frac{\eta_0 \sigma_0}{\varepsilon_0+p_0} \nonumber \\
	\tilde{\textbf{k}}(\tilde{\textbf{k}}\cdot \delta\tilde{\textbf{v}}) 
	=\frac{8\sigma_0}{\xi_0^{B}} v_A^2\left[(\hat{\textbf{n}}\times\delta \tilde{\textbf{B}})
	+\frac{i}{2}(\tilde{\textbf{k}}\times\delta \tilde{\textbf{B}})\times \hat{\textbf{n}}\right]. \label{delta v}
\end{gather}
%
To write eq. (\ref{eq13}) in dimensionless form, we redefine variables  $\omega$, $\textbf{k}$  and $\delta\textbf{B}$ in the dimensionless form as 
$\tilde \omega =\omega/|\omega_{max}|$, $\tilde{\textbf{k}}=\textbf{k}/k_{max}$ and $\delta \tilde{\textbf{B}}=\delta\textbf{B}/B_0$. We have also defined the background magnetic field $\textbf{B}_0=B_0\,  \hat{{\bf n}}$ (Here $\hat{{\bf n}}$ stands for the direction of the background magnetic field). The background magnetic field can be generated by any of the early Universe processes. 
$v_A=B_0/\sqrt{(\varepsilon_0+p_0)}$ is Alf\'ven velocity. We need another equation to solve eq.\eqref{delta v}.
Second equation can be obtained by using Maxwell's eq.(\ref{maxwell}) and eq.(\ref{vectorcurrent}). The dimensionless form of this is
\begin{gather}
	\left(-i\tilde{\omega}+\tilde{k}^2\right)\delta\tilde{\textbf{B}}=i\frac{2\sigma_0}{\xi_0^B}[\tilde{\textbf{k}}\times(\delta\tilde{\textbf{v}}
	\times\hat{{\bf n}})]+2i (\tilde{\textbf{k}}\times\delta\tilde{\textbf{B}}) \nonumber \\
	-\frac{\xi_0}{B_0}[\tilde{\textbf{k}}\times (\tilde{\textbf{k}}\times \delta\tilde{\textbf{v}})]. \label{diffusitivityequation}
\end{gather}
Dispersion relations can be obtained by solving equations (\ref{delta v}) and (\ref{diffusitivityequation}). We considered two cases for the magnetic field perturbations $\delta \tilde{\textbf{B}}$. First one is $\delta \tilde{\textbf{B}} \parallel \hat{{\bf n}} $ and second one is $\delta \tilde{\textbf{B}}\perp \hat{\textbf{n}}$. 

We have solved these equations in two cases: when \\
{\bf i)} $\tilde{\textbf{k}}\parallel \hat{\textbf{n}}$ or
{\bf ii)} $\tilde{\textbf{k}}\perp \hat{{\bf n}}$ 
for the case of $\delta \tilde{\textbf{B}} \parallel \hat{{\bf n}} $ and $\delta \tilde{\textbf{B}}\perp \hat{\textbf{n}}$. \\ \\
%
\underline{\textbf{Case: (i)}} With modes $\tilde{\textbf{k}}\parallel \hat{\textbf{n}}$ and $\delta \tilde{\textbf{B}} \parallel \hat{{\bf n}}$\\
\textbf{(a)} For $\delta \tilde{\textbf{v}}\parallel \tilde{\textbf{k}}$
\begin{gather}
	\tilde{\omega}=-i \tilde{k}^2	\\
	\tilde{\omega}^2+\frac{i}{3}\frac{\eta_0}{\varepsilon_0+p_0}\tilde{k}^2\tilde{\omega}-\frac{4iv_s^2\sigma_0^2}{\xi_0^{B^2}}\tilde{k}^2=0
\end{gather}
\textbf{(b)} For $\delta \tilde{\textbf{v}}\perp \tilde{\textbf{k}}$
\begin{gather}
	\tilde{\omega}=-i \frac{\eta_0\sigma_0}{\varepsilon_0 +p_0}\tilde{k}^2
\end{gather}
\underline{\textbf{Case:(ii)}}. With modes $\tilde{\textbf{k}}\perp \hat{\textbf{n}}$ and $\delta \tilde{\textbf{B}} \parallel \hat{{\bf n}}$\\
\textbf{(a)} For $\delta \tilde{\textbf{v}}\perp \tilde{\textbf{k}}$\\
In this case, dispersion relations are
\begin{gather}
	\tilde{\omega}=-i \frac{\eta_0\sigma_0}{\varepsilon_0 +p_0}\tilde{k}^2
\end{gather}
\textbf{(b)} For $\delta \tilde{\textbf{v}}\perp \tilde{\textbf{k}}$.
In this case, it is straight forward to obtain dispersion relations, but they are too long to be written here. \\
\underline{\textbf{Case: (iii)}} With modes $\tilde{\textbf{k}}\parallel \hat{\textbf{n}}$ and $\delta \tilde{\textbf{B}} \perp \hat{{\bf n}}$\\
\textbf{(a)} For $\delta \tilde{\textbf{v}}\parallel \tilde{\textbf{k}}$
\begin{gather}
	\tilde{\omega}= -i\tilde{k}^2+ 2i\tilde{k} \label{20}, \\
	\tilde{\omega}^2+i\frac{5}{3}\frac{\sigma_0\eta_0}{\varepsilon_0+p_0}\tilde{k}^2\tilde{\omega}- iv_s^2\frac{4\sigma^2_0}{\xi_0^{B^2}}\tilde{k}^2=0.\label{casei1}
\end{gather}
\textbf{(b)} For $\delta \tilde{\textbf{v}}\perp \tilde{\textbf{k}}$
\begin{gather}
	\tilde{\omega}^2+i\tilde{\omega}\tilde{k}^2-4\sigma_0 v_A^2  \left(\frac{2i\sigma_0 B_0 -\xi_0\xi_0^B}{B_0\xi_0^{B^2}}\right)\tilde{k}=0
	\label{caseib}
\end{gather}
\underline{\textbf{Case:(iv)}} With modes with $\tilde{\textbf{k}}\perp \hat{\textbf{n}}$ and $\delta \tilde{\textbf{B}} \perp \hat{{\bf n}}$\\
\textbf{(a)} For $\delta \tilde{\textbf{v}}\perp \tilde{\textbf{k}}$\\
In this case dispersion relations are 
\begin{gather}
	\tilde{\omega}^2+i\tilde{\omega}\left[\frac{\eta_0\sigma_0}{\varepsilon_0+p_0}\tilde{k}^2+\tilde{k}^2\right]-\frac{\eta_0\sigma_0}{\varepsilon_0+p_0}\tilde{k}^4\pm\frac{4\sigma_0\xi_0v_A^2}{\xi_0^B B_0}=0
\end{gather}
\\
\textbf{(b)} For $\delta \tilde{\textbf{v}}\parallel \tilde{\textbf{k}}$.\\
Here we have obtained two modes, one is $\tilde{\omega}=-i\tilde{k}^2$ and other one is
\begin{gather}
	\tilde{\omega}^2+i\frac{5}{3}\frac{\sigma_0 
		\eta_0}{(\varepsilon_0+p_0)}\tilde{k}^2 \tilde{\omega}-i v_s^2\frac{4\sigma_0\tilde{k}^2}{\xi_0^{B^2}}=0.
	%
	%
\end{gather}
In the result and discussion part, we have given a very brief discussion of the above dispersion relations.
\subsection{Chiral magnetohydrodynamics with second order viscous terms}
Total energy momentum tensor with first order and second order contributions can be written as:
\begin{equation}
	T^{\mu\nu}=T_{(ideal)}^{\mu\nu}+\tau_{(1)}^{\mu\nu}+\tau_{(2)}^{\mu\nu},
\end{equation}
%
where $\tau_{(2)}^{\mu\nu}$ the second order derivative terms of the velocity field and other thermodynamical variables respectively. These are defined in \text{Eqns.} \eqref{energymomentumideal}, \eqref{firstorder1} and \eqref{secondorder}. Similar to the first order, we do a linear analysis of the second order hydro for chiral plasma. Background and perturbations can be written as
\begin{gather}
	T^{\mu\nu}_0 =(\varepsilon_0+p_0)u_0^\mu u_0^\nu+p_0 g^{\mu\nu}
\end{gather}
\begin{align}
	\delta T^{\mu\nu}=& (\varepsilon+p)[u_0^\mu \delta u^\nu+\delta u^\mu u_0^\nu]+ \delta p
	g^{\mu\nu} +(\delta\varepsilon+\delta p)u_0^\mu u_0^\nu \nonumber\\
	&-\eta_0 P^{\mu\alpha}_{0}P^{\nu\beta}_{0}[\nabla_\alpha\delta u_\beta+\nabla_\beta 
	\delta u_\alpha]+\frac{1}{3}\eta_0 P^{\mu\nu}_{0}\nabla_\alpha \delta u^\alpha \nonumber\\
	&+\lambda_{1_{0}} \Pi^{\mu\nu}_{{\alpha\beta}_{0}}\nabla^{\alpha}\delta\omega^\beta
	+\lambda_{4_{0}}\Pi^{\mu\nu}_{{\alpha\beta}_{0}} \nabla^\alpha \delta B^\beta \label{30}
\end{align}
Components of the perturbed energy momentum tensor can be found using eq. (\ref{30}) as
\begin{gather}
	\delta T^{00}=\delta \varepsilon \\
	\delta T^{0i}= (\varepsilon_0+p_0) \delta v^i
	-\frac{\lambda_{{1}_{0}}}{2}\epsilon^{ijk}\partial_0\partial _j \delta v_k -
	\frac{\lambda_{{4}_{0}}}{2}\partial_0 \delta B^i \\
	\delta T^{ij}=\delta p \eta^{ij} -\eta_0\large[g^{ik}\partial_k \delta v^j+g^{jk}\partial_k \delta v^i\large] +\frac{1}{3}\eta_0 g^{ij} \partial_k \delta v^k \nonumber \\
	+ \frac{\lambda_{{1}_{0}}}{2}\large[g^{ik}\epsilon^{jlm}\partial_k \partial _l\delta v_m+g^{jk}
	\epsilon^{ilm}\delta_k \partial _l\delta v_m-\frac{2}{3}g^{ij}\epsilon^{klm}\partial_k\partial_l\delta v_m\large] \nonumber \\
	+\frac{\lambda_{{4}_{0}}}{2}\large[g^{ik}\partial_k \delta B^j+g^{jk}\delta_k \delta B^i-\frac{2}{3}g^{ij}\partial_k\delta B^k\large]
\end{gather}
For $\nu=0$, the conservation equation \eqref{conservT} gives
\begin{equation}
	\partial _0 \delta \varepsilon +(\varepsilon _0 +p_0)\partial_i\delta v^i 
	+\frac{\lambda_{{1}_{0}}}{2}\epsilon^{ilm}\partial_0\partial_i \partial_l \delta v_m + \frac{\lambda_{{4}_{0}}}{2} \partial_0\partial_i\delta B^i  \nonumber \\
	=0\label{secondorderenergy}
\end{equation}
Similarly for $\nu=j$, from the conservation equation (\ref{conservT}) one can obtain the following equation
\begin{gather}
	\partial_0\large[(\varepsilon_0+p_0)\delta v^i -\frac{\lambda_{{1}_{0}}}{2}\epsilon_{ilm}\partial_0\partial_l \delta v_m 
	-\frac{\lambda_{{4}_{0}}}{2}\partial_0 \delta B^i\large]\nonumber \\
	+\partial_i\large[\delta p \eta^{ij}-\eta_0\eta^{il}\eta^{jk}\{\partial_l
	\delta v_k+\partial_k\delta v_l\}+\frac{1}{3}\eta_0 \eta^{ij}\partial_k\delta v^k \nonumber \\
	+\frac{\lambda_{{1}_{0}}}{2}\{\eta^{ik}\epsilon^{jlm}\partial_k \partial_l\delta v_m+\eta^{jk}\epsilon^{ilm}
	\partial_k\partial_l \delta v_m-\frac{2}{3}\eta^{ij}\epsilon^{klm}\partial_k\partial_l\delta v_m\}\nonumber \\
	+\frac{\lambda_{{4}_{0}}}{2}\{\eta^{ik}\partial_k\delta B^j +\eta^{jk}\partial_k \delta B^i-\frac{1}{3}g^{ij}\partial_k 
	\delta B^k\}\large] \nonumber \\
	=\xi_0^B \epsilon^{jkl} B_0^k\delta B^l+ \epsilon^{jkl}B_0^l\delta j^k \label{secondordereuler}
\end{gather}
From equations (\ref{secondorderenergy}) and (\ref{secondordereuler}), in Fourier space we have derived dimensionless equation for a coupled system in terms of  $\delta \textbf{{v}}$ and $\delta \textbf{B}$ as
\begin{gather}
	\left(-i\tilde\omega +\frac{\eta_0 \sigma_0}{\varepsilon_0+p_0}\tilde{k} ^2\right)~\delta \tilde{\textbf{v}}+ iv_s^2 \frac{4\sigma_0^2}{\tilde{\omega}\xi_0^{B^2}}\tilde{\textbf{k}}(\tilde{\textbf{k}}\cdot \delta \tilde{\textbf{v}})
	\nonumber \\
	+\frac{1}{3}\eta_0 \frac{\sigma_0}{\varepsilon_0+p_0} \tilde{\textbf{k}}(\tilde{\textbf{k}}\cdot \delta\tilde{\textbf{v}}) 
	+\frac{\pi \lambda_{4_{0}}\xi_0^B B_0}{2(\varepsilon_0+p_0)}\left(\tilde{\omega}^2-\frac{4\sigma_0^2}{\xi_0^{B^2}}\tilde{k}^2\right)\delta\tilde{\textbf{B}} \nonumber \\
	+i\frac{\lambda_{1_{0}}\xi_0^{B^3}}{8(\varepsilon_0+p_0)\sigma_0}\left(\tilde{\omega}^2-\frac{4\sigma_0^2}{\xi_0^{B^2}}\tilde{k}^2\right)
	(\tilde{\textbf{k}}\times \delta \tilde{\textbf{v}}) \nonumber \\ =\frac{16\pi^2\sigma_0^2}{\xi_0^{B^2}} v_A^2\left[\frac{1}{2}(\hat{\textbf{n}}\times\delta \tilde{\textbf{B}})+i(\tilde{\textbf{k}}\times\delta \tilde{\textbf{B}})\times \hat{\textbf{n}}\right] \label{secondordereulerfourier}
\end{gather}
Now we have two equations to solve- (\ref{diffusitivityequation}) and (\ref{secondorderenergy}). It is quite complicated to solve these equations analytically.
In the case of transverse motion, dispersion relation contains series of terms 
\begin{gather}
	\tilde{\omega}\approx
	\pm
	\frac{8\pi \sigma_0^2 v_s}{\xi_0^{B^2}}
	\tilde{k}-i\frac{\eta_0\sigma_0}{2(\varepsilon_0+p_0)}\tilde{k}^2 
	\pm  \frac{16\pi^2\sigma_0 v_A^2 \xi_0^B}{(\varepsilon_0+p_0)}\lambda_{1_{0}}\tilde{k}^3 +\mathcal{O}(4) \label{36}
\end{gather} 
%
Here we have not considered higher order terms. 
Clearly the study of third term can provide important information about transport
coefficients $\lambda_{1_{0}}$.  In the discussion below, we have shown that one can estimate the order of this transport coefficient, by comparing it with the AdS/CFT results. 
\section{Results and Discussions}\label{sec-2result}
In this chapter we have discussed the transport  phenomenon of viscous parity odd chiral plasma.
We have shown that modes with wave number $k< \xi^B$ grow rapidly. This is known as Chiral Plasma Instability \cite{Akamatsu2013df}.
Modes with $k\sim \xi^B/2$ grows maximally and the growth rate is proportional to the $|\omega|_{max}\approx \pi\xi^{B^2}/2\sigma_0$.
In the case of turbulent plasma 
(for example at the time of EW phase transition, turbulence can be generated), the evolution of the fluid becomes more complicated and one can consider that the  system is in quasi-equilibrium. We did linear analysis of the viscous hydro with first and second order terms.
We analyse two cases where perturbation in the magnetic field is either normal to the background field or it is parallel to the background field. In the first case, when $\delta\tilde{B}\parallel \hat{{\bf n}}$, modes damp because of two dissipative phenomenon: finite conductivity and viscosity. However in the same  case, when $\delta\tilde{v}\parallel\tilde{k}$ in the absence of any viscous force, modes evolve with the sound speed. In the second case when $\delta\tilde{B}\perp \hat{{\bf n}}$, and there is no viscosity, modes move either with the speed of sound or with Alf\'ven speed \cite{Yamamoto2015} depending on the fluid flow direction with respect to the propagation vector. One case is especially very interesting, when $\delta\tilde{B}\perp \hat{{\bf n}}$, and $\tilde{k}\parallel \hat{{\bf n}}$, $\delta\tilde{v}\parallel \tilde{k}$. For the case of $\tilde{k}<1\Rightarrow k< \xi^B$, there will be instability in the plasma (see eq.(\ref{20}) for '+' sign). This kind of phenomenon occurs only in the chiral plasma.

So for we have discussed magnetohydrodynamics of chiral plasma considering only first order derivative expansion terms in the energy momentum tensor. And we have shown that our results are consistent with some other
works for chiral plasma \cite{Son2009, Akamatsu2013df}. Next we did the linear analysis of the second order hydrodynamics for the transverse modes. In eq.(\ref{36}), $\pm$ sign depends on the helicity of the fields. The third term is similar to the form obtained in AdS/CFT computations for N=4 super
Yang-Mills (SYM) with $U(1)_R$ symmetry \cite{Sahoo:2009yq}. First term is the oscillatory term, however higher order terms are clearly damping terms. In comparison to the first order hydro for the transverse motion, we have an additional term proportional to $k^3$. The chiral shear wave was first studied  in refs. \cite{Sahoo:2009yq, Baier2008} for strongly interacting plasma using AdS/CFT correspondence. In Ref. \cite{Baier2008}, authors have discussed a sound mode in a strongly interacting N=4 SYM theory, with $AdS_5\times S_5$ background, and the dispersion relations are
\begin{equation}
	\omega =\pm \frac{k}{\sqrt{3}}+\frac{i k^2}{6\pi T}\pm \frac{3-2 ln 2}{6\sqrt{3}(2\pi T)^2} k^3 +\mathcal{O}(k^4)
\end{equation}
As best values of the transport coefficients were calculated using the AdS/CFT correspondence, it is a good idea to calculate some of the transport coefficients by comparing our dispersion relation with the dispersion relations obtained by AdS/CFT correspondence. So comparing eq.(\ref{36}), one can see that for a neutral plasma
\begin{equation}
	\lambda_{1_{(0)}}\propto \frac{3-2 ln 2}{3\times 2^7 \pi^4 }\left(\frac{\eta_{(0)}}{T}\right)
\end{equation}
We have obtained this proportionality relation, using the condition that shear viscosity to entropy must be larger than or equals to $1/4\pi$. This is the estimates of one of the transport coefficients of the second order hydro. Clearly higher order contributions have very little impact on the shear motion of the fluid. We have seen in our calculations that it is very hard to calculate other transport coefficients simply by comparing two dispersion relations. So we have left it for our future work.

To conclude the chapter, we have shown using first and second order conformal hydrodynamics that our results are consistent with the previous results. In the case of second order conformal hydrodynamics, for the chiral plasma, we have shown that the transport coefficients related with the second order conformal hydrodynamics actually contribute to the dispersion relations. We also calculated one of the transport coefficients related with the second order conformal hydrodynamics by comparing our dispersion relation with the result obtained from AdS/CFT correspondence. The work discussed in this chapter is done using conformal hydro. It is interesting to look at non-conformal theories for the chiral plasma using first and second order hydrodynamics for a parity odd plasma. 
\cleardoublepage
\cleardoublepage
\newpage
\chapter{Summary}\label{ch6}
\section{Summary}
The main goal of this thesis is to study the generation and evolution of the Primordial Magnetic fields in presence of Abelian anomaly in the early Universe. In particular we have focused on the generation and dynamics of magnetic field in a parity violating chiral plasma, where there is finite chiral asymmetry in the number density of the left and right handed particles. Chapter wise summary of the thesis is given below
\\
{\bf Chapter \ref{ch1} :} 
In this chapter, we have briefly reviewed the previous work done on the generation and evolution of the magnetic fields. We have also discussed in brief about observations and current bounds. In the end of this chapter, we discussed our motivation behind this study.\\
{\bf Chapter \ref{ch2} :}
This chapter contains the theoretical foundation of our work. 
In the first half part of this chapter, we gave a very brief theory of the Kinetic theory of the classical plasma as well as relativistic plasma. In the later part of the chapter, kinetic theory modified with the Berry curvature has been given.\\
{\bf Chapter \ref{ch3}:} 
\vspace{0.2cm}
In this chapter, our aim is to discuss the generation of magnetic field above EW  scale in presence of Abelian anomaly, using kinetic theory, modified with Berry curvature. We have shown that the Abelian anomaly can be incorporated in the kinetic theory by considering a modified form of the kinetic equations with Berry curvature. We studied with this modification that, in presence of the chiral asymmetry there can be instability in the chiral plasma and which leads to the generation of turbulence and magnetic fields above a typical temperature $T\sim 80$ TeV in both collision dominated regime and collisionless regime.
\\
{\bf Chapter \ref{ch4}:} In this chapter, we have discussed the generation of magnetic field due to gravitational anomaly which induces a term $\propto T^2$ in the vorticity
current. The salient feature of this seed magnetic field is that it will be produced at high temperature irrespective of the whether fluid is charge or neutral. Once seed magnetic field is generated, we have studied its subsequent evolution in a plasma which have chiral asymmetry. \\
{\bf Chapter \ref{ch5} :} In this chapter we examine the evolution of the magnetic fields in an expanding fluid of matter and radiation with particular interest in the evolution of cosmic magnetic fields.  In the context of chiral plasma at high energies, there exist new kind of collective modes [{\it e.g.}, the chiral Alf\'ven waves] in the MHD limit and these modes exist in addition to the usual modes in the standard parity even plasma. Moreover, it has been shown that chiral plasmas exhibit a new type of density waves in presence of either of  an external magnetic field or vorticity and  they are respectively known as Chiral Magnetic Wave (CMW) and the Chiral Vortical Wave (CVW). In this regard we have investigated the collective modes in a magnetized chiral plasma and the damping mechanisms of these modes using first order and second hydrodynamics. Using first order conformal hydro, we obtained previously derived modes in the chiral plasma. However we show in addition that these modes get split into two modes in presence of the first order viscous term. By using second order conformal magnetohydrodynamics, we show that there are a series of terms in the dispersion relation and these terms are in accordance with the results obtained using AdS/CFT correspondence. We also calculated one of the transport coefficients related with the second order magnetohydrodynamics.  
\cleardoublepage
\phantomsection
\addcontentsline{toc}{chapter}{Bibliography}
\markboth{\MakeUppercase{Bibliography}}{\MakeUppercase{Bibliography}}
\bibliographystyle{jhep}
\bibliography{mythesisbib}
 \bibliographystyle{apsrev4-1}
\cleardoublepage
\phantomsection
\addcontentsline{toc}{chapter}{List of publications}
\markboth{\MakeUppercase{List of publications}}{\MakeUppercase{List of publications}}
\chapter*{List of Publications}
\begin{enumerate}
	\item Jitesh R. Bhatt and \textbf{Arun  Kumar Pandey},
	\textit{Primordial magnetic field and kinetic theory with Berry curvature},
	\href{doi:10.1103/PhysRevD.94.043536}{Phys. Rev. D {\bf 94}, no. 4, 043536 (2016),} [arXiv:1503.01878 (astro-ph.CO)]
	\item  Jitesh R. Bhatt and \textbf{Arun Kumar Pandey}, \textit{Primordial Generation of Magnetic Fields}, 	\href{doi:10.1007/978-3-319-25619-1_62}{Springer Proc.\ Phys.\  {\bf 174}, 409 (2016)}, [arXiv:1507.01795 [gr-qc]].
\end{enumerate}

~\\ {\Large \bf Other Publications (Under Review)}
\begin{enumerate}
	\item  \textbf{Arun Kumar Pandey}, \textit{Study of the collective behavior of chiral plasma using first and second order conformal magnetohydrodynamics}, 
	\href{https://arxiv.org/abs/1609.01848}{arXiv: 1609.01848 [hep-ph]}
	\item Sampurnad, Jitesh R. Bhatt, \textbf{ Arun Kumar Pandey}, \textit{Chiral Battery, scaling laws and magnetic fields}, 	\href{https://arxiv.org/abs/1705.03683}{arXiv: 1705.03683 [astro-ph.CO]}
\end{enumerate}
\cleardoublepage
  
\pagestyle{empty}
\phantomsection
\addcontentsline{7toc}{chapter}{Publications attached with thesis}
\chapter*{Publications attached with the thesis}
\begin{enumerate}
	\item \textbf{Arun Kumar Pandey}, Jitesh R. Bhatt, \textit{Primordial magnetic field and kinetic theory with Berry curvature}, Phys. Rev. D, 94, 043536, doi: \href{https://link.aps.org/doi/10.1103/PhysRevD.94.043536}{10.1103/PhysRevD.94.043536}.
	
	\item \textbf{Arun Kumar Pandey}, Jitesh R. Bhatt, \textit{Primordial Generation of Magnetic Fields}, XXI DAE-BRNS High Energy Physics Symposium: Proceedings, Guwahati, India, December 8-12, 2014, Springer Proceedings in Physics, doi: \href{http://dx.doi.org/10.1007/978-3-319-25619-1\_62}{10.1007/978-3-319-25619-1\_62}.
	
	\item \textbf{Arun Kumar Pandey}, Jitesh R. Bhatt, Sampurnand \textit{`Chiral Battery, scaling laws and magnetic fields}, JCAP {\bf 1707}, no. 07, 051 (2017)\\ 
	doi:\href{https://doi.org/10.1088/1475-7516/2017/07/051}{10.1088/1475-7516/2017/07/051}
\end{enumerate}
\cleardoublepage

\thispagestyle{empty}
\end{document}